\documentclass[11pt]{article}
\pdfoutput=1
\usepackage{jheppub}

\usepackage[...]{youngtab}
\usepackage{xfrac}
\usepackage{appendix}

\usepackage{amsfonts}
\usepackage{natbib}
\usepackage{physics}
\usepackage[usenames,dvipsnames]{xcolor}
\usepackage{tabularx,multirow,array,amsmath,mathalfa}
\usepackage{color,colortbl}
\usepackage{ccaption}
\usepackage{mathtools}
\usepackage{bbold}
\usepackage{kantlipsum}
\usepackage{slashbox}
\usepackage{pdflscape}
\usepackage{adjustbox}
\usepackage{longtable}
\usepackage{caption}
\definecolor{Gray}{gray}{0.9}

\allowdisplaybreaks

\usepackage{changepage}
\usepackage[framemethod=TikZ]{mdframed}
\usepackage{lipsum}
\usepackage{titlesec}
\mdfdefinestyle{sid}{%
    linecolor=black,
    outerlinewidth=1.0pt,
    roundcorner=7pt,
    innerrightmargin=15pt,
    innerleftmargin=15pt,
    backgroundcolor=gray!30
		}	
\definecolor{Red}{rgb}{1,0,0}
\usepackage{graphicx}
\makeatletter

\newcommand*\bigcdot{\mathpalette\bigcdot@{.5}}
\newcommand*\bigcdot@[2]{\mathbin{\vcenter{\hbox{\scalebox{#2}{$\m@th#1\bullet$}}}}}
\makeatother
\usepackage{pgf}
\usepackage{tikz}
\usepackage[utf8]{inputenc}
\usetikzlibrary{arrows,automata}
\usetikzlibrary{positioning}
\tikzset{state/.style={rectangle, rounded corners, draw=black, very thick, minimum height=2em, inner sep=2pt, text centered,},}

\title{\boldmath Semiclassical limit of topological Rényi entropy in $3d$ Chern-Simons theory}

\author[a]{Siddharth Dwivedi}
\author[b]{, Vivek Kumar Singh}
\author[c]{, Abhishek Roy}

\affiliation[a]{Center for Theoretical Physics, College of Physical Science and Technology, Sichuan University,\\
Chengdu, 610064, China}
\affiliation[b]{Department of Physics, Indian Institute of Technology Bombay, Powai, Mumbai 400076, India}
\affiliation[c]{Department of Physics, Indian Institute of Technology Jodhpur, Karwar, Jodhpur 342037, India}

\emailAdd{sdwivedi@scu.edu.cn, vivek.singh@fuw.edu.pl, roy.1@iitj.ac.in }

\abstract{We study the multi-boundary entanglement structure of the state associated with the torus link complement $S^3 \backslash T_{p,q}$ in the set-up of three-dimensional SU(2)$_k$ Chern-Simons theory. The focal point of this work is the asymptotic behavior of the R\'enyi entropies, including the entanglement entropy, in the semiclassical limit of $k \to \infty$. We present a detailed analysis for several torus links and observe that the entropies converge to a finite value in the semiclassical limit. We further propose that the large $k$ limiting value of the R\'enyi entropy of torus links of type $T_{p,pn}$ is the sum of two parts: (i) the universal part which is independent of $n$, and (ii) the non-universal or the linking part which explicitly depends on the linking number $n$. Using the analytic techniques, we show that the universal part comprises of Riemann zeta functions and can be written in terms of the partition functions of two-dimensional topological Yang-Mills theory. More precisely, it is equal to the R\'enyi entropy of certain states prepared in topological $2d$ Yang-Mills theory with SU(2) gauge group. Further, the universal parts appearing in the large $k$ limits of the entanglement entropy and the minimum R\'enyi entropy for torus links $T_{p,pn}$ can be interpreted in terms of the volume of the moduli space of flat connections on certain Riemann surfaces. We also analyze the R\'enyi entropies of $T_{p,pn}$ link in the double scaling limit of $k \to \infty$ and $n \to \infty$ and propose that the entropies converge in the double limit as well.}

\usepackage[normalem]{ulem}  % \sout{old text} for strikeout

\begin{document} 
			\maketitle
		\flushbottom

\section{Introduction}
\label{sec1}
\begin{figure}[htbp]
	\centering
		\includegraphics[width=0.70\textwidth]{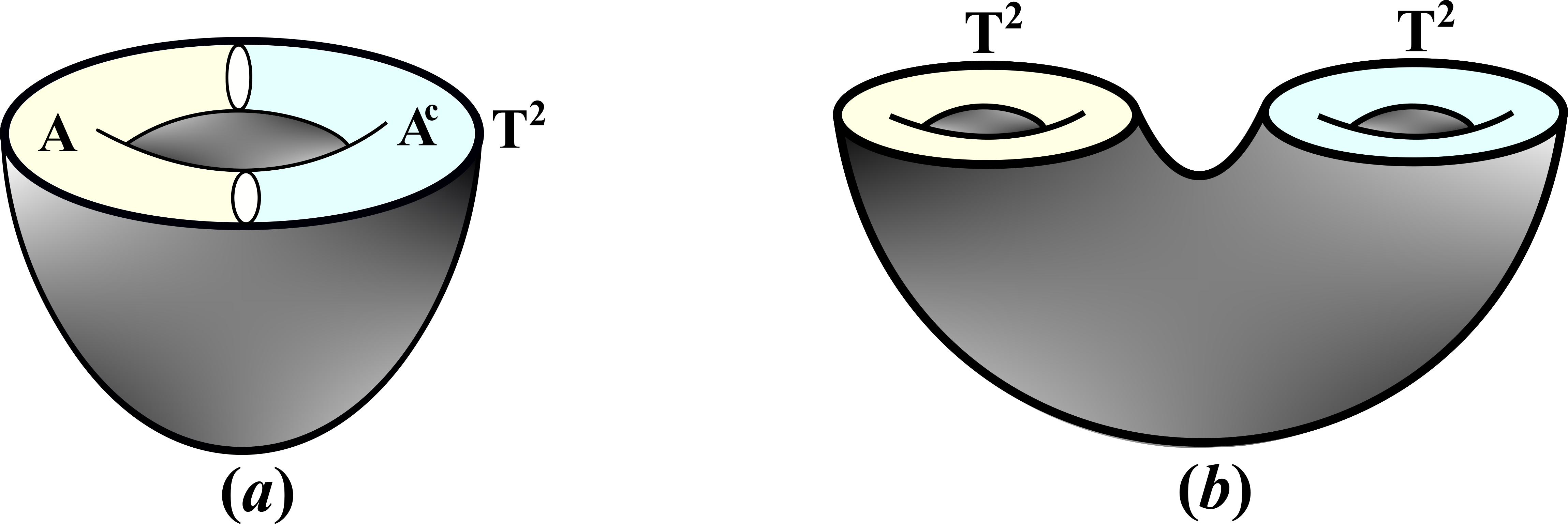}
	\caption{Two ways of computing topological entanglement entropy: the manifold in ($a$) has a single $T^2$ boundary, which is bi-partitioned into spatially connected regions $A$ and $A^c$. The manifold in ($b$) has two disjoint $T^2$ boundaries.}
	\label{T2partition}
\end{figure}

\emph{Quantum entanglement} \cite{Horodecki:2009zz} is one of the important fundamental aspects of the quantum world and is at the core of many revolutionary advances in fields of science and technology. In particular, it has a wide range of applications in quantum information science. It is generally an important question in quantum mechanics and quantum information theory to study the possible patterns of entanglement and its classification that can emerge in quantum field theory. Though analyzing the entanglement structures in a generic QFT is difficult, a simple case where this could be done is for `topological quantum field theories' (TQFT's). A TQFT is a class of field theory in which the physical observables and the correlation
functions do not depend on the spacetime metric. Thus the theory is not sensitive to
changes in the shape of spacetime, and the correlation functions are topological invariants.
Such theories do not have a local dynamics; therefore, all of the entanglement arises from
the topological properties of the underlying manifolds. For example, consider the manifold shown in the figure \ref{T2partition}($a$) having a torus boundary, which is bi-partitioned into spatially connected section $A$ and its complement $A^c$.  
When we trace out the region $A^c$, we are essentially calculating the `topological entanglement entropy', which is independent of the length or area of the region $A$ or $A^c$. Such entropies have been studied in \cite{Kitaev:2005dm,Levin:2006zz,Dong:2008ft} in the context of 2+1 dimensional Chern-Simons theory, which is the best understood TQFT \cite{Witten:1988hf}. Another approach to study the topological entanglement was given in \cite{Balasubramanian:2016sro}, where the authors computed the entanglement entropy of the states obtained by the path integral on a link complement $S^3 \backslash \mathcal{L}$. In this case, we have two or more disjoint torus boundaries similar to the one shown in the figure \ref{T2partition}($b$). The topological entanglement structure, in this case, can be obtained by tracing out one of the boundary components, which is often termed as `multi-boundary entanglement'. We refer the interested readers to \cite{Balasubramanian:2016sro,Dwivedi:2017rnj,Balasubramanian:2018por,Hung:2018rhg,Melnikov:2018zfn,Camilo:2019bbl,Dwivedi:2019bzh,Buican:2019evc,Dwivedi:2020jyx} for the recent developments in this study. 

An important aspect of Chern-Simons theory which is the motivation behind the multi-boundary entanglement is the following decomposition of the Hilbert space associated with a boundary with multiple disjoint components:
\begin{equation}
\mathcal{H}_{\partial M} = \mathcal{H}_{\Sigma_1} \otimes \mathcal{H}_{\Sigma_2} \otimes \ldots \otimes \mathcal{H}_{\Sigma_n} ~,
\end{equation}
where the boundary of the manifold $M$ consists of disjoint components: $\partial M = \Sigma_1 \sqcup \Sigma_2 \sqcup \ldots \sqcup \Sigma_n$ and $\mathcal{H}_{\Sigma_i}$ denotes the Hilbert space associated with the $i^{\text{th}}$ component. The Chern-Simons path integral on such a manifold associates a quantum state $\ket{\Psi}$ to $M$ which lives in the Hilbert space $\mathcal{H}_{\partial M}$. The entanglement properties of $\ket{\Psi}$ can be studied by tracing out a subset of the Hilbert spaces. It was shown in \cite{Balasubramanian:2016sro} that when $M$ is a link complement\footnote{If a link $\mathcal{L}$  is embedded in $S^3$, then the link complement is a three-dimensional manifold which is obtained by removing a tubular neighborhood around $\mathcal{L}$ from $S^3$, i.e $S^3 \backslash \mathcal{L} \equiv S^3 - \text{interior}(\mathcal{L}_{\text{tub}})$.} $S^3 \backslash \mathcal{L}$, the probability amplitudes of the associated state $\ket{\mathcal{L}}$ are precisely the quantum link invariants or the partition functions of $S^3$ in the presence of link $\mathcal{L}$.

In this work, we will study the Rényi entropies, including the entanglement entropy, for the states associated with generic torus link complements $S^3 \backslash T_{p,q}$. These manifolds have `gcd($p,q$)' number of disjoint torus boundaries, and thus the total Hilbert space will be a tensor product of `gcd($p,q$)' copies of $\mathcal{H}_{T^2}$ (the Hilbert space associated with a torus). In particular, we focus on the Chern-Simons theory with SU(2)$_k$ gauge group and analyze the behavior of the Rényi entropies in the semiclassical limit of $k \to \infty$. 

The paper is organized as follows. In section \ref{sec2}, we discuss our set-up, giving a brief review of the study of the multi-boundary entanglement in the Chern-Simons theory and the computation of link states using the topological machinery. In section \ref{sec3}, we discuss the computation of the torus link state for a generic torus link $T_{p,q}$. We use an analytical approach for links of type $T_{p,pn}$ to obtain the explicit expressions for Rényi entropies as a function of $k$. In section \ref{sec4}, we analyze the large $k$ behavior of the Rényi entropies of torus links and study their limiting values as $k \to \infty$. In section \ref{sec5}, we discuss the connection between the large $k$ limit of Rényi entropies of $T_{p,pn}$ links with the Rényi entropies of certain states in topological $2d$ Yang-Mills theory. We finally conclude in section \ref{sec6}.    
%............................................
%............................................
%...................SECTION ends..............
%...................SECTION ends..............
\section{The set-up}
\label{sec2}
Here we will briefly review the multi-boundary entanglement set-up in the Chern-Simons theory. In particular, we discuss how the quantum states can be explicitly computed in certain cases by exploiting the topological property of the Chern-Simons theory using the tools of surgery. We will start with a brief review of the Chern-Simons theory, which will set the tone of this section and fix the notations used in the later sections. 
%............................................
%............................................
\subsection{Chern-Simons theory: A brief review}
The (2+1) dimensional Chern-Simons theory with compact simple gauge group $G$ is defined on a compact 3-manifold $M$ whose action is defined as the integral of the Chern-Simons three-form over the  3-manifold $M$:
\begin{equation}
S(A) = \frac{k}{4 \pi} \int_M \text{Tr}\left(A \wedge dA + \frac{2}{3} A \wedge A \wedge A \right) ~,
\end{equation}
where $A$ denotes the connection $A_i^a$ (Lie algebra valued 1-form with index $a$ running over the basis of the Lie algebra, and $i$ is tangent to $M$) placed on a trivial $G$-bundle over $M$. The variable $k$ is a positive integer that classifies the action $S(A)$. The gauge invariant observables in this theory are `Wilson lines' which are defined as the following functional of the connection $A_i$. Pick an oriented knot $\mathcal{K}$ in $M$ and compute the holonomy (the path ordered exponential) of $A_i$ around $\mathcal{K}$. This will give an element of the gauge group $G$ that is well-defined up to conjugacy. The trace of this element in a particular irreducible representation $R$ of $G$ is defined as the Wilson line:
\begin{equation}
W_{R}(\mathcal{K}) = \text{Tr}_R \,P\exp\left(\oint_{\mathcal{K}} A_i\, dx^i \right) ~.
\end{equation} 
The Feynman path integral is carried out by integrating the exponentiated Chern-Simons action over the space of connections. When $M$ is closed (i.e. without boundary), this path integral will give a complex number which we call as the partition function of $M$ and is given as
\begin{equation}
Z(M) = \int D\mathcal{A} \, e^{i S(A)} ~,
\label{ZM-closed}
\end{equation}  
where $D\mathcal{A}$ represents the integral over all the equivalence classes of connections modulo the gauge transformations. Since the connection has been integrated out, $Z(M)$ will be a topological invariant of $M$.\footnote{Note that $Z(M)$ will also depend on the framing of $M$ as discussed in \cite{Witten:1988hf}. However, as mentioned in \cite{Balasubramanian:2016sro}, these framing factors act as local unitary transformations on the basis states of the Hilbert space and will not affect the entanglement structure. Hence we can safely ignore the framing factors in our computation.} We can further modify this definition to include oriented knots or links in $S^3$. Consider a link $\mathcal{L}$ made of disjoint oriented knot components, i.e. $\mathcal{L} = \mathcal{K}_1 \sqcup \mathcal{K}_2 \sqcup \ldots \sqcup \mathcal{K}_n$, embedded in $S^3$. The path integral in this case can be computed as:
\begin{equation}
Z(M;\mathcal{L}) = \int D\mathcal{A} \, e^{i S(A)}\left(\prod_{i=1}^n W_{R_i}(\mathcal{K}_i)\right) ~,
\label{ZML-closed}
\end{equation} 
where $R_i$ is the irrep associated with the knot $\mathcal{K}_i$. Note that when $M$ does not have any boundary, the partition functions $Z(M)$ and $Z(M;\mathcal{L})$ are complex numbers.

\begin{figure}[htbp]
	\centering
		\includegraphics[width=1.00\textwidth]{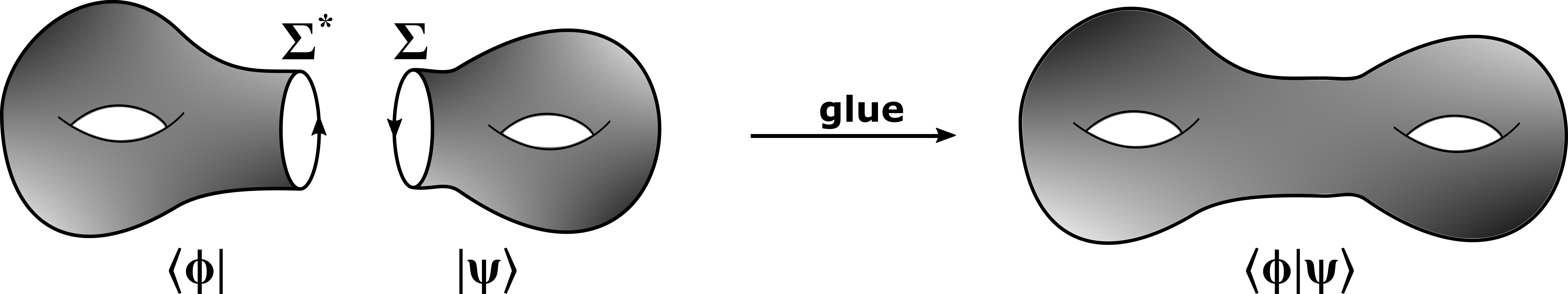}
	\caption{Two manifolds on left with same boundary but opposite orientation. The path integral on these manifolds gives states $\bra{\phi} \in \mathcal{H}_{\Sigma^*}$ and $\ket{\psi} \in \mathcal{H}_{\Sigma}$. The inner product $\bra{\phi}\ket{\psi}$ will be the partition function of manifold shown in right obtained by gluing the two manifolds along the common boundary.}
	\label{gluing}
\end{figure}

Next consider the case when $M$ has a boundary $\Sigma$ where $\Sigma$ is an oriented closed Riemann surface. In such cases, the path integrals defined in \eqref{ZM-closed} or \eqref{ZML-closed} are vectors or states (which we shall denote as $\ket{\Psi}$) and are elements of the Hilbert space $\mathcal{H}_{\Sigma}$ associated with $\Sigma$. From the topological point of view, $\ket{\Psi}$ only depends on the topology of the manifold $M$ and $\mathcal{H}_{\Sigma}$ only depends on the topology of the boundary $\Sigma$. If we consider two topologically different manifolds $M$ and $M'$ with the same boundary $\Sigma$, then the corresponding Chern-Simons partition functions will give two different states $\ket{\Psi}$ and $\ket{\Psi'}$ in the same Hilbert space $\mathcal{H}_{\Sigma}$. Thus, given any 3-manifold, we can associate a quantum state to it. Further note that if we reverse the orientation of $\Sigma$ to get $\Sigma^*$, the associated Hilbert space with $\Sigma^*$ will be dual to that of $\mathcal{H}_{\Sigma}$: 
\begin{equation}
\mathcal{H}_{\Sigma^*} = \mathcal{H}^*_{\Sigma} ~.
\end{equation} 
As a result, there exists a natural pairing, the inner product $\langle \Phi| \Psi \rangle$ for any two states $\ket{\Psi} \in \mathcal{H}_{\Sigma}$ and $\bra{\Phi} \in \mathcal{H}_{\Sigma^*}$. In fact, this technique can be used to compute the partition functions of complicated manifolds by gluing two disconnected pieces along common boundary whose partition functions are already known. This process is shown in the figure \ref{gluing}.

If the boundary of the manifold $M$ has multiple disconnected components, then the total Hilbert space will be a tensor product of the Hilbert spaces associated with each of the boundary components: 
\begin{equation}
\partial M = \Sigma_1 \sqcup \Sigma_2 \sqcup \ldots \sqcup \Sigma_n \,\, \Longrightarrow \,\, \mathcal{H}_{\partial M} = \mathcal{H}_{\Sigma_1} \otimes \mathcal{H}_{\Sigma_2} \otimes \ldots \otimes \mathcal{H}_{\Sigma_n} ~.
\end{equation} 
The path integral for such a manifold will associate a quantum state $\ket{\Psi} \in \mathcal{H}_{\partial M}$. Since the Hilbert space is automatically a tensor product space, we can assign an entanglement structure to the states by tracing out a subset of the Hilbert spaces associated with some of the boundary components. For the present work, we consider $M$ to be a link complement $S^3\backslash \mathcal{L}$. In the following, we will discuss how the topological property of Chern-Simons theory enables one to compute the quantum states associated with link complements by using the surgery (cutting and gluing) techniques.   
%............................................
%............................................
\subsection{Computation of link state associated with link complement}
\begin{figure}[htbp]
\centerline{\includegraphics[width=5.5in]{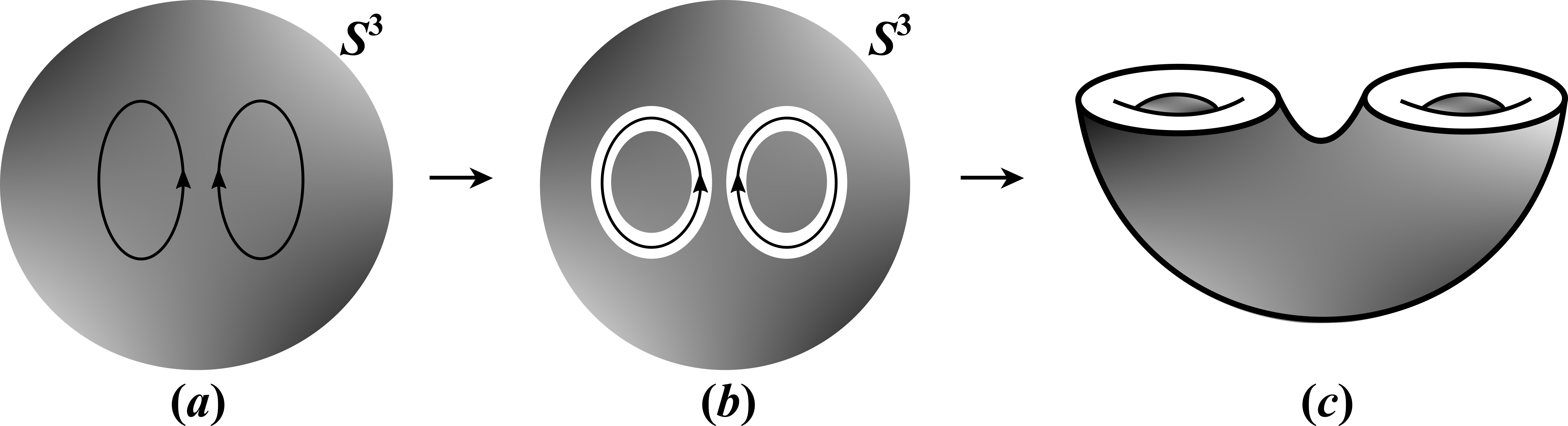}}
\caption[]{Figure showing the construction of link complement manifold $S^3 \backslash T_{0,2}$.}
\label{LinkComplement}
\end{figure}

As a simple example, consider the trivial two component torus link $T_{0,2}$ embedded in $S^3$ as shown in the figure \ref{LinkComplement}($a$). Take a tubular neighborhood around each of the component as given in the figure \ref{LinkComplement}($b$) and remove it from $S^3$. The resulting 3-manifold is the link complement $S^3 \backslash T_{0,2}$ which has two disconnected $T^2$ boundaries as shown in the figure \ref{LinkComplement}($c$).
The Chern-Simons path integral on $S^3 \backslash T_{0,2}$ will give a pure state $\ket{T_{0,2}}\in \mathcal{H}_{T^2} \otimes \mathcal{H}_{T^2}$. To compute this state, first we need to fix the basis of $\mathcal{H}_{T^2}$. As discussed in \cite{Witten:1988hf}, the basis of $\mathcal{H}_{T^2}$ is in one-to-one correspondence with the integrable representations of the affine Lie algebra $\hat{\mathfrak{g}}_k$ at level $k$, where $\mathfrak{g}$ represents the Lie algebra associated with the group $G$. These basis states have a path integral description. Consider a solid torus with a Wilson line carrying an integrable representation $R$ placed in the bulk of a solid torus along its non-contractible cycle. The Chern-Simons path integral on this solid torus will give the basis state $\ket{e_R}$:
\begin{equation}
\begin{array}{c}
\includegraphics[width=0.25\linewidth]{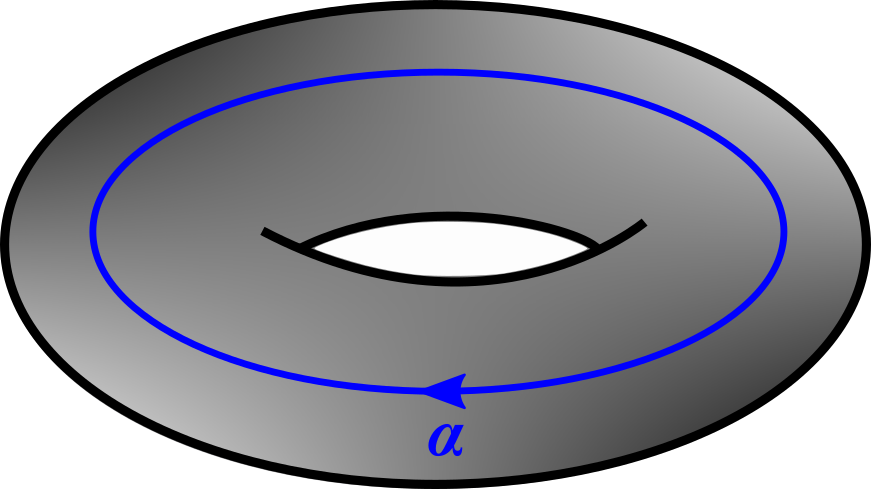}
\end{array} = \ket{e_R} ~.
\end{equation}
The collection of all such states, where $R$ runs over the integrable representations, will define an orthonormal basis of $\mathcal{H}_{T^2}$:
\begin{equation}
\text{basis}(\mathcal{H}_{T^2}) = \{\ket{e_R} : \text{$R$ is an integrable representation} \}~.
\end{equation}
When $G$ is a compact simple gauge group, the Hilbert space $\mathcal{H}_{T^2}$ is finite dimensional. For example, for SU(2) group and level $k$, the integrable representations are those representations of SU(2) whose highest weights are given by the Dynkin labels $[0], [1], \ldots, [k]$. We will label the representation $[a]$ of SU(2) simply by `$a$' where:
\begin{equation}
a =  \underbrace{\yng(4)}_{a} ~.
\end{equation}
Thus the Hilbert space is spanned by the following basis:
\begin{equation}
\text{SU(2)}_k: \quad \text{basis}(\mathcal{H}_{T^2}) = \{\ket{e_0}, \ket{e_1}, \ldots, \ket{e_k}\}~.
\end{equation}
Coming back to the state $\ket{T_{0,2}}$ shown in the figure \ref{LinkComplement}($c$), it can be expanded in the basis of $\mathcal{H}_{T^2} \otimes \mathcal{H}_{T^2}$ as following:
\begin{equation}
\ket{T_{0,2}} = \sum_{a,\,b} C_{ab}\ket{e_a} \otimes \ket{e_b} \equiv \sum_{a,\,b} C_{ab}\ket{e_a, e_b} ~,
\end{equation}
where $C_{ab}$ are complex coefficients which can be fixed as following. Take two solid tori with Wilson lines in their bulk carrying representations $\alpha$ and $\beta$ and glue their boundaries with the two $T^2$ boundaries of $S^3 \backslash T_{0,2}$ manifold. This topological gluing is equivalent to taking inner product of state $\bra{e_{\alpha}, e_{\beta}}$ with $\ket{T_{0,2}}$. Moreover the gluing will give us back the 3-manifold shown in the figure \ref{LinkComplement}($a$) which is $S^3$ with $T_{0,2}$ link embedded in it whose components carry representations $\alpha$ and $\beta$ respectively. Thus we can calculate the value of the inner product $\bra{e_{\alpha}, e_{\beta}}\ket{T_{0,2}}$ as:
\begin{equation}
\bra{e_{\alpha}, e_{\beta}}\ket{T_{0,2}} = \begin{array}{c}
\includegraphics[width=0.45\linewidth]{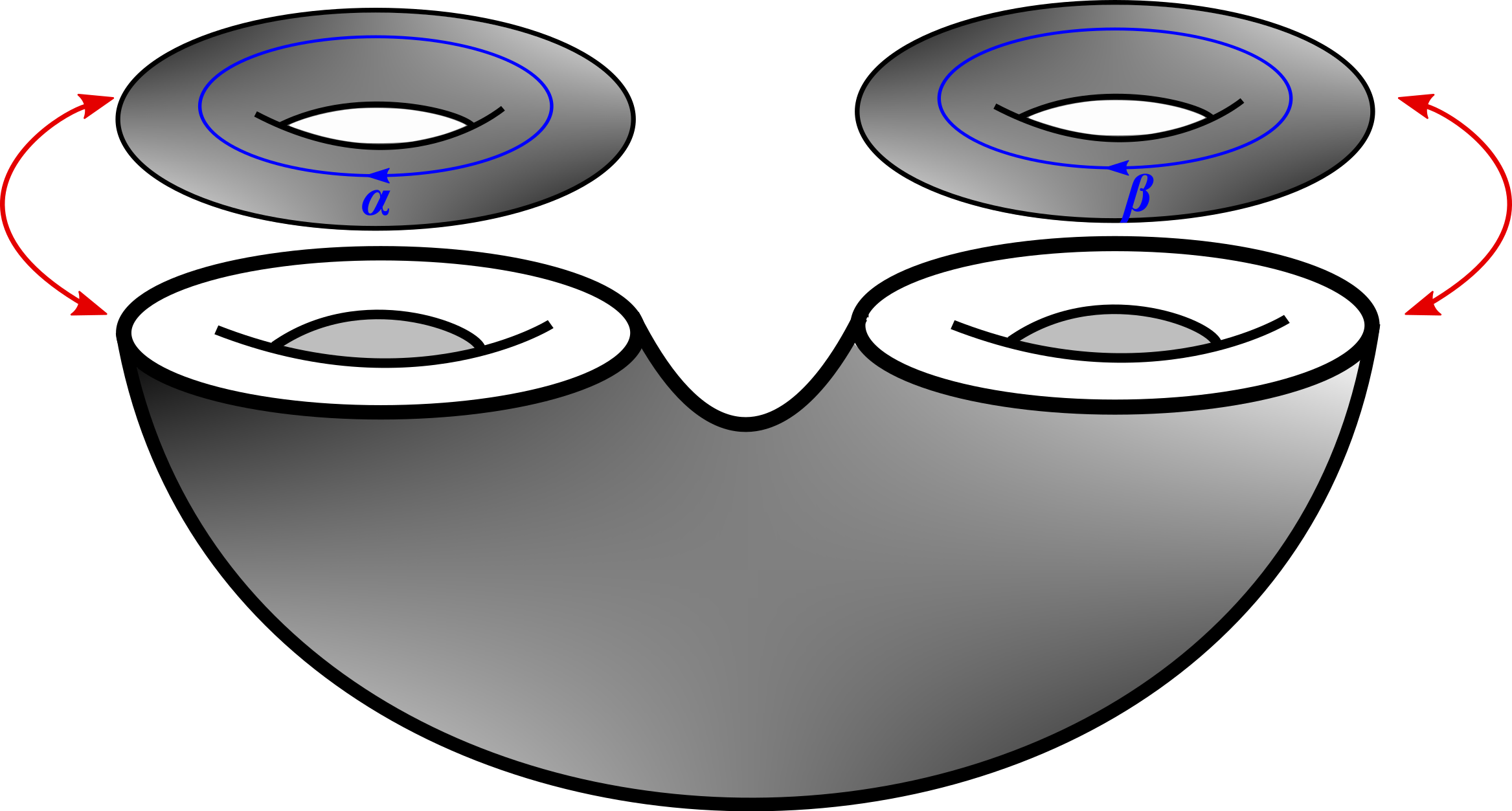}
\end{array} = Z(S^3;\,T_{0,2}[\alpha, \beta]) ~.
\end{equation}
Using the orthonormal property of the basis states, we can fix the coefficients $C_{ab}$ and hence the state is given as,
\begin{equation}
\ket{T_{0,2}} = \sum_{a,\,b} Z(S^3;\,T_{0,2}[a, b]) \ket{e_a, e_b} ~.
\end{equation}
This procedure can be generalized to any link complement $S^3 \backslash \mathcal{L}$. If $\mathcal{L}$ is an $n$-component link, the above steps will give the following link state:
\begin{equation}
\ket{\mathcal{L}} = \sum_{a_1,\ldots,a_d} Z(S^3;\,\mathcal{L}[a_1,\ldots,a_d])\, \ket{e_{a_1},\ldots,e_{a_d}} ~,
\label{linkstate}
\end{equation}
where $Z(S^3;\,\mathcal{L}[a_1,\ldots,a_d])$ is the partition function of $S^3$ in presence of link $\mathcal{L}$ whose components carry representations $a_1,\ldots, a_d$. In the following we will discuss how to study the entanglement properties of this multi-party state.
%............................................
%............................................
\subsection{Entanglement measures associated with link states}
Consider the link state $\ket{\mathcal{L}}$ given in \eqref{linkstate} defined on $d$ copies of $\mathcal{H}_{T^2}$. Let us bi-partition the total Hilbert space as 
\begin{equation}
\mathcal{H} = \underbrace{\mathcal{H}_{T^2} \otimes \mathcal{H}_{T^2} \otimes \ldots \otimes \mathcal{H}_{T^2}}_{x} \,\,\otimes\,\, \underbrace{\mathcal{H}_{T^2} \otimes \mathcal{H}_{T^2} \otimes \ldots \otimes \mathcal{H}_{T^2}}_{d-x} \equiv \mathcal{H}_{A} \otimes \mathcal{H}_{B} ~.
\end{equation}
Given the state $\ket{\mathcal{L}}$, we can associate a projection operator, called as density matrix operator, which acts on $\mathcal{H}_A \otimes \mathcal{H}_B$ given as:
\begin{equation}
\rho_{\text{total}} = \frac{\ket{\mathcal{L}} \bra{\mathcal{L}}}{\braket{\mathcal{L}}} ~,
\end{equation}
where $\braket{\mathcal{L}}$ is the normalization factor that ensures that the density matrix has unit trace. The dual state can be obtained as:
\begin{equation}
\bra{\mathcal{L}} = \sum_{a_1,\ldots,a_d} Z(S^3;\,\mathcal{L}[a_1,\ldots,a_d])^{*} \bra{e_{a_1},\ldots,e_{a_d}} ~.
\end{equation} 
In order to compute the entanglement measures, we must trace out $\mathcal{H}_{B}$ and obtain the reduced density matrix $\rho_A$ acting on $\mathcal{H}_{A}$:
\begin{equation}
\rho_A  = \text{Tr}_{\mathcal{H}_B}(\rho_{\text{total}}) = \sum_{a_{x+1},\ldots, a_{d} } \expval{\rho_{\text{total}}}{a_{x+1},\, a_{x+2},\, \ldots,\, a_{d}} ~.
\end{equation}
From the spectrum $\{\lambda_i \}$ of the reduced density matrix, one can calculate various entanglement measures. For example, the $m^{\text{th}}$ Rényi entropy is given as:
\begin{equation}
\mathcal{R}_m = \frac{1}{1-m}\ln \text{Tr}[(\rho_A)^m] = \frac{1}{1-m} \ln\left(\sum_{i} \lambda_i^m \right) ~.
\end{equation}
The entanglement entropy is given as,
\begin{equation}
\mathcal{E} = -\text{Tr}[\rho_A \ln \rho_A] = \lim_{m \to 1} \mathcal{R}_m \equiv \mathcal{R}_1 ~.
\label{EEm1limit}
\end{equation}
There also exits a minimum entropy which is controlled by the maximum eigenvalue associated with the reduced density matrix and is given as the following limit of the Rényi entropy:
\begin{equation}
\mathcal{R}_{\text{min}} \equiv \mathcal{R}_{\infty} = \lim_{m \to \infty} \mathcal{R}_m ~.
\label{REminfylimit}
\end{equation}
Having discussed the preliminaries and the basic set-up, let us move on to the computation of the entanglement structure of torus link states.
%............................................
%............................................
%...................SECTION ends..............
%...................SECTION ends..............
\section{Entanglement structure of torus link state}
\label{sec3}
In this section, we will study the entanglement structure of the state associated with torus link complement $S^3 \backslash T_{p,q}$. A torus link $T_{p,q}$ is made up of $d = \text{gcd}(p,q)$ number of following torus knots: 
\begin{equation}
T_{p,q} = \bigsqcup_{i=1}^d T_{\frac{p}{d},\frac{q}{d}} ~.
\end{equation}
They are called torus links because they can be drawn on the surface of a torus without self intersections. Each component of this link wraps around the two 1-cycles of the torus with winding numbers $p/d$ and $q/d$ respectively. The linking number between any two components of this link is:
\begin{equation}
n_{l} = \frac{pq}{d^2} ~.
\end{equation}
Using the steps shown earlier, we can obtain the state associated with $S^3 \backslash T_{p,q}$. Restricting to the SU(2) Chern-Simons theory with level $k$, we can write the state as: 
\begin{equation}
\ket{T_{p,q}} = \sum_{a_1=0}^k \ldots \sum_{a_d=0}^k Z(S^3;\,T_{p,q}[a_1, a_2,\ldots, a_d])\, \ket{e_{a_1},e_{a_2},\ldots,e_{a_d}} ~,
\label{state}
\end{equation}
where $Z(S^3;\,T_{p,q}[a_1, a_2,\ldots, a_d])$ is the partition function of $S^3$ in presence of the link $T_{p,q}$ whose components carry SU(2) representations $a_1, \ldots, a_d$. This partition function has been computed in \cite{Stevan:2010jh,Brini:2011wi} and can be given in terms of the unitary representations $\mathcal{S}$ and $\mathcal{T}$ of SL(2,$\mathbb{Z}$) as: 
\begin{equation}
Z(S^3;\,T_{p,q}[a_1, a_2,\ldots, a_d]) = \sum_{\alpha, \beta, \gamma=0}^k  \frac{\mathcal{S}_{\alpha \beta}^* \,\mathcal{S}_{0 \gamma}}{(\mathcal{S}_{0 \alpha})^{d-1}}\, (\mathcal{T}_{\gamma \gamma})^{q/p} \left(\prod_{i=1}^d \mathcal{S}_{a_i \alpha}\right)  X_{\beta\gamma}(p/d) ~,
\label{Z-torus}
\end{equation}
where $\mathcal{S}$ and $\mathcal{T}$ are unitary operators whose matrix elements in the $\mathcal{H}_{T^2}$ basis are given as:
\begin{align}
\mathcal{S}_{ab} &= \langle e_a|\mathcal{S}| e_b \rangle = \sqrt{\frac{2}{k+2}}\, \sin(\frac{(a+1)(b+1)\pi}{k+2}) \nonumber \\ 
\mathcal{T}_{ab} &= \langle e_a|\mathcal{T}| e_b \rangle = \exp(\frac{\pi i\, a(a+2)}{2k+4})\exp(-\frac{\pi i k}{4k+8})\delta_{ab} ~.
\end{align}
The coefficients $X_{\beta \gamma}(y)$ are unique integers which occur in the expansion of the traces of powers of holonomy operator $U$ as:
\begin{equation}
\text{Tr}_{\beta}(U^y) = \sum_{\gamma} X_{\beta \gamma}(y)\, \text{Tr}_{\gamma}(U) ~.
\end{equation}
These coefficients can be obtained by performing the Adams operation on the characters associated with the irreducible representations of the gauge group $G$ (see \cite{Stevan:2014xma} and references therein). For the SU(2) group, the character associated with the representation $\beta$ is given as:
\begin{equation}
\quad \chi_{\beta}(t) = \frac{t^{\beta+2}-t^{-\beta}}{t^2-1} ~,
\end{equation}
where the character is evaluated at variable $t$ which is the fundamental weight fugacity of SU(2). The Adams operation on character is denoted as $\mathcal{A}^y \chi_{\beta}$ and is defined as:
\begin{equation}
(\mathcal{A}^y \chi_{\beta})(t) = \chi_{\beta}(t^y) ~.
\end{equation}
The result of this operation can be expanded in the basis of characters as,
\begin{equation}
(\mathcal{A}^y \chi_{\beta})(t) = \chi_{\beta}(t^y) = \sum_{\gamma} X_{\beta \gamma}(y)\, \chi_{\gamma}(t) ~,
\end{equation}
where $X_{\beta \gamma}(y)$ are unique integers and are precisely the coefficients appearing in \eqref{Z-torus}. The explicit expression of these Adams coefficients for the group SU(2) has been presented in the appendix \ref{appA}. Thus a generic torus link state can be obtained by substituting the partition function of \eqref{Z-torus} into the \eqref{state}. Since the entanglement properties of a state do not change under a local unitary transformation of the basis states, we can further rewrite the state by changing the $i^{\text{th}}$ basis state as: $\ket{e_{a_i}} = \sum_{y_i} \mathcal{S}_{a_i y_i}^{*}\ket{e_{y_i}}$. In this basis, the state can be written as
\begin{equation}
\ket{T_{p,q}} = \sum_{a_1,\ldots,a_d} \, \sum_{y_1,\ldots,y_d} \left(\prod_{i=1}^d \mathcal{S}_{a_i y_i}^*\right) Z(S^3;\,T_{p,q}[a_1,\ldots, a_d]) \ket{e_{y_1},\ldots,e_{y_d}} ~.
\end{equation}
Using the symmetric and unitary property of $\mathcal{S}$ matrix, the above state can be simplified as,
\begin{equation}
\ket{T_{p,q}} =  \sum_{\alpha=0}^k \sum_{\beta=0}^k \sum_{\gamma=0}^k \frac{\mathcal{S}_{\alpha \beta}^* \,\mathcal{S}_{0 \gamma}}{(\mathcal{S}_{0 \alpha})^{d-1}}\, (\mathcal{T}_{\gamma \gamma})^{q/p} \,  X_{\beta \gamma}(p/d) \, \ket{e_{\alpha},e_{\alpha},\ldots,e_{\alpha}} ~.
\end{equation}
We can express this state in a more compact form as following. Collect all the Adams coefficients $X_{\beta \gamma}(y)$ into a matrix $X(y)$ whose rows and columns are labeled by the SU(2) representations $\beta$ and $\gamma$ respectively where $0 \leq \beta, \gamma \leq k$. Thus the state can be written as:
\begin{equation}
\ket{T_{p,q}} =  \sum_{\alpha=0}^k  \dfrac{\left(\mathcal{S}^*X(p/d)\mathcal{T}^{\frac{q}{p}}\mathcal{S}\right)_{\alpha 0}}{(\mathcal{S}_{0 \alpha})^{d-1}}\, \ket{e_{\alpha},e_{\alpha},\ldots,e_{\alpha}} ~.
\end{equation}
We can bi-partition the total Hilbert space into two Hilbert spaces $\mathcal{H}_A$ and $\mathcal{H}_B$, where $\mathcal{H}_A$ is a tensor product of $d_1$ Hilbert spaces and $\mathcal{H}_B$ is the tensor product of remaining $(d-d_1)$ Hilbert spaces. Tracing out $\mathcal{H}_B$ will give the reduced density matrix $\rho$ acting on $\mathcal{H}_A$ which will be a diagonal matrix of order $=\text{dim}\,\mathcal{H}_A = (k+1)^{d_1}$. It has only $(k+1)$ number of non-vanishing eigenvalues and are given as,
\begin{equation}
\boxed{\lambda_{\alpha} = \frac{1}{\braket{T_{p,q}}} (\mathcal{S}_{0 \alpha})^{2-2d} \left(\mathcal{S}^*X(p/d)\mathcal{T}^{\frac{q}{p}}\mathcal{S}\right)_{\alpha 0} \, \left(\mathcal{S}^*X(p/d)\mathcal{T}^{\frac{q}{p}}\mathcal{S}\right)_{\alpha 0}^{*}} ~,
\label{eigen-toruslink}
\end{equation}
where $\alpha=0,1,\ldots,k$. The factor $\braket{T_{p,q}}$ is the normalization factor given as,
\begin{equation}
\braket{T_{p,q}} =  \sum_{\alpha=0}^k \, (\mathcal{S}_{0 \alpha})^{2-2d} \left(\mathcal{S}^*X(p/d)\mathcal{T}^{\frac{q}{p}}\mathcal{S}\right)_{\alpha 0} \, \left(\mathcal{S}^*X(p/d)\mathcal{T}^{\frac{q}{p}}\mathcal{S}\right)_{\alpha 0}^{*} ~.
\end{equation}
Once we have obtained the spectrum of the reduced density matrix, the $m^{\text{th}}$ Rényi entropy can be computed as,
\begin{equation}
\mathcal{R}_m = \frac{\ln \text{Tr}[\rho^{m}]}{1-m}  ~.
\end{equation}
In our analysis, we will bypass the computation of the trace factor $\braket{T_{p,q}}$ by defining the unnormalized eigenvalues as
\begin{equation}
\Lambda_{\alpha} = (\mathcal{S}_{0 \alpha})^{2-2d} \left(\mathcal{S}^*X(p/d)\mathcal{T}^{\frac{q}{p}}\mathcal{S}\right)_{\alpha 0} \, \left(\mathcal{S}^*X(p/d)\mathcal{T}^{\frac{q}{p}}\mathcal{S}\right)_{\alpha 0}^{*}  ~,
\end{equation}
where $\Lambda_{\alpha}$ are the eigenvalues of the unnormalized reduced density matrix $\sigma$. The Rényi entropies in terms of $\sigma$ are given as,
\begin{equation}
\mathcal{R}_m = \frac{1}{1-m} \ln(\frac{\text{Tr}[\sigma^{m}]}{\text{Tr}[\sigma]^m}) = \frac{1}{1-m} \ln(\frac{\Lambda_0^m+\Lambda_1^m+\ldots+\Lambda_k^m}{(\Lambda_0+\Lambda_1+\ldots+\Lambda_k)^m}) ~.
\end{equation}
The two special limits of the Rényi entropy are entanglement entropy and minimum entropy, which can be obtained by taking the $m \to 1$ and $m\to \infty$ limits of the $m^{\text{th}}$ Rényi entropy as mentioned in \eqref{EEm1limit} and \eqref{REminfylimit} respectively. 

A special class of torus links is the family of links of type $T_{p,pn}$ with $p \geq 2$ and $n \geq 1$. A $T_{p,pn}$ link is made up of $p$ number of framed circles $T_{1,n}$ such that the linking number between any pair of framed circles is $n$. A framed circle is an unknot whose planar projection has self-intersection number or writhe equal to $n$, which is also called its framing number. The Chern-Simons partition functions are sensitive to such framing factors, however the entanglement structure in our set-up does not depend on the framing of individual knots (see footnote 2). The $n$-dependence which enters in the computation of the Rényi entropies is due to its global topological feature of being the linking number between two components of the link. In Table \ref{Tppnlinkstable}, we have presented the link diagrams of some of the torus links of type $T_{p,pn}$ where each circle is shown in their standard framing (i.e $T_{1,0}$). We also show the way they can be drawn on the surface of a torus without self intersections.  
\begin{table}
\begin{center}
 %\begin{equation}
$\begin{array}{|c|c|c||c|c|c|} \hline
\rowcolor{Gray}
\text{Link} & \text{Diagram} & \text{drawn on $T^2$} & \text{Link} & \text{Diagram} & \text{drawn on $T^2$} \\ \hline
T_{2,2} & {\begin{array}{c}
\includegraphics[width=0.15\linewidth]{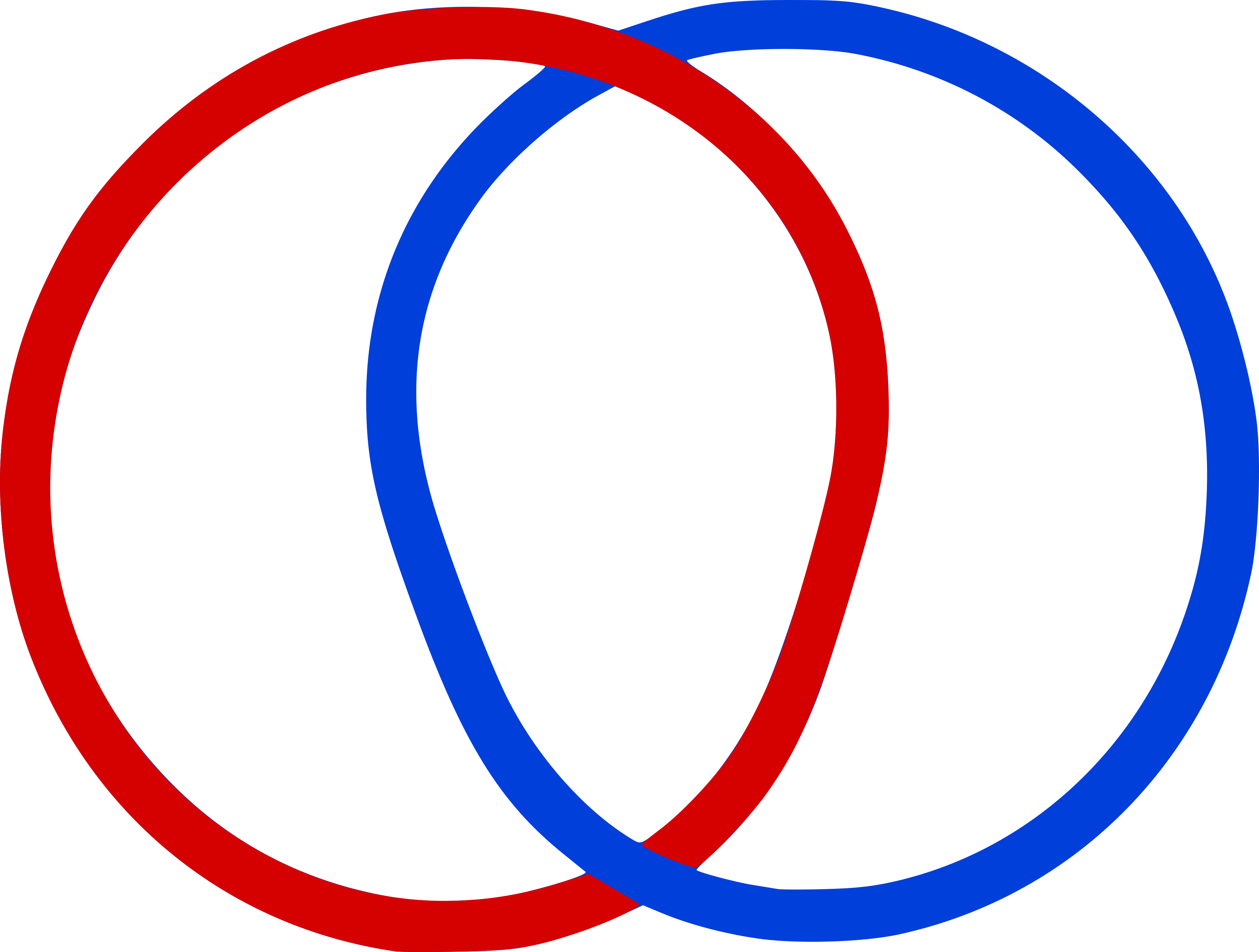}
\end{array}} & {\begin{array}{c}
\includegraphics[width=0.21\linewidth]{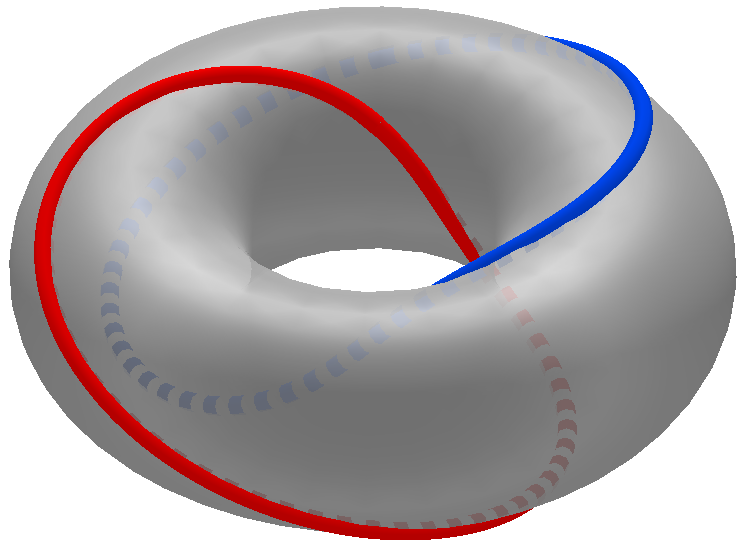}
\end{array}} & T_{2,4} & {\begin{array}{c}
\includegraphics[width=0.15\linewidth]{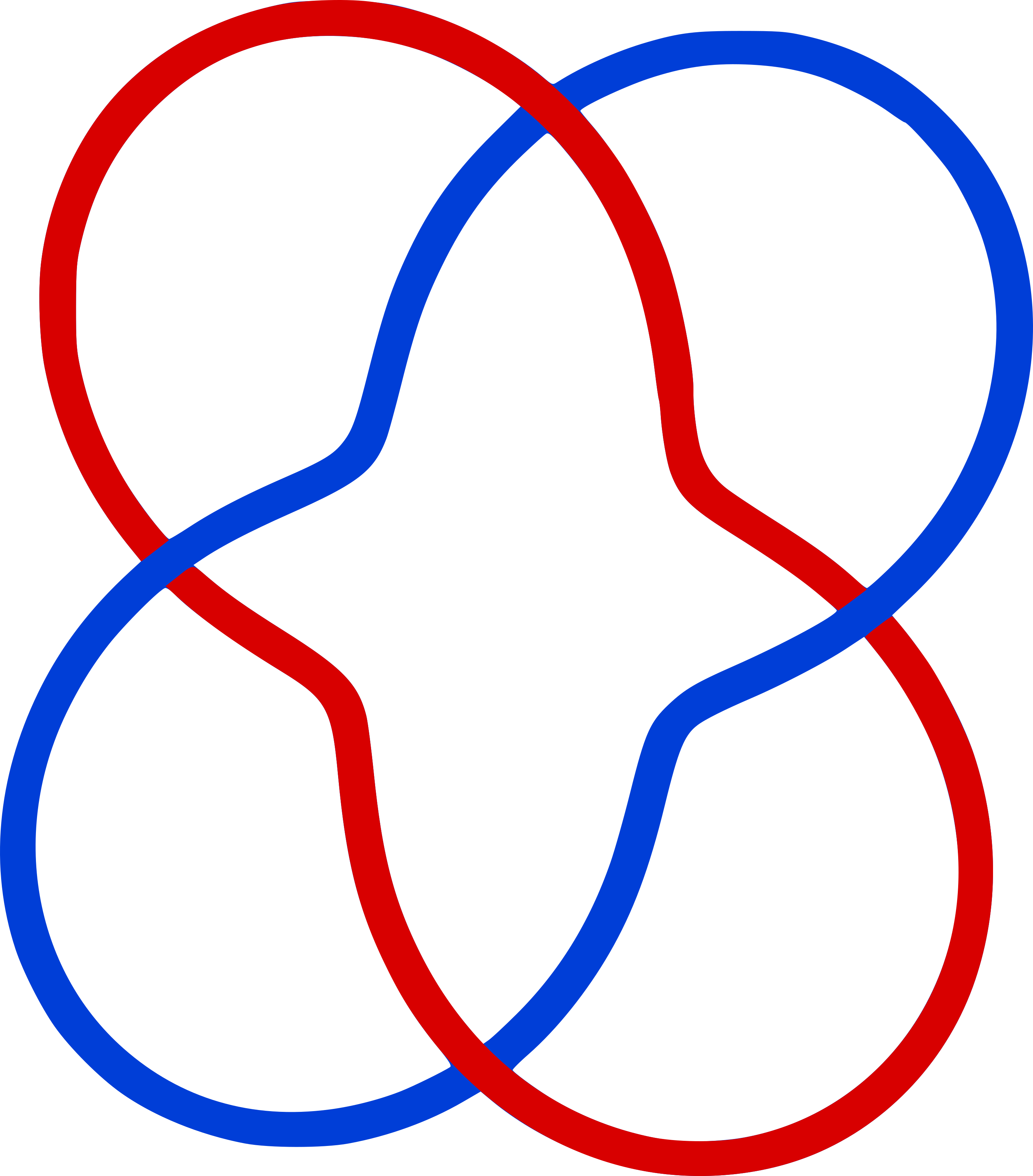}
\end{array}} & {\begin{array}{c}
\includegraphics[width=0.21\linewidth]{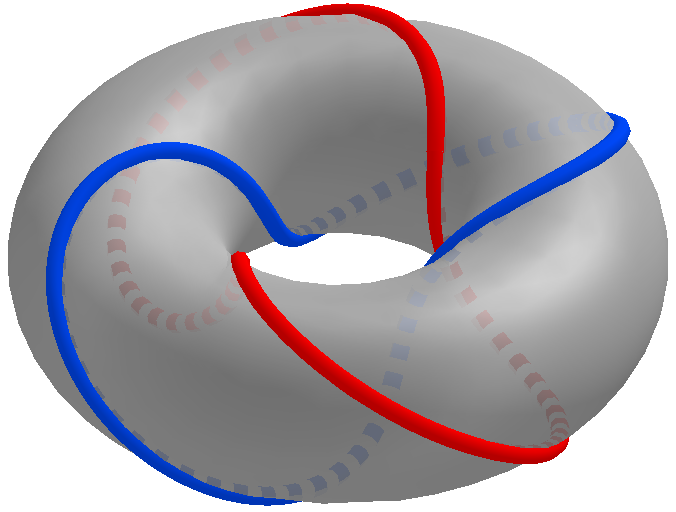}
\end{array}} \\
T_{2,6} & {\begin{array}{c}
\includegraphics[width=0.2\linewidth]{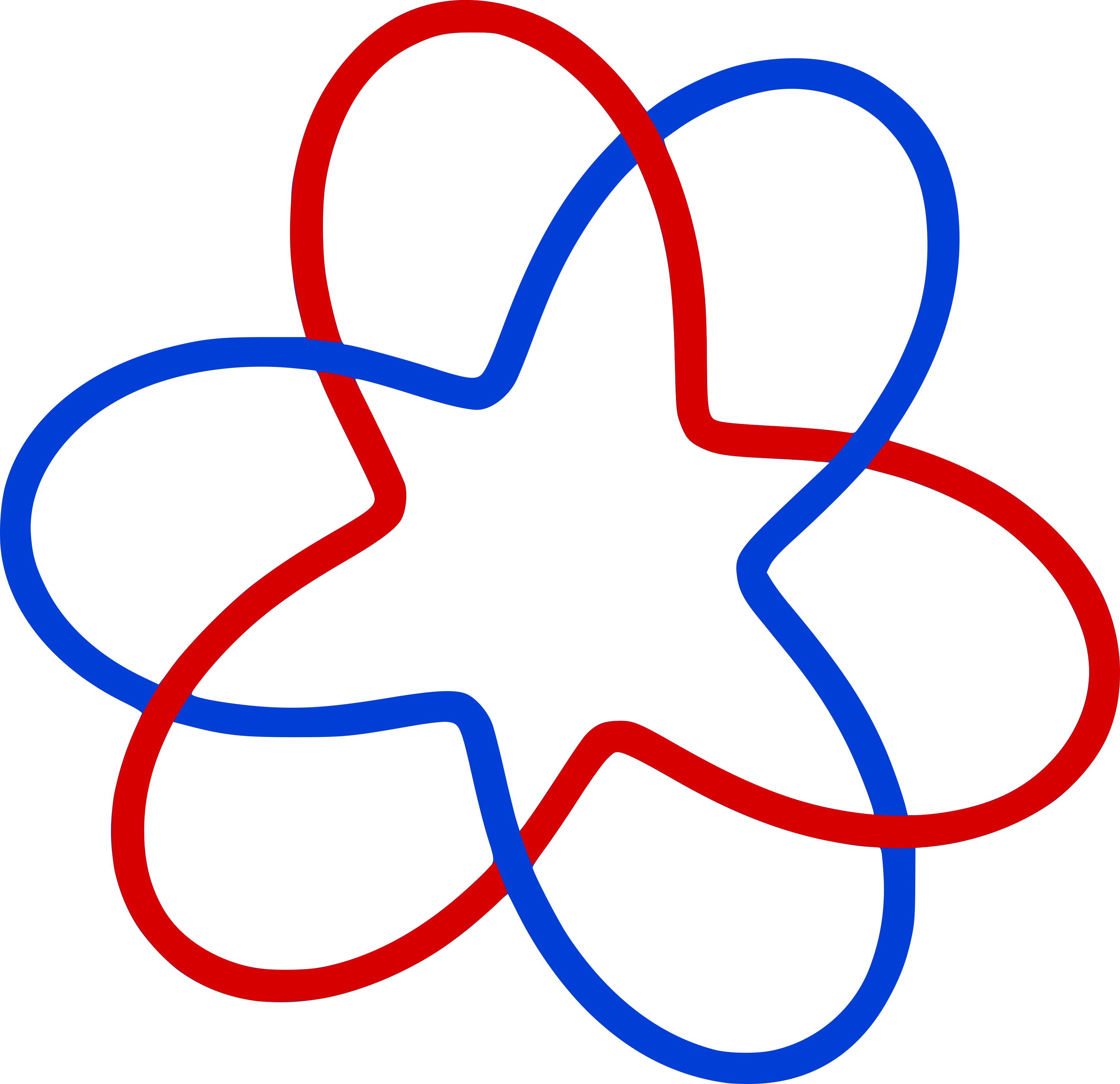}
\end{array}} & {\begin{array}{c}
\includegraphics[width=0.21\linewidth]{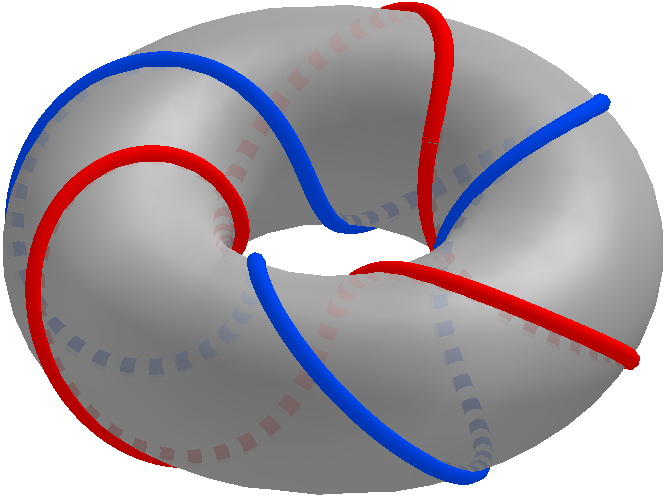}
\end{array}} & T_{2,8} & {\begin{array}{c}
\includegraphics[width=0.2\linewidth]{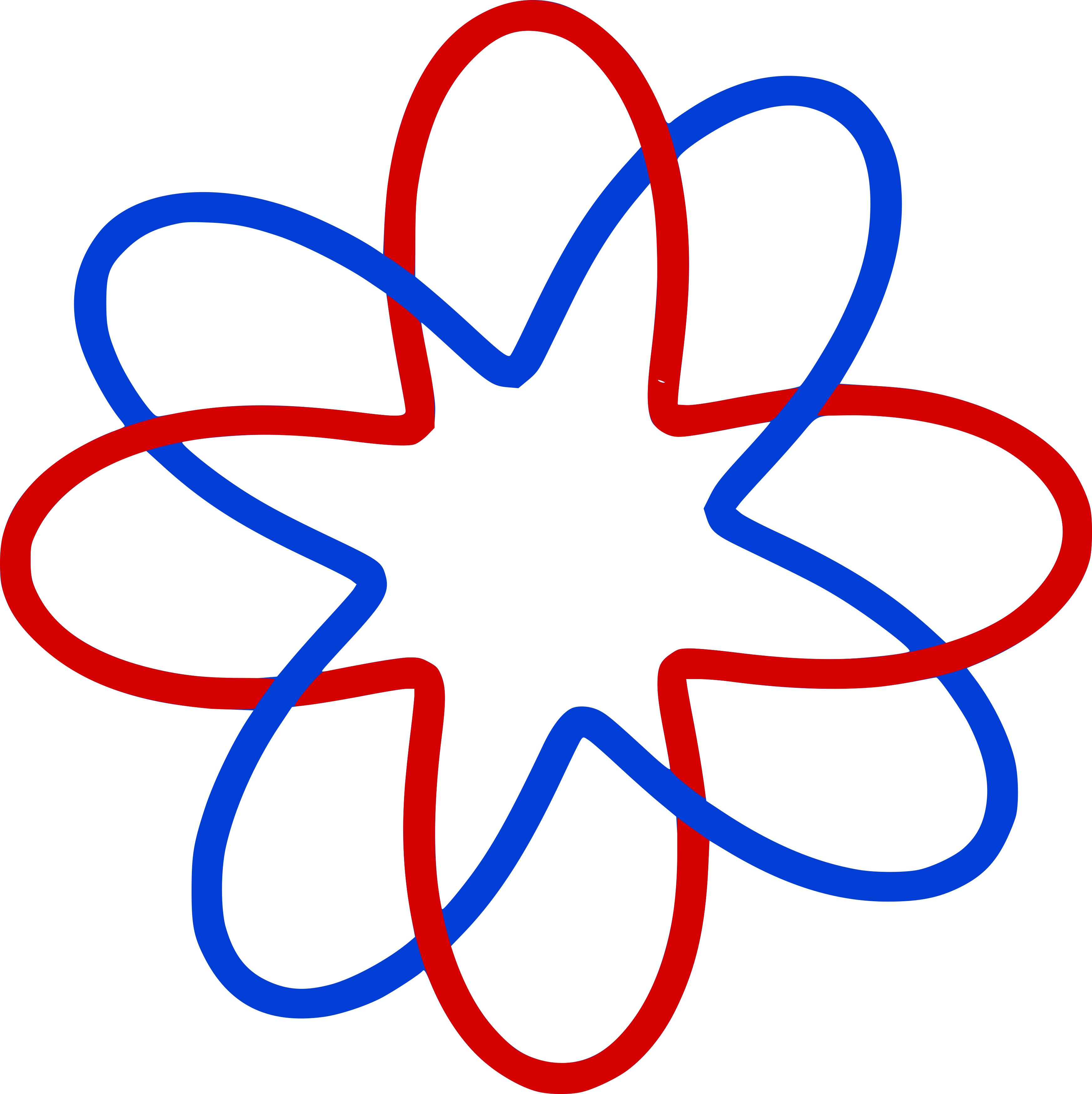}
\end{array}} & {\begin{array}{c}
\includegraphics[width=0.21\linewidth]{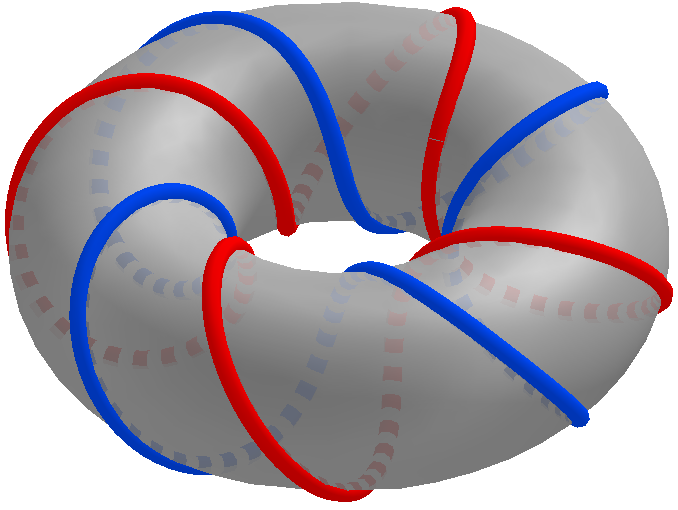}
\end{array}} \\
T_{3,3} & {\begin{array}{c}
\includegraphics[width=0.2\linewidth]{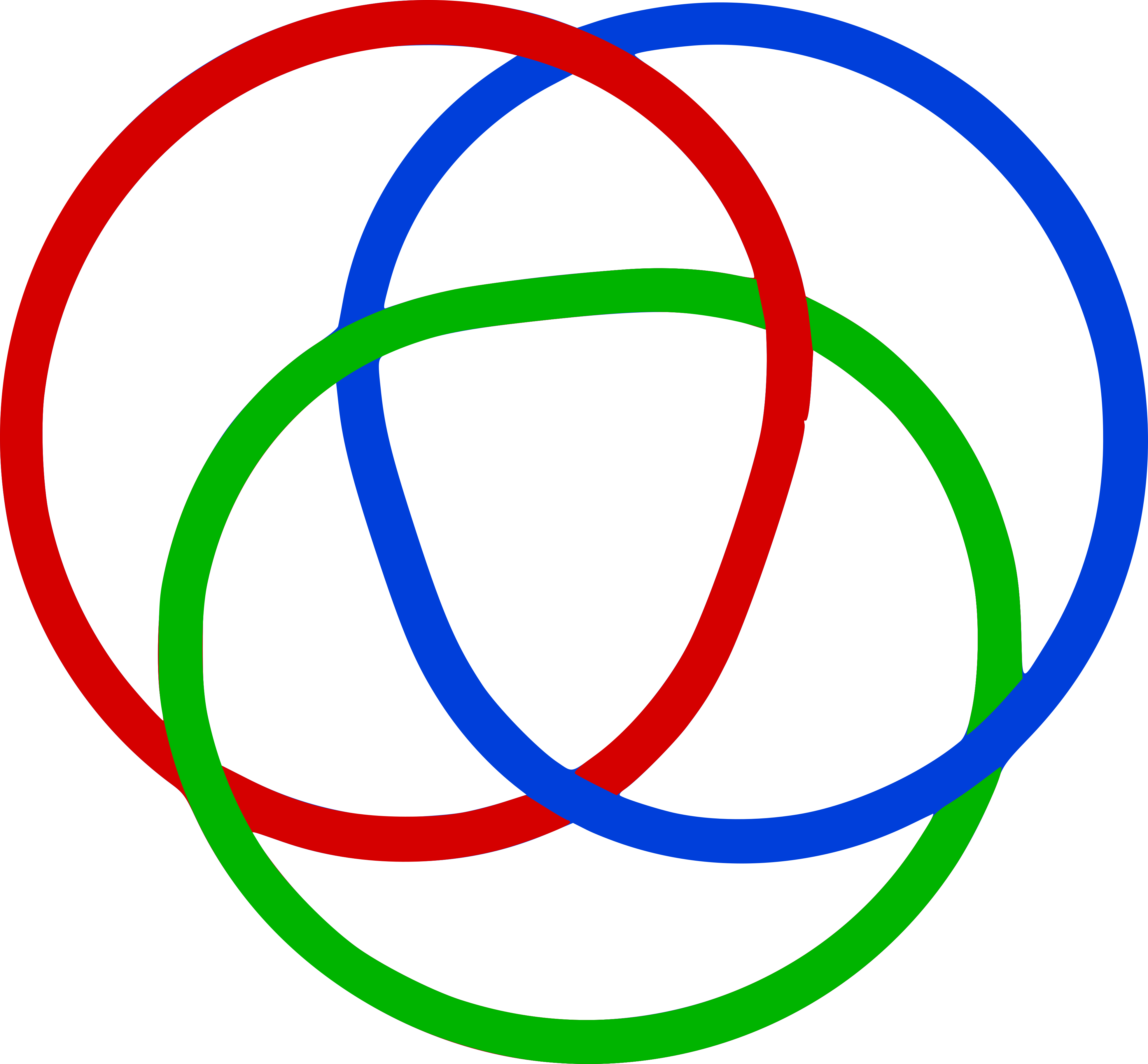}
\end{array}} & {\begin{array}{c}
\includegraphics[width=0.21\linewidth]{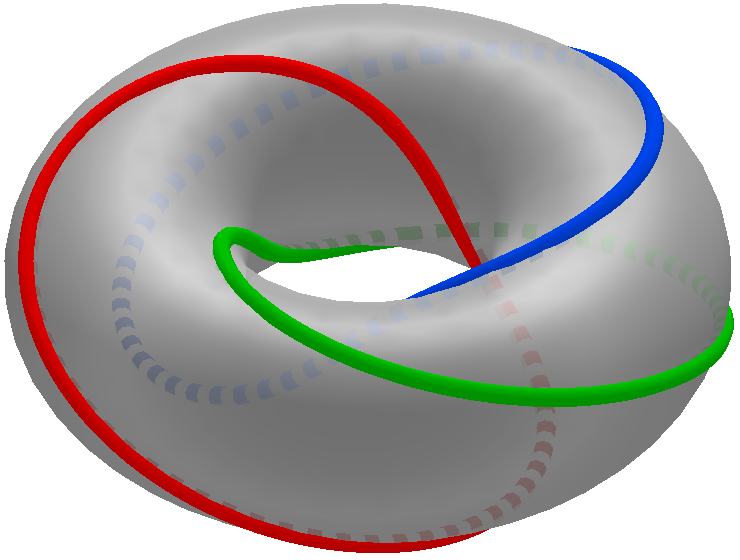}
\end{array}} & T_{3,6} & {\begin{array}{c}
\includegraphics[width=0.2\linewidth]{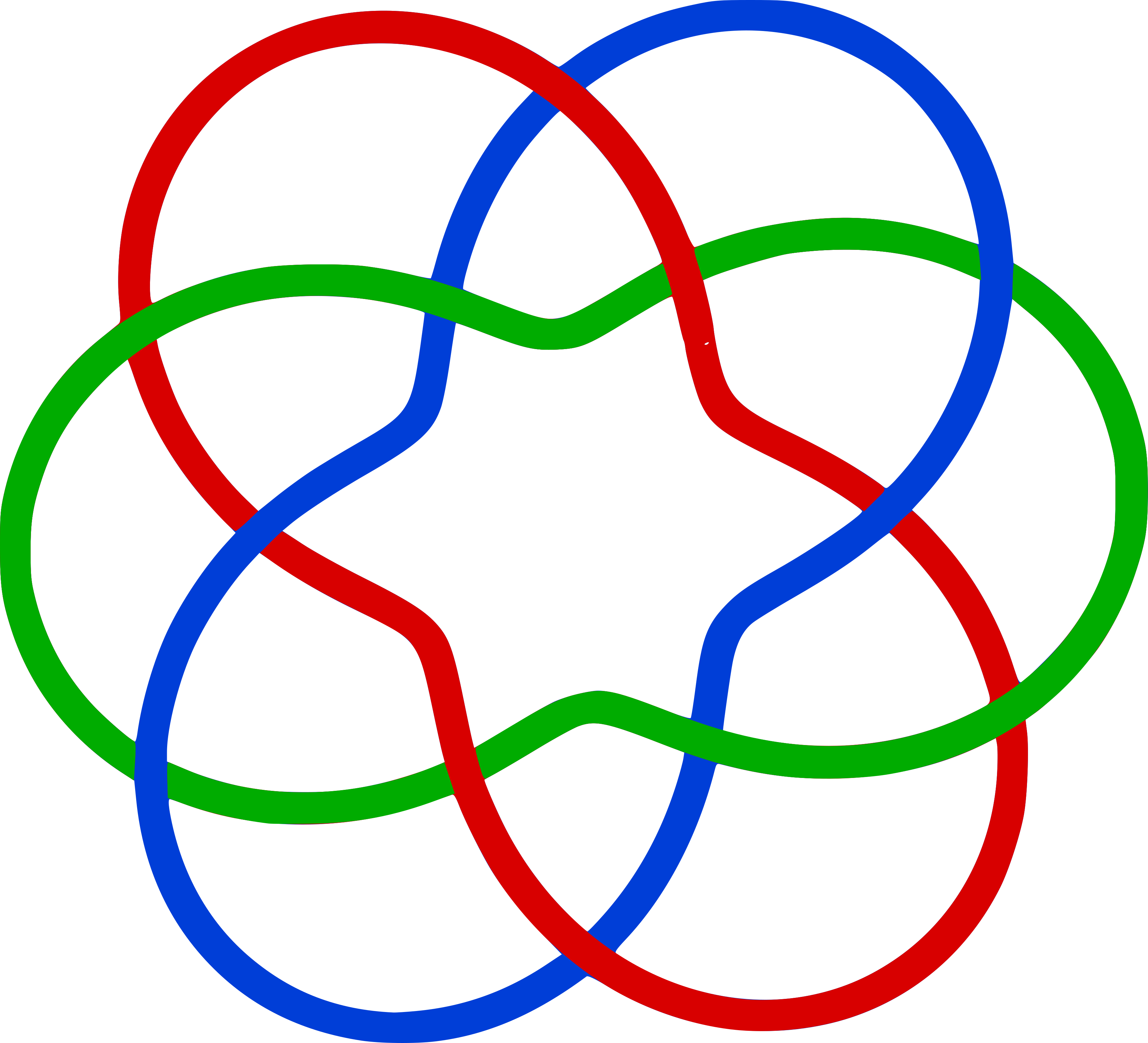}
\end{array}} & {\begin{array}{c}
\includegraphics[width=0.21\linewidth]{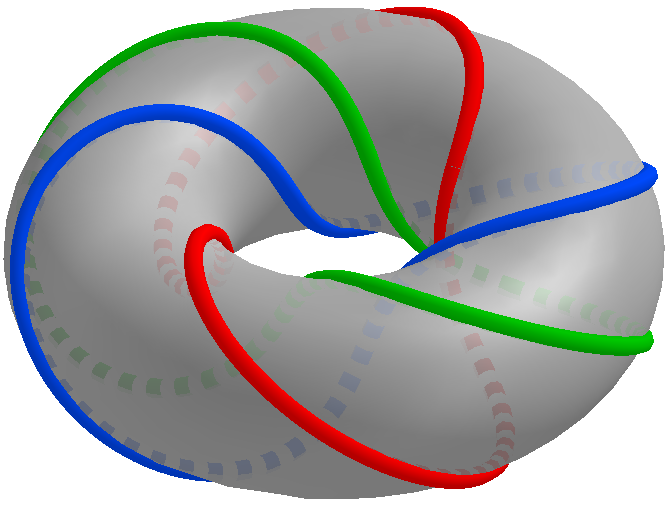}
\end{array}} \\
T_{3,9} & {\begin{array}{c}
\includegraphics[width=0.2\linewidth]{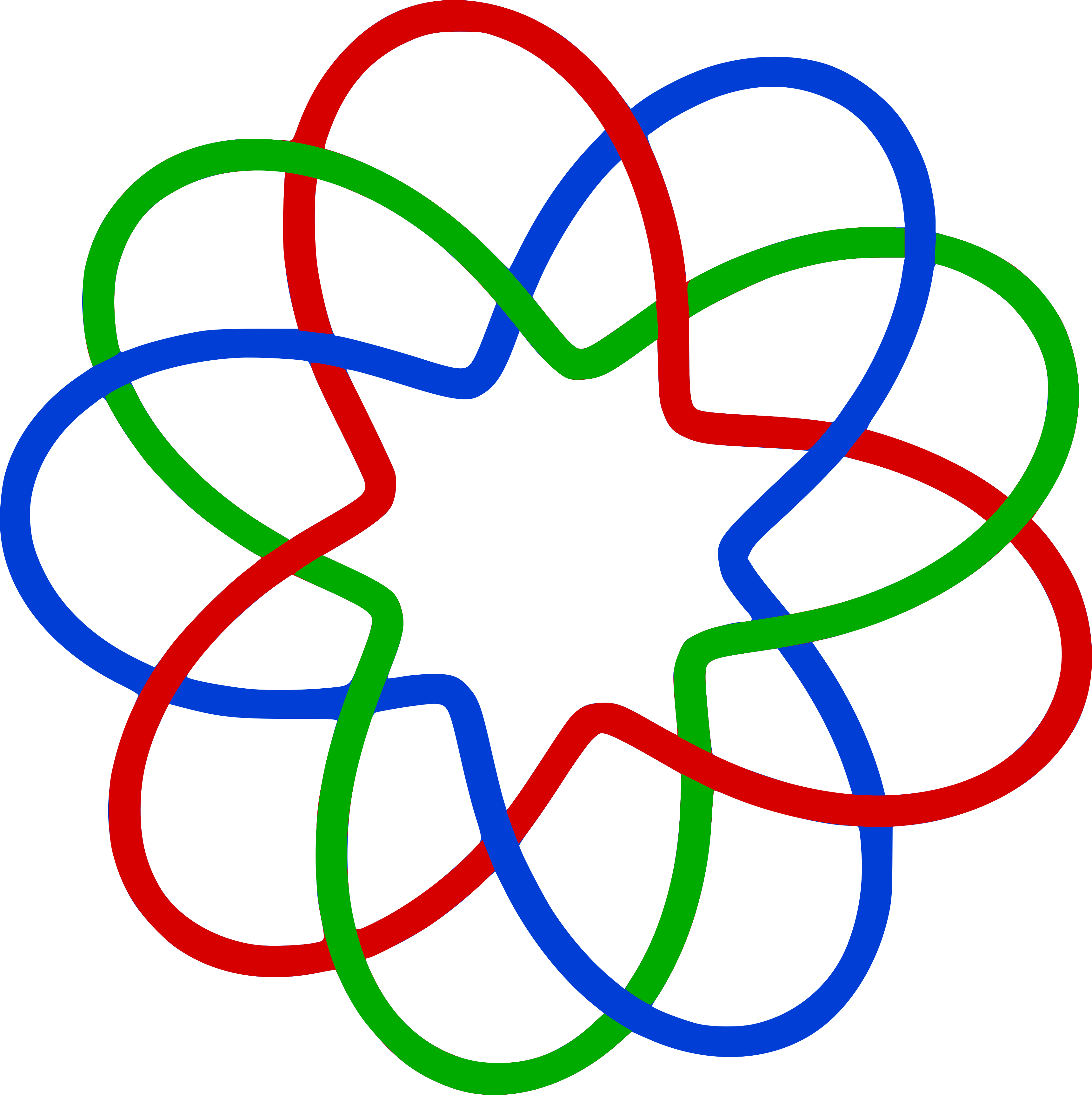}
\end{array}} & {\begin{array}{c}
\includegraphics[width=0.21\linewidth]{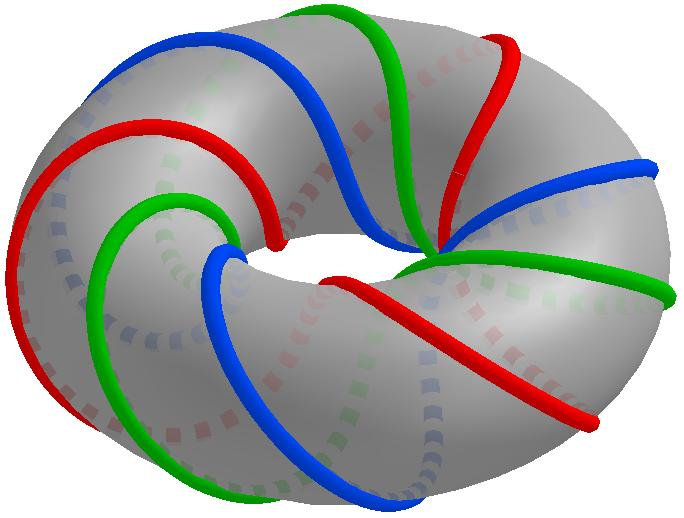}
\end{array}} & T_{3,12} & {\begin{array}{c}
\includegraphics[width=0.2\linewidth]{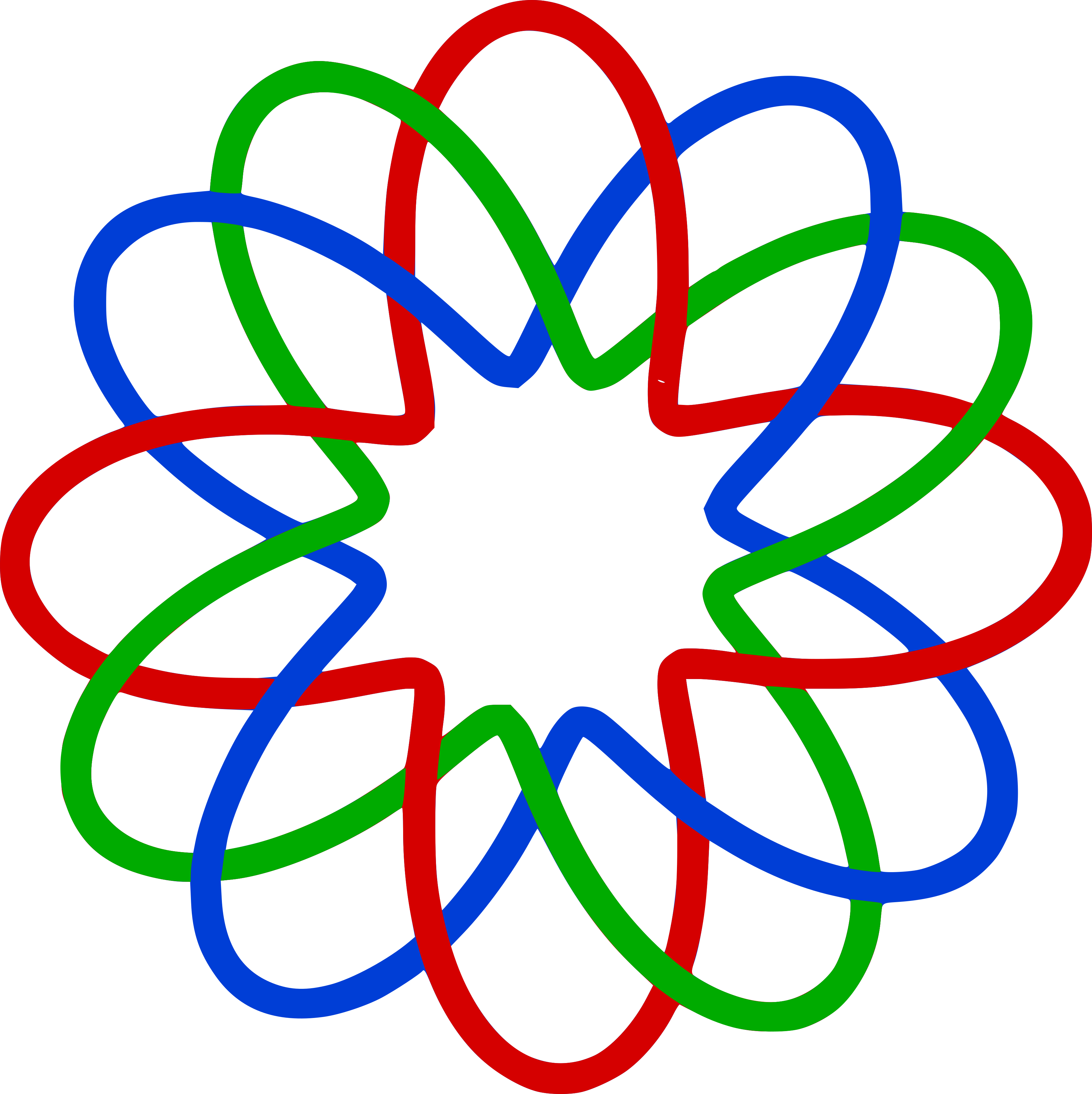}
\end{array}} & {\begin{array}{c}
\includegraphics[width=0.21\linewidth]{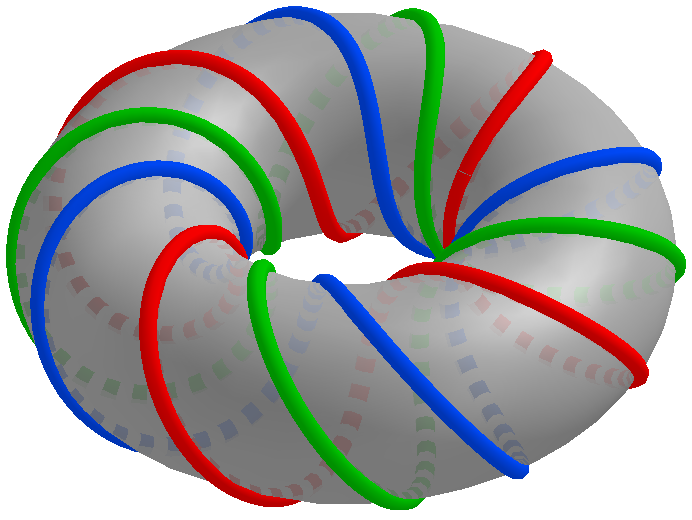}
\end{array}} \\
T_{4,4} & {\begin{array}{c}
\includegraphics[width=0.2\linewidth]{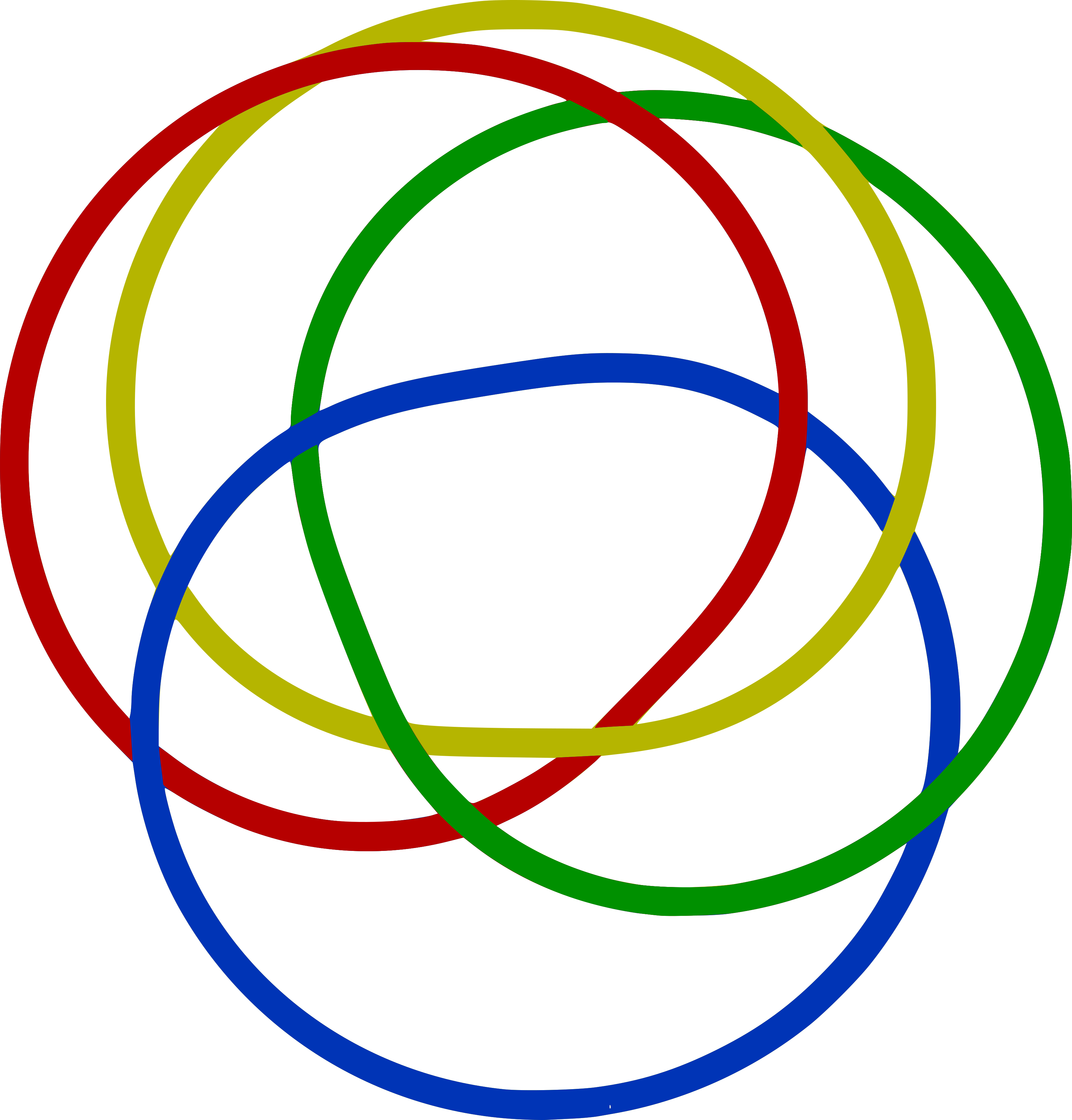}
\end{array}} & {\begin{array}{c}
\includegraphics[width=0.21\linewidth]{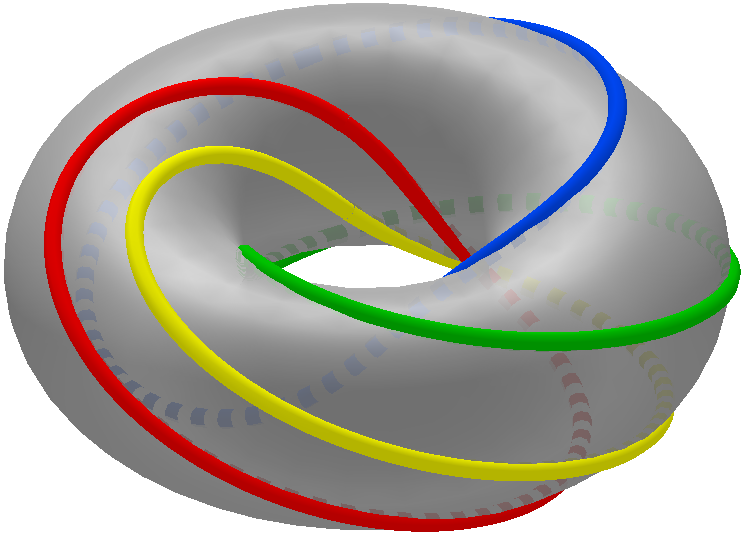}
\end{array}} & T_{4,8} & {\begin{array}{c}
\includegraphics[width=0.2\linewidth]{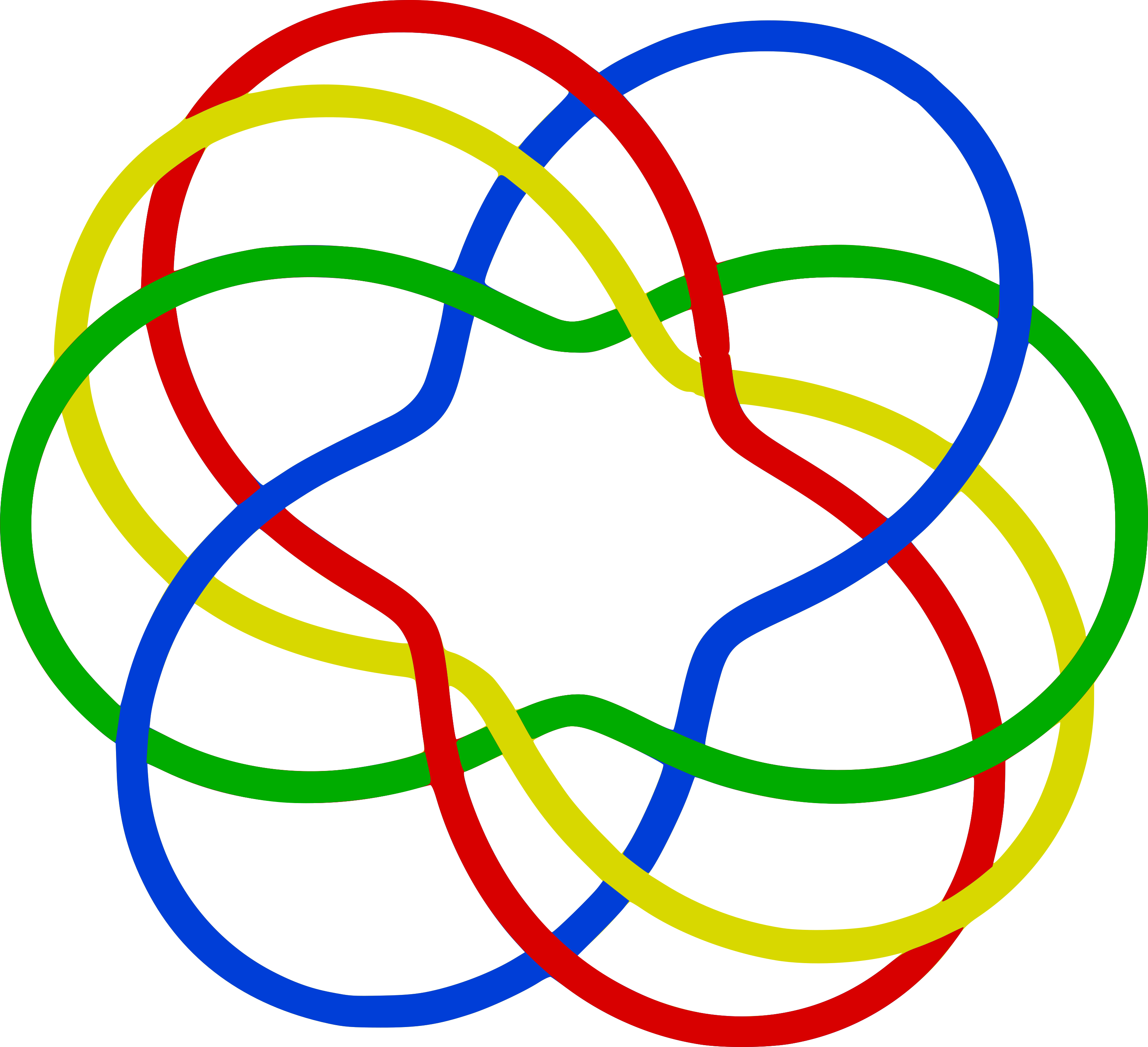}
\end{array}} & {\begin{array}{c}
\includegraphics[width=0.21\linewidth]{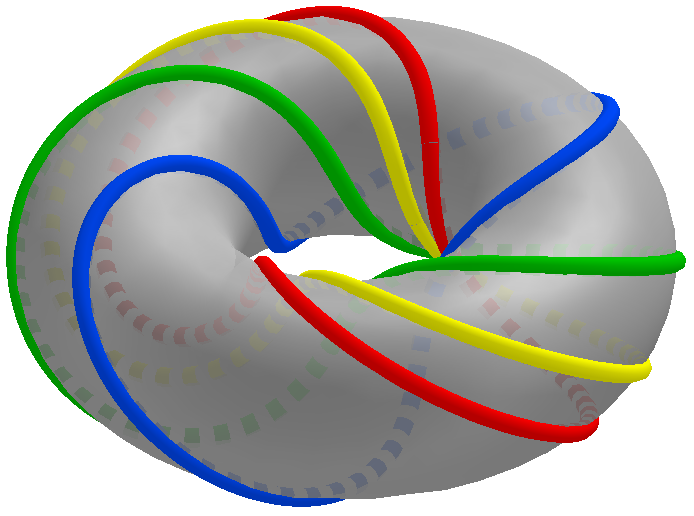}
\end{array}} \\
T_{4,12} & {\begin{array}{c}
\includegraphics[width=0.2\linewidth]{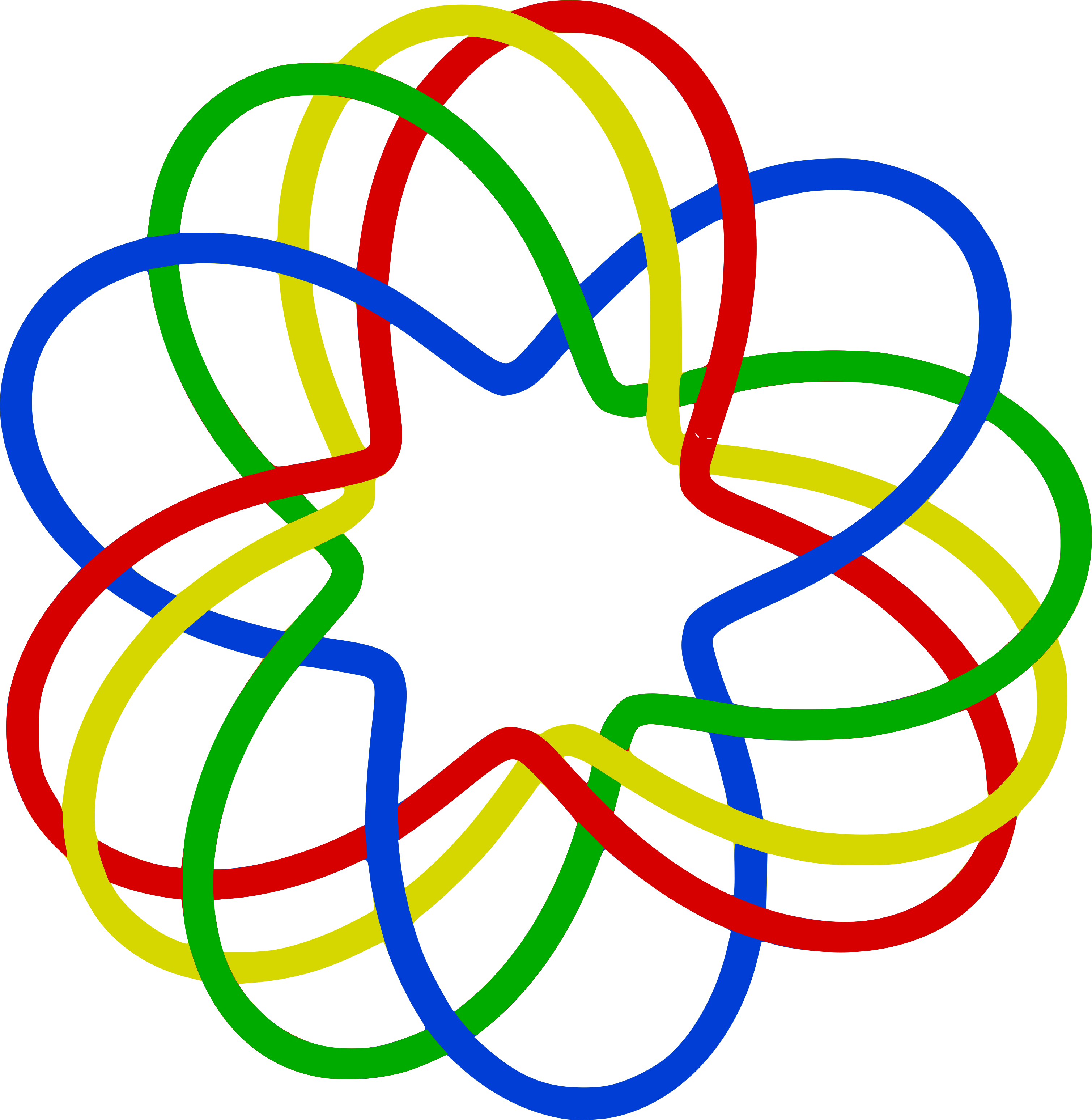}
\end{array}} & {\begin{array}{c}
\includegraphics[width=0.21\linewidth]{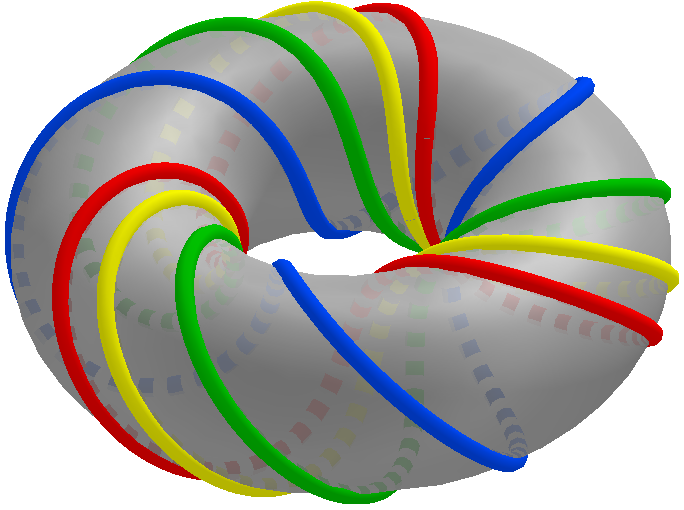}
\end{array}} & T_{4,16} & {\begin{array}{c}
\includegraphics[width=0.2\linewidth]{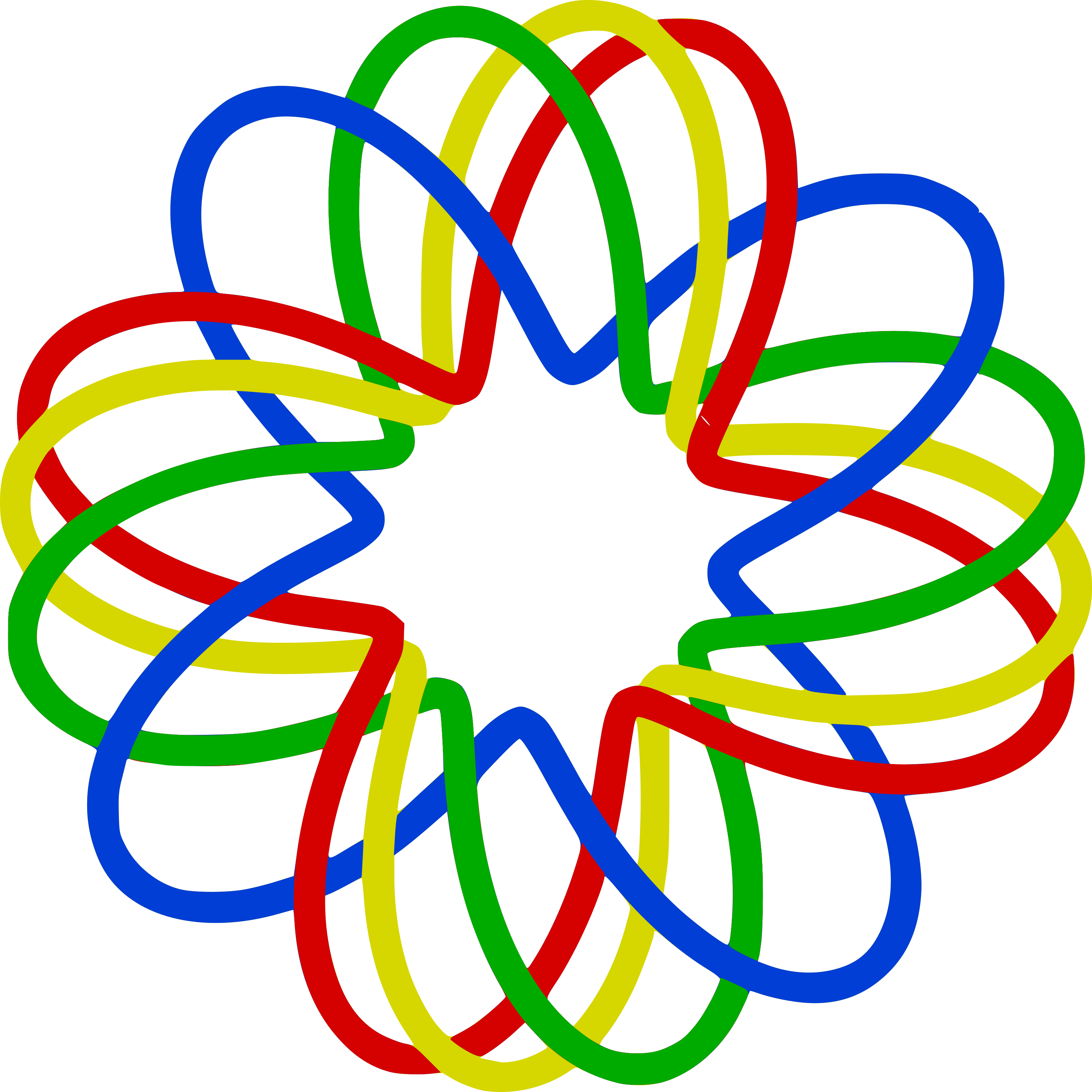}
\end{array}} & {\begin{array}{c}
\includegraphics[width=0.21\linewidth]{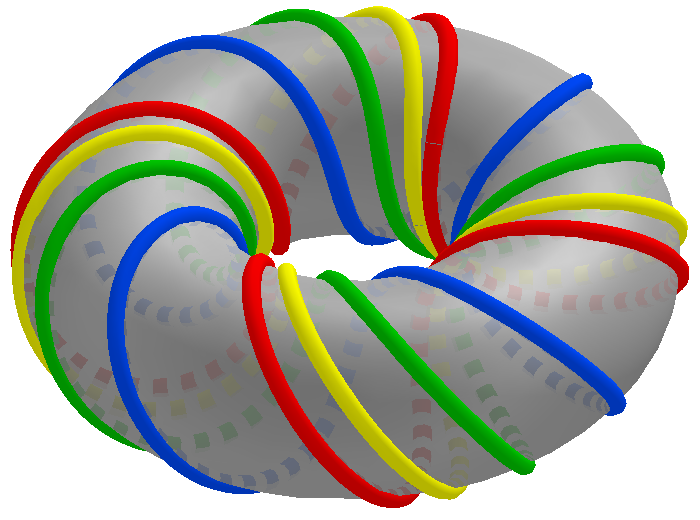}
\end{array}} \\ \hline
\end{array} $ 
%\end{equation}
\end{center}
\caption{Torus links of type $T_{p,pn}$}
\label{Tppnlinkstable}
\end{table}

As far as the reduced density matrix is concerned, we find that the torus links of type $T_{p,pn}$ are unique compared to a generic link in the sense that the traces $\text{Tr}[\sigma^m]$ of the powers of unnormalized reduced density matrices can be written as finite degree polynomial functions in
$k$.\footnote{For $n = 1,2,3,4$, we have computed the traces $\text{Tr}[\sigma^m]$ for several values of $p$ and $m$ and find that they are integers. But
for $n \geq 5$, these traces may not always be integers. However, they can still be written as polynomials in $k$.} This enables us to obtain an analytic expression for the Rényi entropies as a function of $k$ which will help to find the close-form expressions for the large $k$ limits of the entropies.  
In the following subsections, we will discuss such properties in detail.
%............................................
%............................................
\subsection{Rényi entropies for the state $\ket{T_{p,p}}$}
The eigenvalues of the unnormalized reduced density matrix $\sigma$ for the $\ket{T_{p,p}}$ state are given as,
\begin{equation}
\Lambda_{\alpha} = (\mathcal{S}_{0 \alpha})^{2-2p} \left(\mathcal{S}^*\mathcal{T}\mathcal{S}\right)_{\alpha 0} \, \left(\mathcal{S}^*\mathcal{T}\mathcal{S}\right)_{\alpha 0}^{*}  ~,
\end{equation}
and the trace of the powers of unnormalized reduced density matrices can be obtained as,
\begin{equation}
\text{Tr}[\sigma(T_{p,p})^{m}] = \sum_{\alpha=0}^k \Lambda_{\alpha}^m ~.
\end{equation}
Analyzing several cases, we see that these traces are always positive integers and can be expressed as a finite degree polynomial function in $k$ as: 
\begin{equation}
\text{Tr}[\sigma(T_{p,p})^{m}] = \sum_{i=0}^{3mp-6m} C_i(m)\, k^i ~,
\label{degreeTpp}
\end{equation}
where the coefficients $C_i(m) \in \mathbb{Q}$ are positive rational numbers. We have tabulated the traces $\text{Tr}[\sigma(T_{3,3})^{m}]$ as a polynomial function in $k$ for some smaller values of $m$ in Table \ref{PolT33} in the appendix \ref{appB}.  The $m^{\text{th}}$ Rényi entropy therefore, can be expressed as an analytic function of $k$ as,
\begin{equation}
\mathcal{R}_m(T_{p,p}) = \frac{1}{1-m} \ln(\frac{\text{Tr}[\sigma(T_{p,p})^{m}]}{\text{Tr}[\sigma(T_{p,p})]^m}) = \frac{1}{1-m} \ln\left[\frac{\sum_{i=0}^{3mp-6m} C_i(m)\, k^i}{\left(\sum_{i=0}^{3p-6} C_i(1)\, k^i\right)^m}\right] ~.
\end{equation}
Our analysis further shows that the trace of the $m^{\text{th}}$ power of unnormalized reduced density matrix for any torus link of type $T_{p,p}$ can be expressed as the following trace for the $T_{3,3}$ link: 
\begin{equation}
\text{Tr}[\sigma(T_{p,p})^{m}] = \text{Tr}[\sigma(T_{3,3})^{pm-2m}] ~.
\label{traceTppasT33}
\end{equation}
This expression enables one to compute the Rényi entropies of a generic state $\ket{T_{p,p}}$ in terms of the Rényi entropies of $\ket{T_{3,3}}$ as:
\begin{equation}
\boxed{\mathcal{R}_m(T_{p,p}) = \left(\frac{pm-3m}{1-m}\right) \mathcal{R}_{p-2}(T_{3,3})+\left(\frac{2m+1-pm}{1-m}\right) \mathcal{R}_{pm-2m}(T_{3,3})} ~.
\label{RETpp}
\end{equation}
We can immediately see that this relation satisfies the following known results:
\begin{align}
\mathcal{R}_0(T_{p,p}) &= \mathcal{R}_{0}(T_{3,3}) = \ln (k+1) = \ln(\text{dim}\, \mathcal{H}_{T^2}) \\
\mathcal{R}_m(T_{2,2}) &= \mathcal{R}_{0}(T_{3,3}) = \ln (k+1) = \ln(\text{dim}\, \mathcal{H}_{T^2}) ~.
\end{align} 
The equation in the first line gives the $0^{\text{th}}$ Rényi entropy, which by definition, is the maximum entropy. The equation in second line is the well known result that the state $\ket{T_{2,2}}$ associated with the Hopf link is maximally entangled. The relation in \eqref{RETpp} essentially means that it suffices to study and compute the Rényi entropies for the state $\ket{T_{3,3}}$. In the remaining part of this section, we will analyze the Rényi entropies of $\ket{T_{3,3}}$.

\begin{figure}[htbp]
\centerline{\includegraphics[width=6.0in]{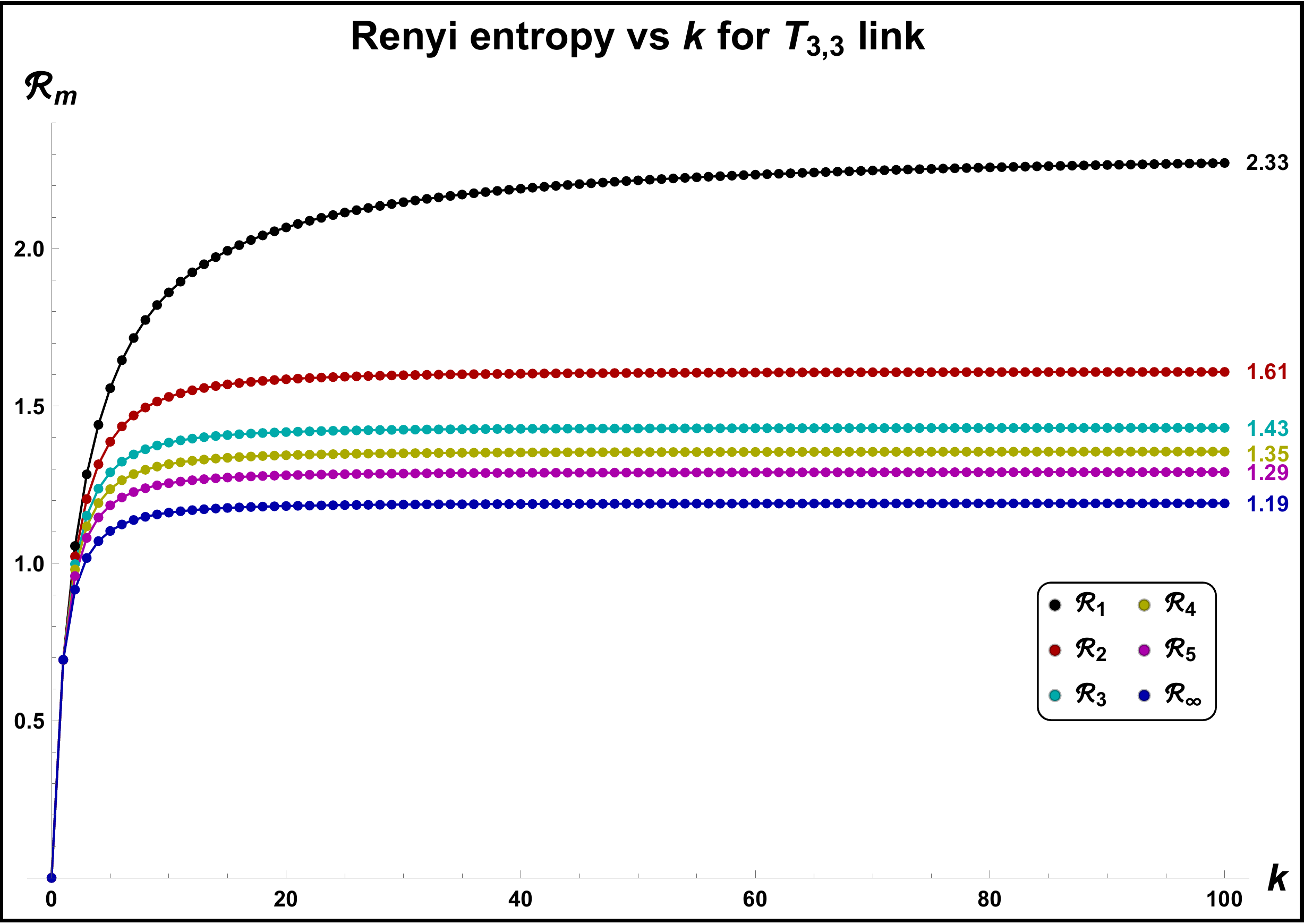}}
\caption[]{The variation of the Rényi entropy $\mathcal{R}_m$ with $k$ computed for SU$(2)_k$ group. The state under consideration is $\ket{T_{3,3}}$. The values mentioned against the plots are the approximate values (up to two decimal places) at which these entropies converge when $k \to \infty$. The exact asymptotic values are given in \eqref{RenyiTpp} and \eqref{EERminTpp}.}
\label{RenyivskT33}
\end{figure}

The trace of the $m^{\text{th}}$ power of unnormalized reduced density matrix for the state $\ket{T_{3,3}}$ can be conveniently written as,
\begin{equation}
\text{Tr}[\sigma(T_{3,3})^{m}] = (k+1) (k+2)^{m} (k+3)\, P_m(k^2+4k) ~,
\end{equation}
where $P_m(x)$ is a polynomial of degree $(m-1)$ in variable $x$ with rational coefficients. We have tabulated the polynomials for some smaller values of $m$ below:
\begin{equation}
\begin{array}{|c|c|} \hline
\rowcolor{Gray}
 m & P_m(x) \\ \hline
 1 & \dfrac{1}{6} \\[0.3cm]
 2 & \dfrac{1}{180}x+\dfrac{1}{12} \\[0.3cm]
 3 & \dfrac{1}{3780}x^2+\dfrac{13}{2520}x+\dfrac{1}{24} \\[0.3cm]
4 & \dfrac{1}{75600}x^3+\dfrac{79}{226800}x^2+\dfrac{11}{3024}x+\dfrac{1}{48} \\[0.3cm]
5 & \dfrac{1}{1496880}x^4+\dfrac{67}{2993760}x^3+\dfrac{311}{997920}x^2+\dfrac{17}{7392}x+\dfrac{1}{96} \\[0.3cm]
6 & \dfrac{691}{20432412000}x^5 +\dfrac{149}{108108000}x^4+\dfrac{109777}{4540536000}x^3+\dfrac{71327}{302702400}x^2+\dfrac{113}{82368}x+\dfrac{1}{192} \\[0.2cm] \hline
\end{array} \nonumber
\end{equation}
With this analysis, we can now give our general result which is valid for torus links of type $T_{p,p}$. The $m^{\text{th}}$ Rényi entropy can be obtained as,
\begin{equation}
\boxed{\mathcal{R}_m(T_{p,p}) = \ln(k+1) +\ln(k+3) + \frac{1}{1-m} \ln\left[\frac{P_{pm-2m}}{(P_{p-2})^m}\right]} ~,
\end{equation}
where it is understood that the polynomials $P_m$ are evaluated at variable $k^2+4k$. For a given value of $m$, the polynomial $P_m$ can be explicitly obtained and thus the Rényi entropies can be computed as an analytical function of $k$. For example, the second Rényi entropy for various links are given below (where variable $x=k^2+4k$):
\begin{equation}
\begin{array}{|c|c|} \hline
\rowcolor{Gray}
 \mbox{} & \mathcal{R}_2 \\ \hline
 T_{2,2} & \ln(k+1) \\[0.2cm]
 T_{3,3} & \ln(\dfrac{5 x+15}{x+15}) \\[0.4cm]
 T_{4,4} & \ln(\dfrac{7 x^2+126 x+315}{3 x^2+34 x+315}) \\[0.4cm]
T_{5,5} & \ln(\dfrac{2860 x^5+120120 x^4+2323035 x^3+23532795 x^2+123648525 x+212837625}{1382 x^5+56322 x^4+987993 x^3+9629145 x^2+56062125 x+212837625}) \\[0.4cm]
\hline
\end{array} ~. \nonumber
\end{equation}
The entanglement entropy can be obtained by taking the $m \to 1$ limit of the $m^{\text{th}}$ Rényi entropy:
\begin{equation}
\boxed{\mathcal{E}(T_{p,p}) = \ln(k+1) +\ln(k+3) + \ln(P_{p-2}) - (p-2)\frac{P_{p-2}'}{P_{p-2}} } ~,
\end{equation}
where we have defined,
\begin{equation}
P_{p-2}'(x) \equiv \left.\frac{d P_m(x)}{dm}\right|_{m=p-2} ~.
\label{}
\end{equation}
Similarly, the minimum entropy can be obtained by taking the $m \to \infty$ limit:
\begin{equation}
\boxed{\mathcal{R}_{\text{min}}(T_{p,p}) = \ln(k+1) +\ln(k+3) + \lim_{m \to \infty}\, \frac{1}{1-m} \ln\left[\frac{P_{pm-2m}}{(P_{p-2})^m}\right]} ~.
\end{equation}
Note that though the $\mathcal{R}_m$ can be computed for a given value of $m$, the computation of $\mathcal{E}$ and $\mathcal{R}_{\text{min}}$ requires a close form expression of the polynomial $P_m(x)$ as a function of $m$, which we do not have at the moment. In the figure \ref{RenyivskT33}, we have shown the variation of some of the Rényi entropies as a function of $k$ for the link $T_{3,3}$. 
This completes our analysis for the Rényi entropies of the torus links of type $T_{p,p}$.
%............................................
%............................................
\subsection{Rényi entropies for state $\ket{T_{p,pn}}$}
\begin{figure}[htbp]
\centerline{\includegraphics[width=6.0in]{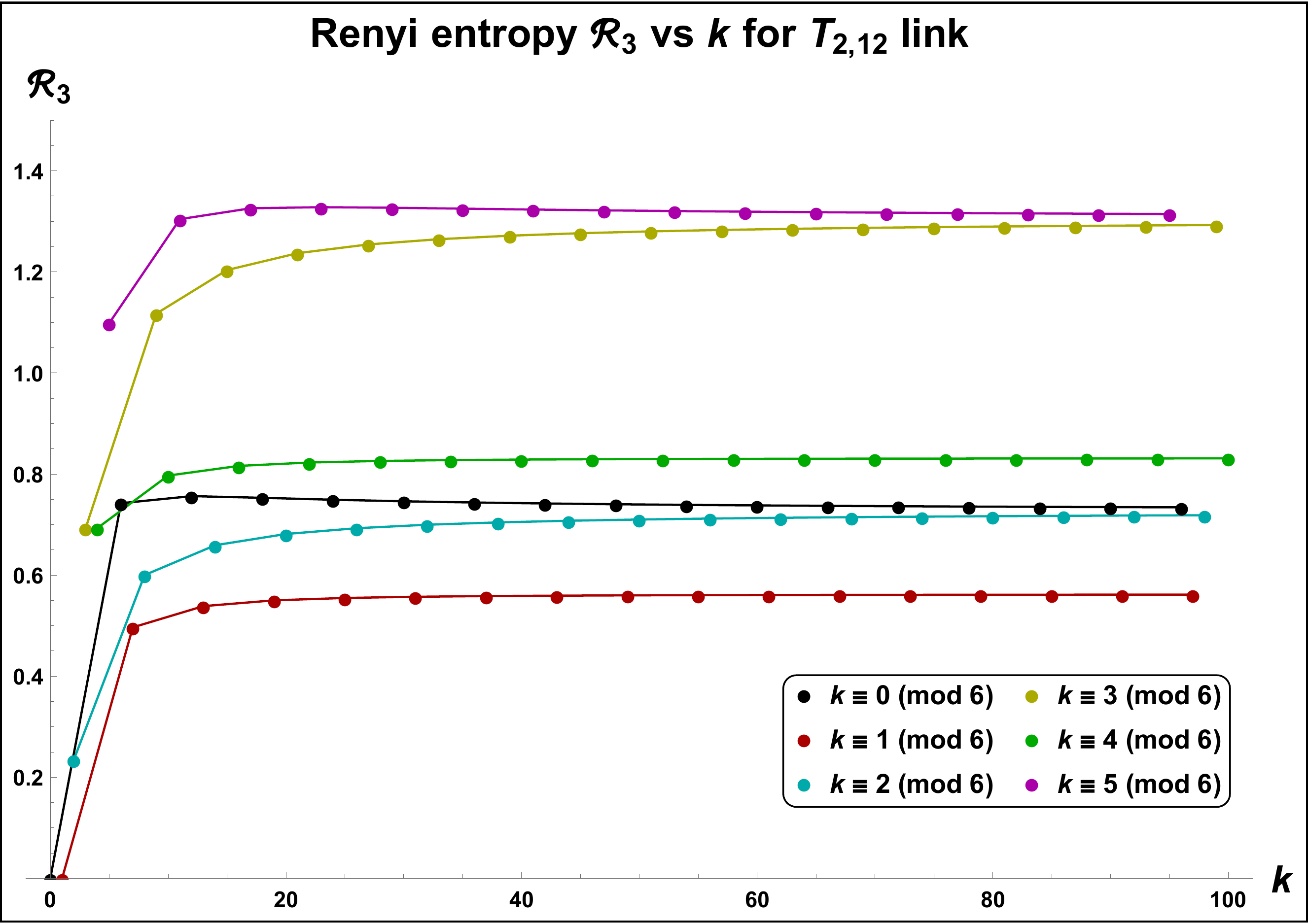}}
\caption[]{The variation of the third Rényi entropy $\mathcal{R}_3$ with $k$ computed for SU$(2)_k$ group. The state under consideration is $\ket{T_{2,12}}$. The entropy has been plotted for various values of $k$ modulo 6.}
\label{R3vskT212}
\end{figure}

The eigenvalues of the unnormalized reduced density matrix $\sigma$ for $T_{p,pn}$ torus links are given as,
\begin{equation}
\Lambda_{\alpha} = (\mathcal{S}_{0 \alpha})^{2-2p} \left(\mathcal{S}^*\mathcal{T}^n\mathcal{S}\right)_{\alpha 0} \, \left(\mathcal{S}^*\mathcal{T}^n\mathcal{S}\right)_{\alpha 0}^{*}  ~,
\end{equation}
and the trace of the powers of unnormalized reduced density matrices can be obtained as,
\begin{equation}
\text{Tr}[\sigma^{m}] = \sum_{\alpha=0}^k \Lambda_{\alpha}^m ~.
\end{equation}
We find that $\text{Tr}[\sigma^{m}]$ can be written as a polynomial expression in $k$ where the coefficients are periodic functions of $k$ with maximum period $n$. For example, consider the simplest link $T_{2,4}$. The trace of the corresponding reduced density matrix can be written as
\begin{equation}
\text{Tr}[\sigma] = \frac{1}{4}k^2 + k + \frac{7+(-1)^k}{8} ~,
\end{equation} 
where the last term is a periodic function of $k$ with periodicity 2. Thus if we consider the even and odd values of $k$, we can write the trace as a second degree polynomial in $k$ with fixed rational coefficients: 
\begin{equation}
\text{Tr}[\sigma] = \begin{cases} \frac{1}{4}k^2+k+1, & \text{for $k=$ even} \\ \frac{1}{4}k^2+k+\frac{3}{4},& \text{for $k=$ odd} \end{cases} ~.
\end{equation} 
We give another example of $T_{2,8}$ link where the trace can be given as: 
\begin{equation}
\text{Tr}[\sigma] =  \left(\frac{(-1)^k}{32}+\frac{7}{32}\right) k^2 +\left(\frac{(-1)^k}{8}+\frac{1}{8} \sin \left(\frac{\pi  k}{2}\right)+\frac{7}{8} \right)k + \left( \frac{5 (-1)^k}{32}+\frac{1}{4} \sin \left(\frac{\pi  k}{2}\right)+\frac{27}{32} \right) \nonumber ~,
\end{equation}
where the coefficients are periodic functions of $k$ with maximum period 4. Considering $k=4\mathbb{Z}, 4\mathbb{Z}+1,4\mathbb{Z}+2,4\mathbb{Z}+3$, we can write the trace as second degree polynomials in $k$:  
\begin{equation}
\text{Tr}[\sigma] = \begin{cases} \frac{1}{4}k^2+k+1 & \text{, for $k \equiv 0$ (mod 4)} \\ \frac{3}{16}k^2+\frac{7}{8}k+\frac{15}{16} & \text{, for $k \equiv 1$ (mod 4)} \\ \frac{1}{4}k^2+k+1 & \text{, for $k \equiv 2$ (mod 4)} \\ \frac{3}{16}k^2+\frac{5}{8}k+\frac{7}{16} & \text{, for $k \equiv 3$ (mod 4)} \end{cases} ~.
\end{equation} 
When $n$ is large or when $n$ is prime, the coefficients appearing in $\text{Tr}[\sigma^m]$ can be some complicated periodic function of $k$ and obtaining the polynomial in such cases is usually difficult and time consuming. However analyzing several links, we observe that since the maximum periodicity in the coefficients can be $n$, it is easier to obtain the polynomial expression by computing the traces at those values of $k$ which are fixed modulo $n$. Given a value of $n$, we define an integer $\ell$ which is the remainder when $k$ is divided by $n$. In other words,
\begin{equation}
k \equiv \ell \,(\text{mod}\,\,n) \quad;\quad  \ell \in \{0,1,\ldots,n-1 \} ~.
\end{equation}
We find that for each fixed value of $\ell$, the traces for $T_{p,pn}$ link can be written as a finite degree polynomial function of $k$:
\begin{equation}
k = n \mathbb{Z} + \ell \quad \Longrightarrow \quad \text{Tr}[\sigma^{m}]_{\ell} = \sum_{i=0}^{3mp-4m} C_i^{\ell}(m)\, k^i ~,
\label{degreeTppn}
\end{equation}
where the notation $\text{Tr}[\sigma^{m}]_{\ell}$ simply means that these traces are valid only when $k = n \mathbb{Z} + \ell$. Listing the traces for each value of $\ell \in \{0,1,\ldots,n-1\}$, we can exhaust all possible integer values of $k$. Some examples for traces of the unnormalized reduced density matrix for the $T_{2,2n}$ link has been tabulated below for some smaller values of $n$. 
\begin{equation}
\begin{array}{|c|c|c|c|c|} \hline
\rowcolor{Gray}
   & n=2 & n=3 & n=4 & n=5 \\ \hline
 \ell =0 & \frac{1}{4} \left(k^2+4 k+4\right) & \frac{1}{9} \left(2 k^2+9 k+9\right) & \frac{1}{16} \left(4k^2+16 k+16\right) & \frac{1}{25} \left(6 k^2+25 k+25\right) \\[0.1cm]
 \ell =1 & \frac{1}{4} \left(k^2+4 k+3\right) & \frac{1}{9} \left(2k^2+8 k+8\right) & \frac{1}{16} \left(3 k^2+14 k+15\right) & \frac{1}{25} \left(6 k^2+23 k+21\right) \\[0.1cm]
 \ell =2 & \textendash & \frac{1}{9} \left(2 k^2+7 k+5\right) & \frac{1}{16} \left(4k^2+16 k+16\right) & \frac{1}{25} \left(4 k^2+19 k+21\right) \\[0.1cm]
 \ell =3 & \textendash & \textendash & \frac{1}{16} \left(3 k^2+10 k+7\right) & \frac{1}{25} \left(\sqrt{5}+5\right) \left(k^2+4 k+4\right) \\[0.1cm]
 \ell =4 & \textendash & \textendash & \textendash  & \frac{1}{25} \left(4 k^2+13 k+9\right) \\ \hline
\end{array} \nonumber
\end{equation}   
For higher values of $m$ and $n$ for a given link, the evaluation of the polynomials becomes more and more computationally involved. In appendix \ref{appB}, we have listed the polynomial expressions of $\text{Tr}[\sigma^m]$ for the links $T_{2,2n}, T_{3,3n}, T_{4,4n}$ for various values of $m$ and some small values of $n$. For $n=2$, the polynomials are listed in Table \ref{PolT24}, Table \ref{PolT36} and Table \ref{PolT48} respectively. For $n=3$, the polynomials can be found in  Table \ref{PolT26}, Table \ref{PolT39} and Table \ref{PolT412}, while for $n=4$, they are given in Table \ref{PolT28}, Table \ref{PolT312} and Table \ref{PolT416} respectively. Equipped with the values of various traces, we can obtain the $m^{\text{th}}$ Rényi entropy associated with the link $T_{p,pn}$ as an analytic expression in $k$. For example, we have tabulated the third Rényi entropy $\mathcal{R}_3$ for the link $T_{2,12}$ as a function of $k$ modulo $6$ below and have plotted them in the figure \ref{R3vskT212}.
\begin{equation}
\begin{array}{|c|c|} \hline
\rowcolor{Gray}
 T_{2,12}  & \mathcal{R}_3 \\ \hline
 \ell =0 & \dfrac{1}{2} \ln \left(\dfrac{1215 k^6+14580 k^5+72900 k^4+194400 k^3+291600 k^2+233280 k+77760}{284 k^6+2968 k^5+15780 k^4+48960 k^3+107136 k^2+145152 k+77760}\right) \\[0.5cm]
 \ell =1 & \dfrac{1}{2} \ln \left(\dfrac{40 k^6+480 k^5+2400 k^4+6400 k^3+9600 k^2+7680 k+2560}{13 k^6+156 k^5+915 k^4+3160 k^3+7332 k^2+10704 k+6880}\right) \\[0.5cm]
 \ell =2 & \dfrac{1}{2} \ln \left(\dfrac{1215 k^6+14580 k^5+72900 k^4+194400 k^3+291600 k^2+233280 k+77760}{284 k^6+3848 k^5+24580 k^4+92160 k^3+225536 k^2+347392 k+241600}\right) \\[0.5cm]
 \ell =3 & \dfrac{1}{2} \ln \left(\dfrac{625 k^6+9000 k^5+53325 k^4+166320 k^3+287955 k^2+262440 k+98415}{46 k^6+762 k^5+6300 k^4+30780 k^3+117774 k^2+283338 k+262440}\right) \\[0.5cm]
 \ell =4 & \dfrac{1}{2} \ln \left(\dfrac{480 k^6+5760 k^5+28800 k^4+76800 k^3+115200 k^2+92160 k+30720}{91 k^6+1092 k^5+6360 k^4+21760 k^3+48624 k^2+67008 k+40960}\right) \\[0.5cm]
 \ell =5 & \dfrac{1}{2} \ln \left(\dfrac{625 k^6+6000 k^5+23325 k^4+46880 k^3+51315 k^2+29040 k+6655}{46 k^6+342 k^5+2100 k^4+6980 k^3+42174 k^2+101478 k+64480}\right) \\ \hline
\end{array} \nonumber 
\end{equation}
We close this section by stressing on the fact that given a torus link of type $T_{p,pn}$, we can, in principle compute the close-form $k$ dependence of any $m^{\text{th}}$ Rényi entropy using the trick prescribed here. For obtaining an analytic expression (in $k$) of the entanglement entropy and the minimum Rényi entropy, we need to obtain the functional dependence of $\text{Tr}[\sigma(T_{p,pn})^{m}]$ on $m$. In the present work, we will not attempt this and instead our focus will be to extract the large $k$ behavior of various Rényi entropies, including the entanglement entropy and the minimum Rényi entropy. For this, it is sufficient to obtain the $m$ dependence of the leading order coefficient appearing in the polynomial expression of $\text{Tr}[\sigma(T_{p,pn})^{m}]$, which we elaborate in the next section.   
%............................................
%............................................
%...................SECTION ends..............
%...................SECTION ends..............
\section{Large $k$ behavior of SU$(2)_k$ Rényi entropies}
\label{sec4}
In this section we will analyze the large $k$ behavior of SU$(2)_k$ Rényi entropies for torus links. For the links of type $T_{p,pn}$, we will use analytical techniques to find the exact limiting values of entropies. For generic torus links which are not of the type $T_{p,pn}$, we carry out the numerical computation to find the variation of entropy with $k$. With enough analytical and numerical analysis, we propose the following:
\begin{mdframed}[style=sid]
\textbf{Proposition 1.} \emph{All the Rényi entropies ($\mathcal{R}_1,\mathcal{R}_2,\ldots,\mathcal{R}_{\infty}$) associated with the torus links (excluding the Hopf link), computed within the context of SU(2)$_k$ Chern-Simons theory, converge to a finite value in the semiclassical limit of $k \to \infty$.}
\end{mdframed}
To compute the large $k$ limit of the entropies for the $T_{p,pn}$ links, we essentially use the same trick of computing the trace of the $m^{\text{th}}$ power of the unnormalized reduced density matrix as prescribed in the previous section. In particular, we use the fact that these traces can be expressed as a finite degree polynomial function in $k$:\footnote{We would like to remind that we construct these polynomials by considering the values of $k$ which are constant modulo $n$. In other words, we do the analysis for a fixed $\ell$ where $k \equiv \ell \,(\text{mod}\,\,n)$ and simply quote the result for various values of $\ell \in \{0,1,\ldots,n-1\}$.}  
\begin{equation}
\text{Tr}[\sigma^m] = \sum_{i=0}^{\textsf{y}(m)} C_i(m) \, k^i ~,
\label{trpoly}
\end{equation} 
where $\textsf{y}(m)$ is the degree of the polynomial for a given $m$ and $C_i(m)$ are positive numbers. The crucial point to note is that for $m \geq 1$, the degree $\textsf{y}(m)$ is proportional to $m$ (see the analysis in previous section):
\begin{equation}
\textsf{y}(m)=\begin{cases}
(3p-6)m,  & \text{for $T_{p,p}$ link} \\
(3p-4)m, & \text{for $T_{p,pn}$ link with $n>1$}
\label{degreePolypnlink}
\end{cases} ~.
\end{equation}
As a result, in the large $k$ regime, we can carry out the following approximation: 
\begin{equation}
\frac{\text{Tr}[\sigma^m]}{\text{Tr}[\sigma]^m} \sim \frac{C_{\textsf{lead}}^m \, k^{\textsf{y}(m)}\left(1 + \mathcal{O}\left(1/k\right) + \ldots\right)}{(C_{\textsf{lead}}^1)^m \, k^{\textsf{y}(m)}\left(1 + \mathcal{O}\left(1/k\right) + \ldots\right)}  ~,
\label{}
\end{equation} 
where we denote $C_{\textsf{lead}}^m$ to be the leading order coefficient in \eqref{trpoly}. Thus, in the limiting case of $k \to \infty$, we see that the Rényi entropies converge to the following limit:\footnote{Note that this argument is not valid for $m=0$ since $\text{Tr}[\sigma^0] = k+1 \Rightarrow \textsf{y}(0)=1$. The zeroth Rényi entropy diverges as $k \to \infty$ since $\mathcal{R}_0 = \ln \text{dim}\, \mathcal{H}_{T^2} = \ln (k+1)$.}
\begin{equation}
\boxed{\lim_{k \to \infty} \mathcal{R}_m = \frac{1}{1-m} \ln\left[\frac{C_{\textsf{lead}}^m}{(C_{\textsf{lead}}^1)^m}\right]} ~.
\end{equation} 
The large $k$ limits of the entanglement entropy and the minimum entropy can also be obtained by taking the following limits of above expression:
\begin{align}
\boxed{\lim_{k \to \infty} \mathcal{E} = \lim_{m \to 1} \, \frac{1}{1-m} \ln\left[\frac{C_{\textsf{lead}}^m}{(C_{\textsf{lead}}^1)^m}\right]} \quad;\quad
\boxed{\lim_{k \to \infty} \mathcal{R}_{\text{min}} = \lim_{m \to \infty} \, \frac{1}{1-m} \ln\left[\frac{C_{\textsf{lead}}^m}{(C_{\textsf{lead}}^1)^m}\right]} ~.
\end{align} 
The above analysis means that we only need the leading order coefficients in \eqref{trpoly} to obtain the large $k$ limits of Rényi entropies. We computed these limits for various torus links (some examples are given in later subsections) and we propose the following:
\begin{mdframed}[style=sid]
\textbf{Proposition 2.} \emph{The large $k$ limit of the SU(2)$_k$ Rényi entropies for the torus links of type $T_{p,pn}$ (except the Hopf link) is the sum of two parts: a universal part which is independent of $n$ and comprises of Riemann zeta functions, and a non-universal part (or the linking part) which depends explicitly on the linking number $n$.}
\end{mdframed}
To be more precise, the Proposition 2 can be stated as,
\begin{equation}
\boxed{\lim_{k \to \infty} \mathcal{R}_m = \frac{1}{1-m} \ln\left[\frac{\zeta(m \times s_p)}{\zeta(s_p)^m}\right] + \frac{1}{1-m} \ln\left[\frac{a_{m}}{a_{1}^m}\right]\equiv \mathcal{R}_m^{\text{uni}} + \mathcal{R}_m^{\text{link}}} ~.
\label{Renyi2pieces}
\end{equation} 
Here the first term, which we call as a universal term depends on Riemann zeta function evaluated at positive even integers:
\begin{equation}
\zeta(s) = \sum_{x=1}^{\infty} \frac{1}{x^{s}} ~,
\end{equation}
and the integer $s_p$ is given as,
\begin{equation}
s_p =\begin{cases}
2p-4, & \text{where $p \geq 3:$ valid for $n=1$} \\
2p-2, & \text{where $p \geq 2:$ valid for $n>1$}
\end{cases} ~.
\end{equation}
The argument that there is a universal term comprising of Riemann zeta functions in the large $k$ limit of Rényi entropy is based on the observation that the leading order coefficient of $k$ in $\text{Tr}[\sigma^m]$ will contain a multiplicative factor which is given by Riemann zeta function evaluated at even integers as given below:
\begin{equation}
C_{\textsf{lead}}^m = \begin{cases}
\zeta(2pm-4m) \times a_m, & \text{where $p \geq 3:$ valid for $n=1$} \\
\zeta(2pm-2m) \times a_m, & \text{where $p \geq 2:$ valid for $n>1$}
\end{cases} ~.
\end{equation}
For the links of type $T_{p,p}$, we explicitly show the computation of $C_{\textsf{lead}}^m$. For the links $T_{p,pn}$ with $n>1$, we provide empirical evidence for the appearance of a universal term $\zeta(2pm-2m)$ in the leading order coefficient while performing the large $k$ asymptotics. Note that this term occurs universally and contributes as the first piece in \eqref{Renyi2pieces} which we call as the universal part in the entropy. The structure of $a_m$ explicitly depends on the linking number $n$ and it contributes as the second piece in \eqref{Renyi2pieces} which we denote as the non-universal or the linking part. At the moment we do not have a general form of $a_m$ and we compute it on a case-by-case basis. We give some examples in the remaining part of this section.

As a part of Proposition 2, we can also write the limits of entanglement entropy as
\begin{equation}
\boxed{\lim_{k \to \infty} \mathcal{E} = \mathcal{E}^{\text{uni}} + \mathcal{E}^{\text{link}} = \left(\ln \zeta(s_p) - s_p\frac{\zeta'(s_p)}{\zeta(s_p)}\right) + \left(\ln a_{m} - \frac{d}{dm}\ln a_{m} \right)_{m=1}} ~,
\label{EEunivtop}
\end{equation}   
where the second term is evaluated at $m=1$ and $\zeta'(s)$ is the derivative of zeta function evaluated at $s$ and is given as,
\begin{equation}
\zeta'(s) = -\sum_{x=1}^{\infty} \frac{\ln x}{x^{s}} ~.
\end{equation} 
Similarly, the large $k$ limit of the minimum Rényi entropy can be given as,
\begin{equation}
\boxed{\lim_{k \to \infty} \mathcal{R}_{\text{min}} = \mathcal{R}_{\text{min}}^{\text{uni}} + \mathcal{R}_{\text{min}}^{\text{link}} = \ln \zeta(s_p)  + \lim_{m \to \infty} \frac{1}{1-m} \ln \left[\frac{a_{m}}{a_{1}^m}\right] } ~.
\label{Rminunivtop}
\end{equation}
In the following subsections, we will analyze the large $k$ limits for some of the links supporting the two propositions.
%............................................
%............................................
\subsection{For torus links of type $T_{p,p}$}
For these links, the eigenvalues of the unnormalized reduced density matrix are given as: 
\begin{equation}
\Lambda_{\alpha} = (\mathcal{S}_{0 \alpha})^{2-2p}\, (\mathcal{S}\mathcal{T}\mathcal{S})_{\alpha 0} \, (\mathcal{S}\mathcal{T}\mathcal{S})_{\alpha 0}^{*} ~.
\end{equation}
We know that $\mathcal{S}$ and $\mathcal{T}$ matrices for SU(2) group satisfy $\mathcal{S}^2 = 1$ and $(\mathcal{S}\mathcal{T})^3 = 1$. Using these, we arrive at the following identities:
\begin{equation}
\mathcal{S}\mathcal{T}\mathcal{S} = \mathcal{T}^{-1}\mathcal{S}\mathcal{T}^{-1} \quad;\quad (\mathcal{S}\mathcal{T}^{-1}\mathcal{S}) = \mathcal{T}\mathcal{S}\mathcal{T} ~.
\end{equation}
Thus we can write a close-form expression for the following matrix elements:
\begin{align}
(\mathcal{S}\mathcal{T}\mathcal{S})_{\alpha 0} &= (\mathcal{T}^{-1})_{\alpha \alpha}\,\mathcal{S}_{\alpha 0}\,(\mathcal{T}^{-1})_{00} = \sqrt{\frac{2}{k+2}}\, e^{\frac{i \pi  (k-\alpha^2+2\alpha)}{2 (k+2)}} \sin \left(\frac{\pi +\pi \alpha}{k+2}\right) \nonumber \\ 
(\mathcal{S}\mathcal{T}\mathcal{S})_{\alpha 0}^{*} &= \mathcal{T}_{\alpha \alpha}\,\mathcal{S}_{\alpha 0}\,\mathcal{T}_{00} = \sqrt{\frac{2}{k+2}}\, e^{\frac{-i \pi  (k-\alpha^2+2\alpha)}{2 (k+2)}} \sin \left(\frac{\pi +\pi \alpha}{k+2}\right) ~.
\end{align}
Using this, the eigenvalues can be simplified as,
\begin{equation}
\Lambda_{\alpha} = (\mathcal{S}_{0 \alpha})^{2-2p}\, \mathcal{T}_{\alpha \alpha}^{*}\,\mathcal{S}_{\alpha 0}\,\mathcal{T}_{00}^{*} \,\mathcal{T}_{\alpha \alpha}\,\mathcal{S}_{\alpha 0}\mathcal{T}_{00} = (\mathcal{S}_{0 \alpha})^{4-2p} ~.
\end{equation}
We note that the large $k$ expansion of $(\mathcal{S}_{0 \alpha})^{-2}$ for a fixed $\alpha$ is given as,
\begin{equation}
\frac{1}{(\mathcal{S}_{0 \alpha})^2} = \frac{k+2}{2} \csc^2\left(\frac{\pi+\pi \alpha}{k+2}\right) \sim \frac{k^3}{2 \pi ^2 (1+\alpha)^2} +\mathcal{O}(k^2) ~.
\end{equation}
Thus we can approximate $\Lambda_{\alpha}^m$ for a fixed $\alpha$ by the leading order coefficent in $k$:
\begin{equation}
\Lambda_{\alpha}^m \sim \frac{k^{3pm-6m}}{2^{pm-2m} \pi^{2pm-4m} (1+\alpha)^{2pm-4m}} ~.
\end{equation}
Moreover, because of the symmetry $\mathcal{S}_{0 \alpha} = \mathcal{S}_{0 (k-\alpha)}$, the eigenvalues also obey $\Lambda_{\alpha}=\Lambda_{k-\alpha}$. Thus the above large $k$ expansion of $\Lambda_{\alpha}^m$ holds for both fixed $\alpha$ and fixed $k-\alpha$. The two regions of $\alpha<<k$ and $k-\alpha<<k$ contribute equally and their total contribution can be combined to give the leading order term in the traces:
\begin{equation}
\text{Tr}[\sigma^m] = \sum_{\alpha=0}^k \Lambda_{\alpha}^m \sim \sum_{\alpha=0}^{\infty} 2\left(\frac{k^{3pm-6m}}{2^{pm-2m} \pi^{2pm-4m} (1+\alpha)^{2pm-4m}}\right) ~.
\end{equation}
The summation over $\alpha$ gives a Riemann zeta function which can be separated from the remaining terms. As a result the leading order coefficient in the series expansion of $\text{Tr}[\sigma^m]$ can be written as,
\begin{equation}
C_{\textsf{lead}}^m = \zeta(2pm-4m) \times \left( \frac{2^{1+2m-pm}}{\pi^{2pm-4m}} \right) \equiv \zeta(2pm-4m) \times a_m  ~.
\label{CleadforTpp}
\end{equation} 
The large $k$ limit of the Rényi entropy, therefore can be separated as,
\begin{equation}
\boxed{\lim_{k \to \infty} \mathcal{R}_m = \frac{1}{1-m}\ln\left[\frac{\zeta (2 p m-4 m)}{\zeta (2 p-4)^m}\right] + \ln 2} ~,
\label{RenyiTpp}
\end{equation} 
where the first term forms the universal part and is due to the presence of the multiplicative factor $\zeta(2pm-4m)$ in the leading order coefficient of $\text{Tr}[\sigma^m]$. The `$\ln 2$' contribution is the non-universal part and is coming from the multiplicative factor $a_m$ in the leading order coefficient of $\text{Tr}[\sigma^m]$. Since Hopf link is excluded from this analysis, the value of $p \geq 3$. Thus the Riemann zeta term reads as $\zeta(2m), \zeta(4m), \zeta(6m),\cdots$ for first few links.
From this result, we can also compute the large $k$ limits of the entanglement entropy and the minimum entropy:
\begin{alignat}{2}
&\lim_{k \to \infty} \mathcal{E} &&= \left(\ln \zeta(2p-4) - (2p-4)\frac{\zeta'(2p-4)}{\zeta(2p-4)}\right) + \ln 2 \nonumber \\
&\lim_{k \to \infty} \mathcal{R}_{\text{min}} &&= \ln \zeta(2p-4) + \ln 2 ~.
\label{EERminTpp}
\end{alignat}
One can also see this convergence from the plots presented in the figure \ref{RenyivskT33} for the link $T_{3,3}$. 
%............................................
%............................................
\subsection{For torus links of type $T_{p,pn}$}
For links $T_{p,pn}$ with $n >1$, the eigenvalues are given as,
\begin{equation}
\Lambda_{\alpha} = (\mathcal{S}_{0 \alpha})^{2-2p}\, (\mathcal{S}\mathcal{T}^n\mathcal{S})_{\alpha 0} \, (\mathcal{S}\mathcal{T}^n\mathcal{S})_{\alpha 0}^{*} ~.
\end{equation} 
Unfortunately, for $n>1$, we were not able to write a close-form expression for the matrix element $(\mathcal{S}\mathcal{T}^n\mathcal{S})_{\alpha 0}$. Thus we were not able to carry out the asymptotic analysis similar to what we did for the $T_{p,p}$ link. However motivated by the $T_{p,p}$ case, we tried to see if the leading order coefficients in $\text{Tr}[\sigma^m]$ contain a Riemann zeta function and indeed we found that there is a multiplicative factor $\zeta(2pm-2m)$ sitting in the coefficient where $p \geq 2$. For first few links (i.e. $p=2,3,4,5,\ldots$), the Riemann zeta term is $\zeta(2m), \zeta(4m), \zeta(6m), \zeta(8m)$. We will give one example of $T_{2,4}$ on how we were able to observe the $\zeta(2m)$ term in the leading order coefficient in $\text{Tr}[\sigma^m]$. A similar analysis can then be applied for all other links  $T_{p,pn}$ with $n >1$. Thus, we propose that in the asymptotic limit of $k \to \infty$, the leading order coefficient in the series expansion of $\text{Tr}[\sigma^m]$ can be written as,
\begin{equation}
C_{\textsf{lead}}^m = \zeta(2pm-2m) \times a_m  ~.
\label{CleadforTppn}
\end{equation} 
Thus the large $k$ limit of the Rényi entropy can be separated as,
\begin{equation}
\lim_{k \to \infty} \mathcal{R}_m = \frac{1}{1-m}\ln\left[\frac{\zeta (2 p m-2 m)}{\zeta (2 p-2)^m}\right] + \frac{1}{1-m}\ln\left[\frac{a_m}{(a_1)^m} \right] ~.
\label{RenyiTppn}
\end{equation} 
The first term in the above equation is the universal part which is same for any $n>1$. The second term in the Rényi entropy is non-universal because it comes from the multiplicative factor $a_m$ which is different for different values of $n$.  

As mentioned earlier, we obtain the traces as $\text{Tr}[\sigma^{m}]_{\ell}$ where this notation means that such traces are evaluated for $k = n \mathbb{Z} + \ell$ where $0 \leq \ell < n$ is a fixed integer. When we do the large $k$ analysis, we take $k$ to be very large such that $k = n \mathbb{Z} + \ell$ with fixed $\ell$. Thus the leading order coefficients and hence the limiting values of the Rényi entropies will depend on $\ell$ and we must quote the result for all values of $\ell \in \{ 0,1,\ldots, n-1\}$. We give some examples below. 
%............................................
%............................................
\subsubsection{$T_{p,2p}$ links}
For these links we find that the leading order coefficients ($C_{\textsf{lead}}^m$) of both $\text{Tr}[\sigma^m]_{\ell=0}$ and $\text{Tr}[\sigma^m]_{\ell=1}$ are same (this can be seen in some of the examples presented in Table \ref{PolT24}, Table \ref{PolT36} and Table \ref{PolT48} respectively). Thus the Rényi entropy should converge to a unique value irrespective of whether we consider $k$ to be even or odd. We also observe that the term $\zeta(2pm-2m)$ is present as a multiplicative factor in the leading order coefficients. We give $T_{2,4}$ link as an example on how we got this term in $C_{\textsf{lead}}^m$.

\begin{figure}[htbp]
\centerline{\includegraphics[width=6.0in]{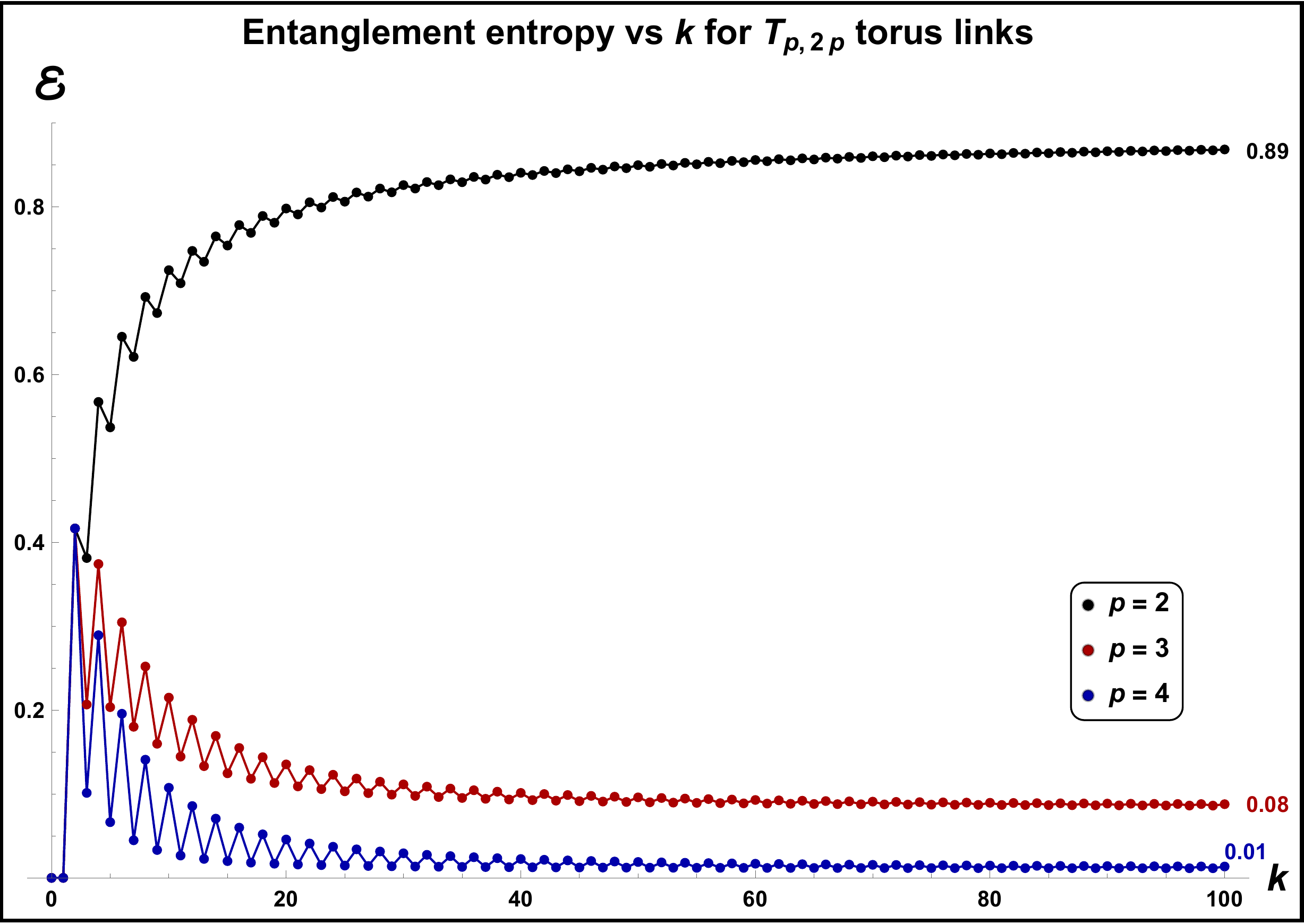}}
\caption[]{The variation of the SU$(2)_k$ entanglement entropy with $k$ for $T_{p,2p}$ torus links. The entropy shows two distinct patterns depending upon whether $k$ is even or odd. As $k$ becomes large, the two patterns converge to a unique value given in \eqref{EETp2p}. The corresponding numerical values up to two decimal places have been indicated against each plot.}
\label{EEvskTp2p}
\end{figure}

Referring to Table \ref{PolT24} for the polynomial expressions of $\text{Tr}[\sigma^m]$ for the link $T_{2,4}$, the leading order coefficients for first few values of $m$ are given below in their simplest form:
\begin{equation}
\{C_{\textsf{lead}}^m\} = \left\{\frac{1}{4},\frac{1}{24},\frac{1}{120},\frac{17}{10080},\frac{31}{90720},\frac{691}{9979200},\frac{5461}{389188800},\frac{929569}{326918592000},\ldots\right\} \nonumber
\end{equation} 
The numerators of these coefficients 
\begin{equation}
\text{Num}[\{C_{\textsf{lead}}^m\}] = \left\{1,1,1,17,31,691,5461,929569,\ldots\right\}
\end{equation}
form the OEIS sequence A276592 \cite{OEISA276592} which is the numerator of following:
\begin{equation}
\text{Num}[\{C_{\textsf{lead}}^m\}] = \text{Num}\left[\sum_{i=1}^{\infty} \frac{\pi^{-2m}}{(2i-1)^{2m}}\right] = \text{Num}\left[\left\{\pi^{-2m}(1-4^{-m}) \zeta(2m)\right\}\right] ~.
\end{equation}
Dividing $C_{\textsf{lead}}^m$ by $\pi^{-2m}(1-4^{-m}) \zeta(2m)$, we obtain the following sequence:
\begin{equation}
\left\{2,4,8,16,32,64,128,256,\ldots \right\} = 2^m ~. \nonumber
\end{equation} 
Thus, we get
\begin{equation}
C_{\textsf{lead}}^m = 2^{-m} \left(4^m-1\right) \pi ^{-2 m} \zeta (2 m) ~.
\end{equation} 
Computing for various values of $p$ in such a fashion, we can generalize the above expression for $T_{p,2p}$ link as,
\begin{equation}
C_{\textsf{lead}}^m = \zeta (2pm-2m) \times \frac{2^{3 m-m p}-2^{5 m-3 m p}}{\pi^{2pm-2m}} = \zeta (2pm-2m) \times a_m ~.
\label{leadcoeffTp2p}
\end{equation} 
The large $k$ limit of the Rényi entropy for $T_{p,2p}$ links can be separated as, 
\begin{equation}
\boxed{\lim_{k \to \infty} \mathcal{R}_m = \frac{1}{1-m} \ln\left[\frac{\zeta(2mp-2m)}{\zeta(2p-2)^m}\right] + \frac{1}{1-m} \ln\left[\frac{(2^{2mp}-2^{2m})}{(2^{2p}-2^2)^m}\right]} ~.
\end{equation}
Using this expression, the large $k$ limits of the entanglement entropy and the minimum entropy can be computed by taking the $m \to 1$ and $m \to \infty$ limits and are given as:
\begin{equation}
\boxed{\lim_{k \to \infty} \mathcal{E} = \left(\ln \zeta(2p-2) - (2p-2)\frac{\zeta'(2p-2)}{\zeta(2p-2)}\right) + \left(\ln \left(4^p-4\right)-\frac{\left(4^p p-4\right)}{4^p-4}\ln 4\right)} ~.
\label{EETp2p}
\end{equation}
\begin{equation}
\boxed{\lim_{k \to \infty} \mathcal{R}_{\text{min}} = \ln \zeta(2p-2)  + \ln(\frac{4^p-4}{4^{p}})} ~.
\end{equation}
As an illustration of these results, we present the numerical plots in the figure \ref{EEvskTp2p} showing the variation of the entanglement entropy with $k$ for some of the $T_{p,2p}$ links. One can clearly see that the entropy admits two distinct patterns depending on whether $k$ is even or odd as discussed earlier. As $k$ increases, the two patterns merge and as $k \to \infty$, the two patterns converge to a unique value, which is evident from the fact that the leading order coefficient in \eqref{leadcoeffTp2p} is independent of the value of $\ell$.   

We further checked the leading order coefficients in $\text{Tr}[\sigma^m]$ for torus links $T_{p,pn}$ for several values of $n$ and $p$ and find that the term $\zeta(2pm-2m)$ appears as a  multiplicative factor in all the cases. This term contributes as the universal part in the large $k$ limit of the Rényi entropy as mentioned in \eqref{Renyi2pieces}. The remaining factor is the term $a_m$ which is different for different values of $n$. This term contributes as the non-universal or the linking part in \eqref{Renyi2pieces}. In the following subsections, we give the expression of $a_m$ and obtain the asymptotic limit of the Rényi entropies for some more links.
%............................................
%............................................
\subsubsection{$T_{p,3p}$ links}
\begin{figure}[htbp]
\centerline{\includegraphics[width=6.0in]{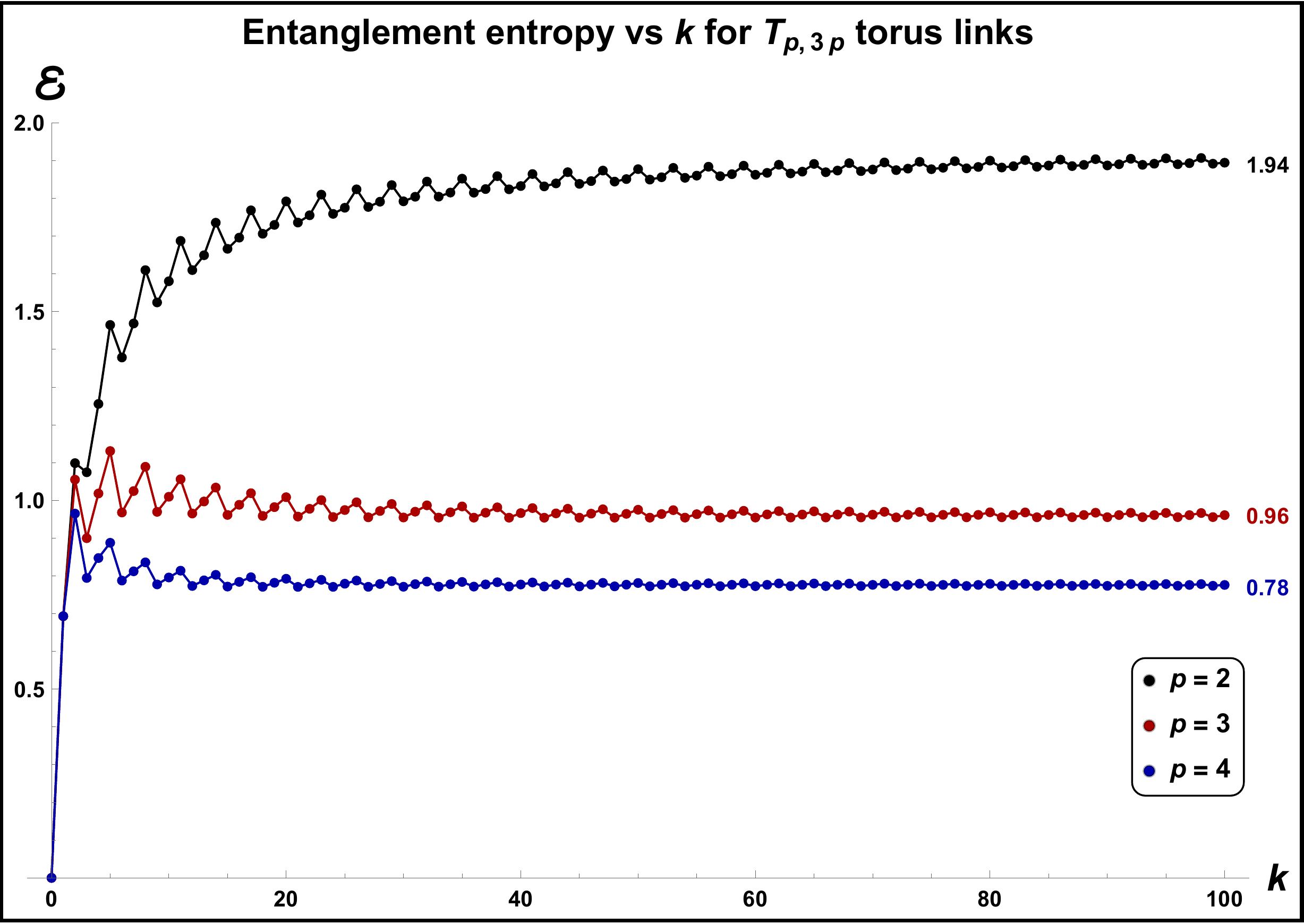}}
\caption[]{The variation of the SU$(2)_k$ entanglement entropy with $k$ for $\ket{T_{p,3p}}$ torus links. For each plot, we see three distinct patterns of entropy depending upon the values of $\ell \in \{ 0,1,2 \}$. As $k$ becomes large, the three patterns converge to a unique value which is given in the \eqref{EETp3p}. The corresponding values up to 2 decimal places have been indicated against each plot. }
\label{EEvskTp3p}
\end{figure}

The leading order coefficient is given as,
\begin{equation}
C_{\textsf{lead}}^m = 2^{1-mp}\times 3^{m-2mp} \times(3^{2mp}-3^{2m}) \times \frac{\zeta(2mp-2m)}{\pi^{2mp-2m}} ~.
\label{leadcoeffTp3p}
\end{equation}
Thus we obtain the large $k$ limit of the Rényi entropies as,
\begin{equation}
\boxed{\lim_{k \to \infty} \mathcal{R}_m = \frac{1}{1-m} \ln\left[2^{1-m}\times \frac{3^{2mp}-3^{2m}}{(3^{2p}-3^2)^m} \times \frac{\zeta(2mp-2m)}{\zeta(2p-2)^m}\right]}~.
\end{equation}
Using this expression, the large $k$ limits of the entanglement entropy and the minimum entropy can be computed by taking the $m \to 1$ and $m \to \infty$ limits and are given as:
\begin{equation}
\boxed{\lim_{k \to \infty} \mathcal{E} = \ln \zeta(2p-2) - (2p-2)\frac{\zeta'(2p-2)}{\zeta(2p-2)} + \left(\ln(9^{p}-9) - \frac{\left(9^p p-9\right)\ln 9}{9^p-9}  + \ln 2 \right)} ~.
\label{EETp3p}
\end{equation}
\begin{equation}
\boxed{\lim_{k \to \infty} \mathcal{R}_{\text{min}} = \ln \zeta(2p-2) + \ln(\frac{9^p-9}{9^p}) + \ln 2} ~.
\end{equation}
As an example, we show numerical plots in the figure \ref{EEvskTp3p} showing the variation of the entanglement entropy with $k$ for some of the $T_{p,3p}$ links. One can clearly see that the entropy follows three distinct patterns for $\ell = 0, 1, 2$ as discussed in earlier sections. As $k$ increases, the three patterns start to merge and as $k \to \infty$, the patterns converge to a unique value, which can also be seen from the leading order coefficient in \eqref{leadcoeffTp3p} which is independent of the value of $\ell$.   
%............................................
%............................................
\subsubsection{$T_{p,4p}$ links}
\begin{figure}[htbp]
\centerline{\includegraphics[width=6.0in]{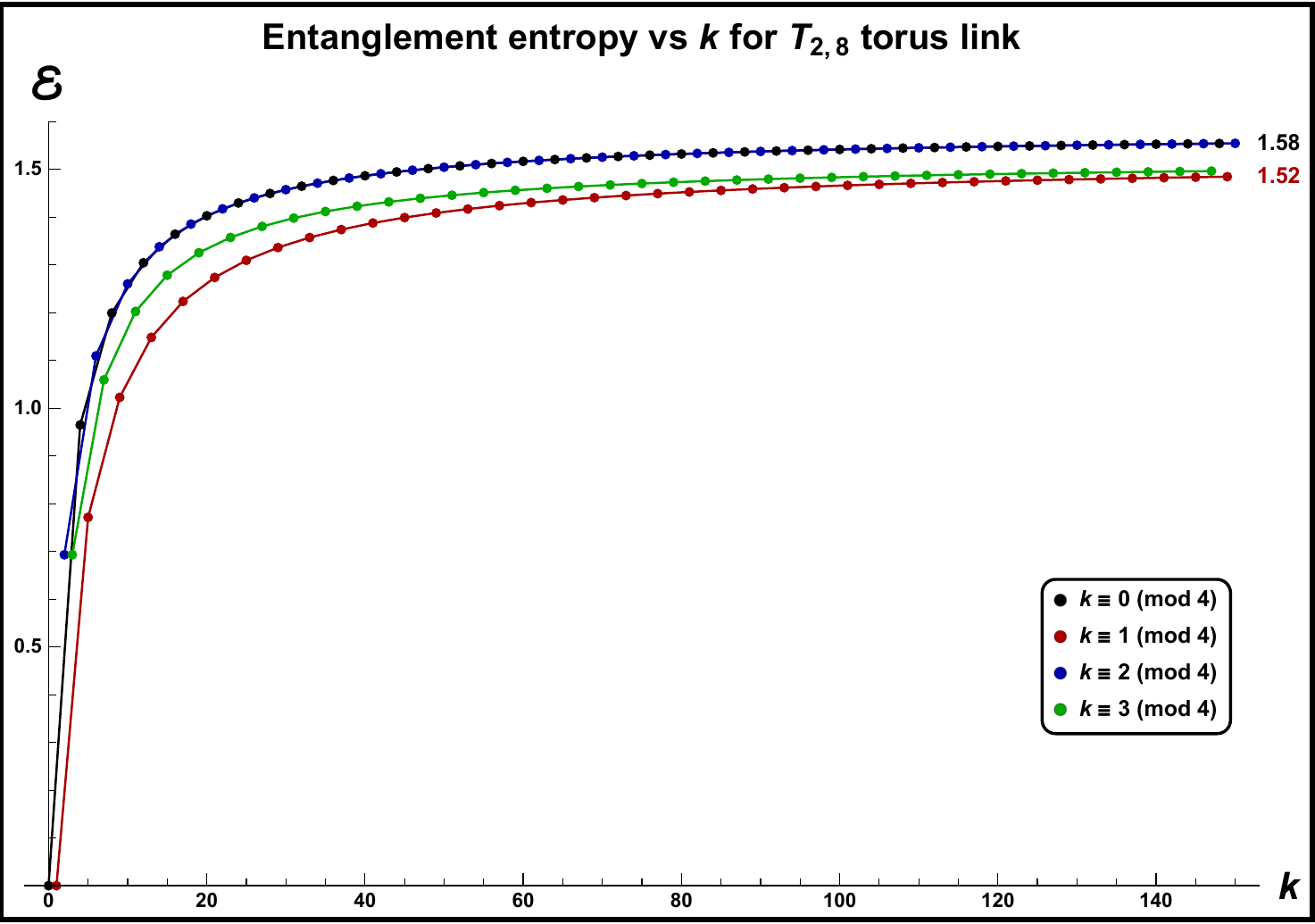}}
\caption[]{The variation of the SU$(2)_k$ entanglement entropy with $k$ for $\ket{T_{2,8}}$ torus link. The plots are shown for values of $\ell=0,1,2,3$ where $k \equiv \ell \,(\text{mod}\,\,4)$. As $k$ becomes large, the entropies for $\ell = 0,2$ and $\ell = 1,3$ respectively converge to same values as given in \eqref{EET28}. The corresponding asymptotic values up to 2 decimal places have been indicated against the plots.}
\label{EEvskT28}
\end{figure}

The leading order coefficients for this case are given as,
\begin{equation}
C_{\textsf{lead}}^m = 
\begin{cases}
\dfrac{2^{1+2m-3mp}\times \left(4^{mp}-4^m\right)\, \zeta (2mp-2m)}{\pi^{2mp-2m}} & \text{, for $\ell = 0$ or 2} \\[7.5pt]
\dfrac{2^{2m-5mp}\times \left(4^{mp}+8^m\right) \left(4^{mp}-4^m\right)\, \zeta (2mp-2m)}{\pi^{2mp-2m}} & \text{, for $\ell = 1$ or 3}
\end{cases} ~.
\end{equation}
Thus the large $k$ limit of the Rényi entropies are given as:
\begin{equation}
\boxed{\lim_{k \to \infty} \mathcal{R}_m = 
\begin{cases}
\dfrac{1}{1-m} \ln\left[2^{1-m}\dfrac{(4^{mp}-4^m)\,\zeta(2mp-2m)}{(4^p-4)^m\,\zeta(2p-2)^m}\right] & \text{, for $\ell = 0$ or 2} \\[13.5pt]
\dfrac{1}{1-m} \ln\left[\dfrac{(4^{mp}+8^m)(4^{mp}-4^m)\,\zeta(2mp-2m)}{(4^{p}+8)^m(4^p-4)^m\,\zeta(2p-2)^m}\right] & \text{, for $\ell = 1$ or 3}
\end{cases}} ~.
\end{equation}
Using this expression, the large $k$ limit of the entanglement entropy can be computed by taking the $m \to 1$ limit as:
\begin{equation}
\boxed{\lim_{k \to \infty} \mathcal{E} = 
\begin{cases}
\mathcal{X}_{2p-2} - \dfrac{4^p}{4^p-4}\ln(4^{p-1}) + \ln(\dfrac{4^p-4}{4}) + \ln 2 & \text{, for $\ell = 0$ or 2} \\[8.5pt]
\mathcal{X}_{2p-2} - \alpha_p\ln 16 + \ln(4^p-4)+\ln(4^p+8) & \text{, for $\ell = 1$ or 3}
\end{cases}}  ~,
\label{EET28}
\end{equation}
where we have defined 
\begin{equation}
\mathcal{X}_{2p-2} = \ln \zeta(2p-2) - (2p-2)\frac{\zeta'(2p-2)}{\zeta(2p-2)} \quad;\quad \alpha_p = \frac{(16^p + 2^{2p+1})p+4^{p+1}-40}{(4^p-4)(4^p+8)} ~.
\end{equation}
Similarly the large $k$ limit of the minimum entropy is given as:
\begin{equation}
\boxed{\lim_{k \to \infty} \mathcal{R}_{\text{min}} = 
\begin{cases}
\ln \zeta(2p-2) + \ln(2-2^{3-2 p})  & \text{, for $\ell = 0$ or 2} \\[4.5pt]
\ln \zeta(2p-2) + \ln(\dfrac{\left(4^p-4\right) \left(4^p+8\right)}{16^p}) & \text{, for $\ell = 1$ or 3}
\end{cases}} ~.
\end{equation}
In figure \ref{EEvskT28}, we plot the variation of the entanglement entropy with $k$ for $T_{2,8}$ link for different values of $\ell$. One can clearly see that at large $k$, the entropies for $\ell=0$ and $\ell=2$ converge to same value, while the entropies for $\ell=1$ and $\ell=3$ converge to same value.   
%............................................
%............................................
\subsubsection{$T_{p,5p}$ links}
\begin{figure}[htbp]
\centerline{\includegraphics[width=6.0in]{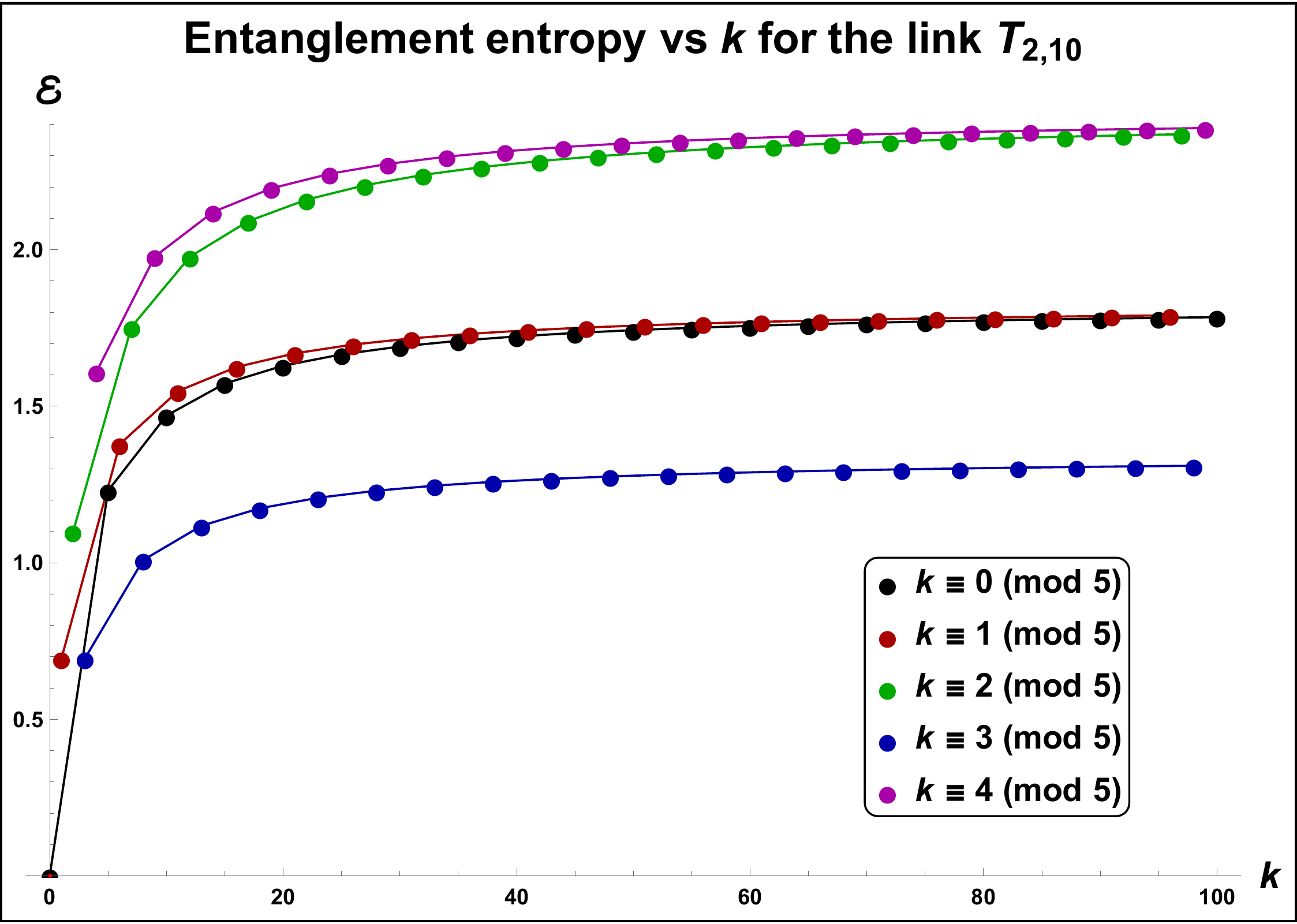}}
\caption[]{The variation of the SU$(2)_k$ entanglement entropy with $k$ computed for the state $\ket{T_{2,10}}$. The entropies follow different patterns for $\ell \in \{0,1,2,3,4\}$. In the large $k$ limit, patterns for $\ell=0,1$ and $\ell=2,4$ merge which is clear from \eqref{coeffleadn5}.}
\label{EEvskT210}
\end{figure}

In this case, we have
\begin{equation}
\text{Tr}[\sigma^m] = \sum_{i=0}^{3mp-4m} C_i^{\ell}(m)\, k^i.
\end{equation}
We checked that the leading order coefficients will form different patterns for different values of $\ell \in \{0,1,2,3,4\}$ such that:
\begin{equation}
C_{\textsf{lead}}^m(\ell=0) = C_{\textsf{lead}}^m(\ell=1) \quad;\quad C_{\textsf{lead}}^m(\ell=2) = C_{\textsf{lead}}^m(\ell=4)   ~.
\label{coeffleadn5}
\end{equation} 
Unfortunately, we were not able to find the close-form expression of the coefficients $a_m$ appearing in the \eqref{Renyi2pieces}. However, we present the numerical plot for the variation of the entanglement entropy of $T_{2,10}$ link with $k$ in the figure \ref{EEvskT210}. We see that the entropy is converging as $k$ becomes large which is in agreement with our Proposition 1. 
%............................................
%............................................
\subsubsection{$T_{p,6p}$ links}
As our last example, we consider the torus link $T_{p,6p}$. The trace of the powers of the unnormalized reduced density matrix can be written as the following polynomial expression in $k$:
\begin{equation}
\text{Tr}[\sigma^m] = \sum_{i=0}^{3mp-4m} C_i^{\ell}(m)\, k^i,
\end{equation}
where the coefficients $C_i^{\ell}(m)$ are rational numbers. We find that the leading order coefficients will form different patterns for different values of $\ell \in \{0,1,2,3,4,5\}$ such that:
\begin{equation}
C_{\textsf{lead}}^m(\ell=0) = C_{\textsf{lead}}^m(\ell=2) \quad;\quad C_{\textsf{lead}}^m(\ell=3) = C_{\textsf{lead}}^m(\ell=5)   ~.
\end{equation}
The analytic expression of the leading order coefficients are given below:
\begin{equation}
\begin{array}{|c|c|} \hline
\rowcolor{Gray}
 \ell & C_{\textsf{lead}}^m  \\ \hline
 \ell=0  & \dfrac{(4^{mp-m}-1)(9^{mp-m}+4^m+3^{2mp-m}-3^m-1)\,\zeta(2mp-2m)}{72^{mp-m} \times \pi^{2mp-2m}} \\[0.5cm]
 \ell=1  & \dfrac{(9^{mp}-9^m)\,\zeta(2mp-2m)}{2^{mp-m} \times 3^{2mp-m} \times \pi^{2mp-2m}}  \\[0.5cm]
 \ell=2  & \dfrac{(4^{mp-m}-1)(9^{mp-m}+4^m+3^{2mp-m}-3^m-1)\,\zeta(2mp-2m)}{72^{mp-m} \times \pi^{2mp-2m}}  \\[0.5cm]
 \ell=3  & \dfrac{(36^{mp-m}-9^{mp-m}-4^{mp-m}+3^{2mp-m}+4^{mp}-4^m-3^m+1)\,\zeta(2mp-2m)}{72^{mp-m} \times \pi^{2mp-2m}}  \\[0.5cm]
 \ell=4  & \dfrac{(9^{mp}-9^m)(4^{mp}-4^m)\,\zeta(2mp-2m)}{2^{3mp-m-1} \times 3^{2mp-m} \times \pi^{2mp-2m}}  \\[0.5cm]
 \ell=5  & \dfrac{(36^{mp-m}-9^{mp-m}-4^{mp-m}+3^{2mp-m}+4^{mp}-4^m-3^m+1)\,\zeta(2mp-2m)}{72^{mp-m} \times \pi^{2mp-2m}}  \\[0.3cm] \hline
\end{array}
\end{equation}
Thus the Rényi entropies converge and can be evaluated using the leading order coefficients. As an example, we plot the variation of the entanglement entropy with $k$ for the link $T_{2,12}$ in the figure \ref{EEvskT212}, where the limiting values of the entropy as $k \to \infty$ are given below:
\begin{equation}
\boxed{\lim_{k \to \infty} \mathcal{E} = 
\begin{cases}
\ln \zeta(2) - 2\dfrac{\zeta'(2)}{\zeta(2)} + \dfrac{1}{9} \ln\left(\dfrac{27}{256}\right) \approx 1.3877  &,\quad \text{for $\ell = 0$ or 2} \\[7.5pt]
\ln \zeta(2) - 2\dfrac{\zeta'(2)}{\zeta(2)} + \dfrac{1}{8} \ln \left(\dfrac{16777216}{387420489}\right) \approx 1.24519 &,\quad \text{for $\ell = 1$} \\[7.5pt]
\ln \zeta(2) - 2\dfrac{\zeta'(2)}{\zeta(2)} + \dfrac{1}{5} \ln \left(\dfrac{3125}{729}\right) \approx 1.92873 &,\quad \text{for $\ell = 3$ or 5} \\[7.5pt]
\ln \zeta(2) - 2\dfrac{\zeta'(2)}{\zeta(2)} + \dfrac{1}{12} \ln \left(\dfrac{65536}{14348907}\right) \approx 1.18855  &,\quad \text{for $\ell = 4$}
\end{cases}} ~.
\label{EET212limit}
\end{equation}
Similarly, the large $k$ limit of the minimum entropy for $T_{2,12}$ link will be given as,
\begin{equation}
\boxed{\lim_{k \to \infty} \mathcal{R}_{\text{min}} = 
\begin{cases}
\ln \zeta(2) \approx 0.4977  &,\quad \text{for $\ell = 0$ or 2} \\
\ln \zeta(2) + \ln 8 - \ln 9  \approx 0.37992 &,\quad \text{for $\ell = 1$} \\
\ln \zeta(2) + \ln 5 - \ln 3 \approx 1.0085  &,\quad \text{for $\ell = 3$ or 5} \\
\ln \zeta(2) + \ln 4 - \ln 3 \approx 0.78538  &,\quad \text{for $\ell = 4$}
\end{cases}} ~.
\end{equation} 
\begin{figure}[htbp] 
\centerline{\includegraphics[width=6.0in]{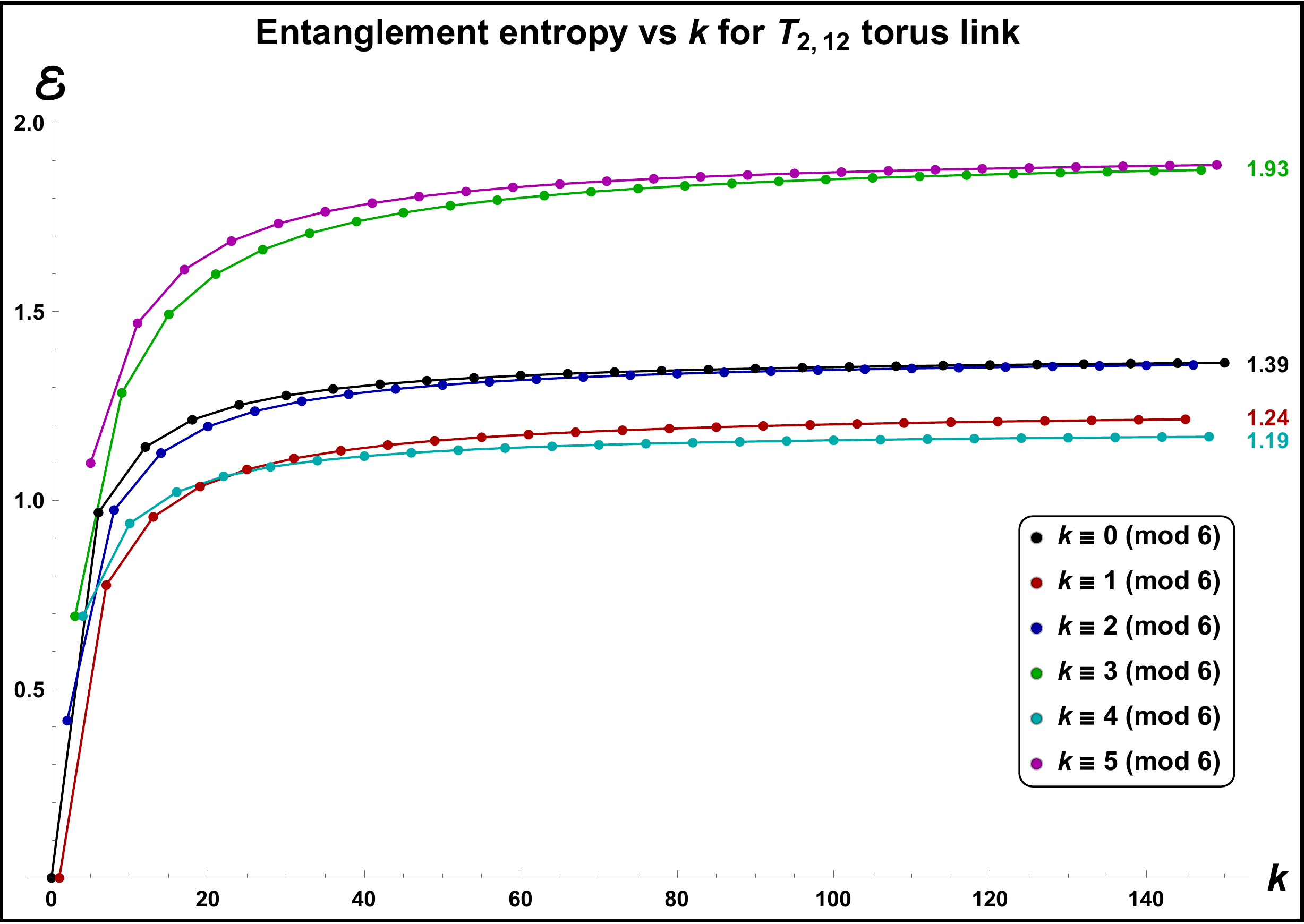}}
\caption[]{The variation of the SU$(2)_k$ entanglement entropy with $k$ computed for the link $T_{2,12}$. The entropies have been plotted for $k \equiv \ell \,(\text{mod}\,\,6)$. The asymptotic limits are given in \eqref{EET212limit} and are indicated against each plot in the figure.}
\label{EEvskT212}
\end{figure}
So far we have presented the large $k$ analysis of the Rényi entropies associated with the links $T_{p,pn}$ for various values of $n$, which provides enough evidence to support our two propositions. Finding the large $k$ analysis and obtaining the precise limits of entropies for a generic $n$ is a difficult exercise and we will not attempt this in this work. However, we will return to this question later in the section, where we show that it might be possible to extract the large $n$ behavior of the leading order coefficients and we present some exact results for $T_{2,2n}$ links.     
%............................................
%............................................
\subsection{Numerical results for generic torus links $T_{p,q}$}
\begin{figure}[htbp] 
\centerline{\includegraphics[width=6.5in]{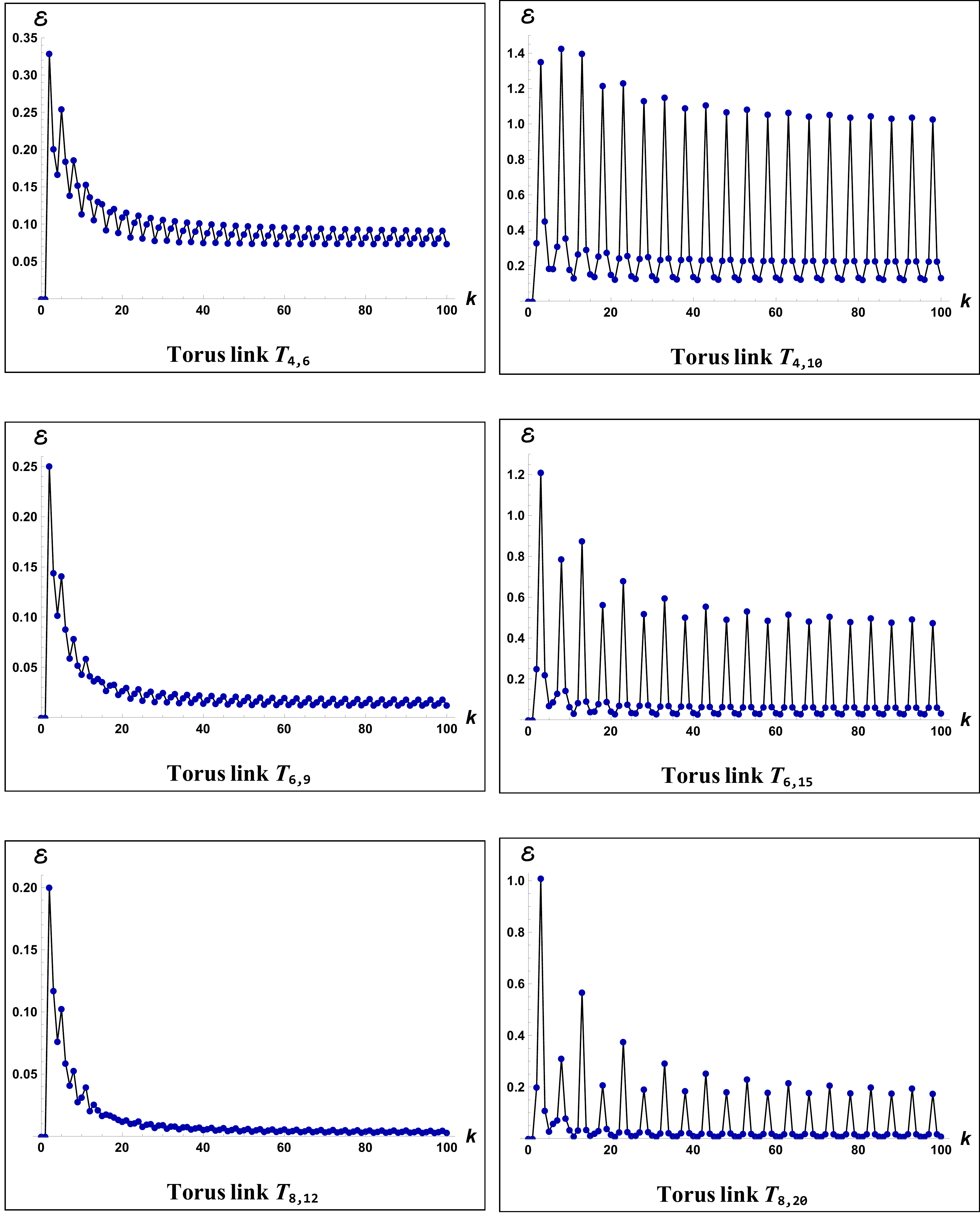}}
\caption[]{The variation of the SU$(2)_k$ entanglement entropy with $k$ for some of the torus links $T_{p,q}$ which are not of the form of $T_{p,pn}$. The black line is some interpolation curve.}
\label{EEvskTpq}
\end{figure}

For generic torus links $T_{p,q}$ which are not of the type $T_{p,pn}$, the eigenvalues of the unnormalized reduced density matrix are given as,
\begin{equation}
\Lambda_{\alpha} = (\mathcal{S}_{0 \alpha})^{2-2d} \left(\mathcal{S}^*X(p/d)\mathcal{T}^{\frac{q}{p}}\mathcal{S}\right)_{\alpha 0} \, \left(\mathcal{S}^*X(p/d)\mathcal{T}^{\frac{q}{p}}\mathcal{S}\right)_{\alpha 0}^{*}  ~,
\end{equation}
where $d=\text{gcd}(p,q)$ is the number of components in the link. The additional information required to compute these eigenvalues is the matrix $X(y)$ which is the matrix consisting of the Adams coefficients $X_{\beta \gamma}(y)$ associated with the SU(2) group as discussed earlier. A computational friendly formula to evaluate these coefficients is presented in \eqref{Adams-formula}. Unfortunately, for such links, the trace of the $m^{\text{th}}$ power of the unnormalized reduced density matrix, i.e.
\begin{equation}
\text{Tr}[\sigma^m] = \Lambda_0^m+\Lambda_1^m+\ldots+\Lambda_k^m
\end{equation}
is an irrational number for $k>1$. This makes it difficult to find a visible pattern in the traces as a function of $k$. However, we have numerically computed the Rényi entropies for several torus links which are not of the type $T_{p,pn}$ and observe that all the Rényi entropies, including the entanglement entropy converge as $k \to \infty$ which supports our Proposition 1. We present the variation of entanglement entropy with $k$ for some of the links in figure \ref{EEvskTpq}.  
This concludes our large $k$ analysis of SU(2)$_k$ Rényi entropies for torus links.
%............................................
%............................................
\subsection{The double limit: Large $k$ and large $n$ limits for torus links $T_{p,pn}$}
\begin{figure}[htbp] 
\centerline{\includegraphics[width=6.0in]{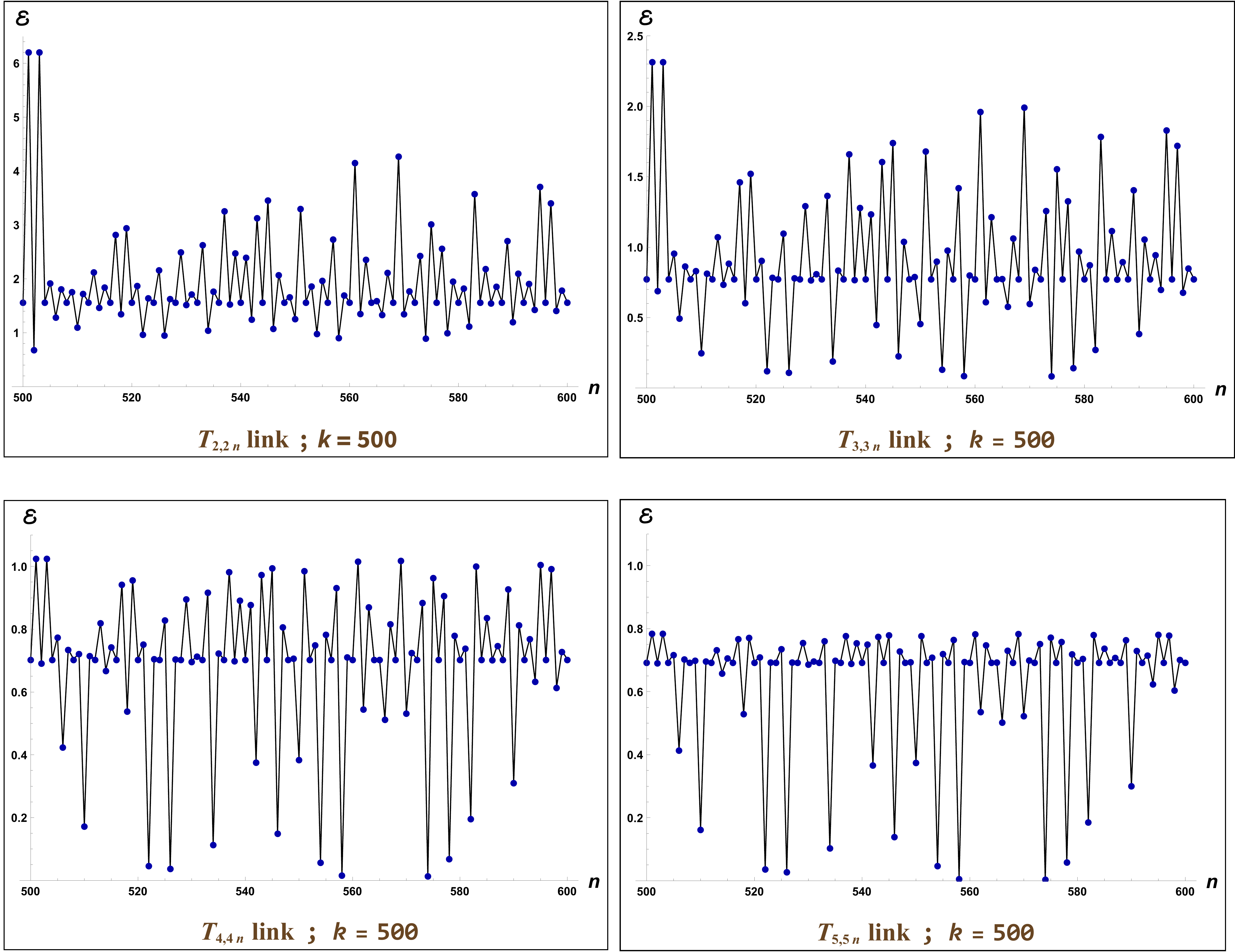}}
\caption[]{The variation of the entanglement entropy with $n$ for large values of $k$ and $n$.}
\label{EEvsnlargekn}
\end{figure}

In the previous analysis, we obtained the large $k$ limits of Rényi entropies for $T_{p,pn}$ links and proposed that the limiting value has two parts: a universal part which is independent of the linking number $n$ and comprises of Riemann zeta function. The other part has an intricate dependence on $n$ (see Proposition 2):
\begin{equation}
\lim_{k \to \infty} \mathcal{R}_m = \mathcal{R}_m^{\text{uni}} + \mathcal{R}_m^{\text{link}} ~.
\end{equation} 
In earlier subsections, we obtained a close-form expression for $\mathcal{R}_m^{\text{link}}$ for $T_{p,pn}$ links for some smaller values of $n$. Though obtaining a close-form expression for $\mathcal{R}_m^{\text{link}}$ for a generic $n$ is a non-trivial exercise, analyzing its asymptotic behavior when $n \gg 1$ is relatively easier. With enough analysis, which we present shortly, we propose the following:
\begin{mdframed}[style=sid]
\textbf{Proposition 3.} \emph{The SU(2)$_k$ Rényi entropies associated with the $T_{p,pn}$ type torus links converge to a finite value in the double scaling limit of $k \to \infty$ and $n \to \infty$ where $k = n\mathbb{Z} + \ell$ with $\ell$ being a finite number and the limiting values depend on the choice of $\ell$.}
\end{mdframed}
This proposition can be more precisely stated as,
\begin{equation}
\lim_{k \to \infty} \, \lim_{n \to \infty} \mathcal{R}_m = \lim_{n \to \infty} \, \lim_{k \to \infty} \mathcal{R}_m = \mathcal{R}_m^{\text{uni}} + \mathcal{P}_m ~,
\end{equation}
where the $\mathcal{P}_m$ part is the large $n$ limit of $\mathcal{R}_m^{\text{link}}$ and converges as $n \to \infty$. Note that $\mathcal{P}_m$ may have different values depending on how we chose to vary $n$ and $k$. For the current analysis, we consider $k = n \mathbb{Z} + \ell$ where $k,n,\ell \geq 0$. Now we take both $k \to \infty$ and $n \to \infty$ such that $\ell$ is a finite number. In such an asymptotic limit, the Rényi entropy converges to a finite number whose value depends upon the choice of $\ell$. As an example, we refer to \eqref{exnmodding} where the values of $\mathcal{P}_m$ are mentioned for $\ell=0$ and $\ell=1$. As an evidence of the Proposition 3, we present some numerical plots of entanglement entropy in figure \ref{EEvsnlargekn} for sufficiently large values of $k$ and $n$ for some of the $T_{p,pn}$ links.

\begin{figure}[htbp] 
\centerline{\includegraphics[width=6.0in]{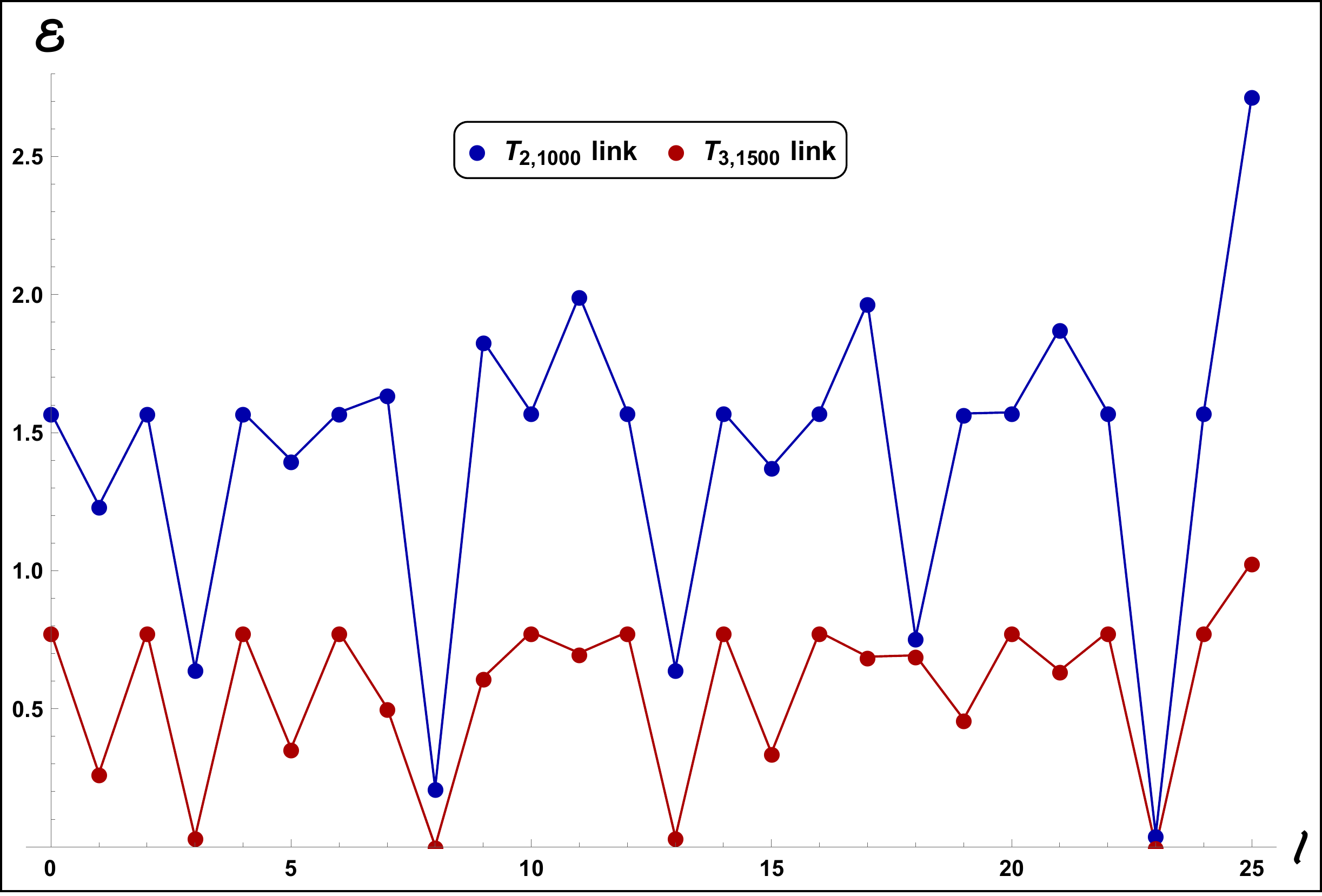}}
\caption[]{The variation of the entanglement entropy with $\ell$ for large values of $k$ and $n$. The two plots are respectively for $T_{2,2n}$ and $T_{3,3n}$ links with $n=500$ and the value of $k$ is given as $k = 500 + \ell$.}
\label{EElargekn}
\end{figure}

To find the double scaling limit, we first observe that the trace of $m^\text{th}$ power of unnormalized reduced density matrix is a finite degree polynomial in $k$ whose coefficients are Laurent polynomials in $n$. For a generic link $T_{p,pn}$ with $k=n \mathbb{Z} + \ell$ and a fixed value of $\ell$, we will have:
\begin{equation}
\text{Tr}[\sigma^m]_{\ell} = \sum_{i=0}^{3mp-4m}\left(\sum_{j=0}^{3mp-4m} \frac{C_{ij}(n)}{n^j}\right) k^i ~,
\label{largekn}
\end{equation} 
where the coefficients $C_{ij}$ may have a periodic dependence on $n$. For example, for the $T_{2,2n}$ link and $\ell=0$, we have
\begin{equation}
\text{Tr}[\sigma]_{\ell = 0} = 1 + k + \left(\frac{b}{8 n^2}+\frac{1}{4}\right)k^2 ~,
\end{equation}
and
\begin{align}
\text{Tr}[\sigma^2]_{\ell = 0} &= 1 + \left(-\frac{b}{3n}+\frac{4}{3}\right)k \, + \left(-\frac{a}{2 n^2}-\frac{a}{6 n}+\frac{2}{3}\right)k^2 \, + \left(\frac{a}{3 n^3}-\frac{a}{2 n^2}+\frac{1}{6}\right)k^3 \nonumber \\
&+ \left(\frac{b}{32 n^4}+\frac{a}{6 n^3}-\frac{a}{8 n^2}+\frac{1}{48}\right)k^4 ~,
\end{align}
where we have defined,
\begin{equation}
a = \cos \left(\frac{\pi  n}{2}\right)-(-1)^n \quad;\quad b = (-1)^n-1 ~.
\end{equation}
From the structure of \eqref{largekn}, we can see that when both $k$ and $n$ are large, we can perform the following approximation:
\begin{equation}
\frac{\text{Tr}[\sigma^m]_{\ell}}{(\text{Tr}[\sigma]_{\ell})^m} \sim \frac{A_{\text{lead}}^m}{(A_{\text{lead}}^1)^m} ~,
\end{equation} 
where we define the $A_{\text{lead}}^m$ to be the coefficient of $n^0k^{3mp-4m}$. Note that the coefficient $A_{\text{lead}}^m$ may contain periodic functions of $n$ but is finite as $n \to \infty$. Thus the Rényi entropies of $T_{p,pn}$ links converge in the double scaling limit of $k \to \infty$ and $n \to \infty$ and are given as,
\begin{equation}
\lim_{\substack{k \to \infty \\n \to \infty}} \mathcal{R}_m = \frac{1}{1-m} \ln(\frac{A_{\text{lead}}^m}{(A_{\text{lead}}^1)^m}) ~.
\end{equation}
As mentioned before, this limit has a universal part and a constant part:
\begin{equation}
\lim_{\substack{k \to \infty \\n \to \infty}} \mathcal{R}_m = \mathcal{R}_m^{\text{uni}} + \mathcal{P}_m ~,
\end{equation} 
where $\mathcal{R}_m^{\text{uni}}$ is the term containing the Riemann zeta functions (see \eqref{Renyi2pieces}) and the other piece is given as,
\begin{equation}
\mathcal{P}_m = \lim_{n \to \infty} \mathcal{R}_m^{\text{link}} ~.
\end{equation}
Obtaining a close form expression for $\mathcal{P}_m$ is difficult in general and we present numerical values for various links. However, with enough data, one can obtain $\mathcal{P}_m$ on a case by case basis. For some simple cases like $T_{2,2n}$ link and small values of $\ell$, the analytic results are given below:
\begin{equation}
\begin{array}{|c|c|c|} \hline
\rowcolor{Gray}
 T_{2,2n}  & \ell =0 & \ell =1 \\ \hline
 \mathcal{P}_m & \dfrac{1}{1-m}\ln(\dfrac{4\sinh (m \ln 2)}{3^m}) & \dfrac{1}{1-m}\ln(\dfrac{\left(9^m-1\right) \left(\left(2^m-2\right) (-1)^n+2^m+2\right)}{2^{4 m+1}}) \\ \hline
\end{array} ~.
\label{exnmodding}
\end{equation}
For these cases, the large $k$ and large $n$ limit of entanglement entropy will be given as,
\begin{equation}
\boxed{\lim_{\substack{k \to \infty \\n \to \infty}} \mathcal{E} = \begin{cases} 24 \ln A - 2\gamma -\dfrac{14}{3} \ln 2 & \text{,\, for $\ell=0$} \\[0.3cm] 24 \ln A - 2\gamma -\dfrac{13}{4} \ln 3 + \ln\left([n]_2+1\right) & \text{,\, for $\ell=1$} \end{cases}}  ~,
\label{EEl01}
\end{equation} 
where $A$ and $\gamma$ are the Glaisher's constant and Euler's constant respectively and $[n]_2$ denotes the value of $n$ modulo 2. Thus for $\ell =1$, there are two different limits depending upon whether we chose to study large $n$ limit with $n$ even or odd. The double scaling limit of the minimum entropy in this case is given as,
\begin{equation}
\boxed{\lim_{\substack{k \to \infty \\n \to \infty}} \mathcal{R}_{\text{min}} = \begin{cases} \ln\left(\dfrac{\pi^2}{4}\right) & \text{,\, for $\ell=0$} \\[0.4cm] \ln\left(\dfrac{4\pi^2}{27}\right) + \ln\left([n]_2+1\right) & \text{,\, for $\ell=1$} \end{cases}}  ~.
\end{equation} 
In figure \ref{EElargekn}, 
we plot the large $k$ and large $n$ limiting values of entanglement entropy for $T_{2,2n}$ and $T_{3,3n}$ links for various values of $\ell$, where we set $n=500$ and $k=500+\ell$. For $\ell=0$ and $\ell=1$, the values of $\mathcal{E}$ in the plot for $T_{2,1000}$ are the approximate values of the exact results given in \eqref{EEl01}.
%............................................
%............................................
\section{Large $k$ limit of torus link Rényi entropy and $2d$ Yang-Mills theory}
\label{sec5}
It is well known that the two-dimensional Yang-Mills theory (YM$_2$) is closely related to the large $k$ limit of the three-dimensional Chern-Simons theory. In our computations, we observed that the universal part appearing in the large $k$ limit of the Rényi entropy for $T_{p,pn}$ links \eqref{Renyi2pieces} can be written as the partition functions of the topological $2d$ Yang-Mills theory and hence can be equated to the Rényi entropies associated with certain states prepared in the topological Yang-Mills theory. Further, we can write the universal part of the large $k$ limit of the entanglement entropy and the minimum entropy in terms of the symplectic volume of the moduli space of flat connections on certain Riemann surfaces. We will briefly discuss these results in the remainder of this section.       

An interesting feature of the YM$_2$ is that there are no propagating degrees of freedom and we can investigate the theory on spacetimes with non-trivial topology, for example $\Sigma_g$ which is a genus $g$ Riemann surface. The Yang-Mills action can be defined as,
\begin{equation}
S_{\text{YM}} = -\frac{1}{2e^2} \int_{\Sigma_g} d^2x\sqrt{\text{det } g_{ij}}\, \text{Tr} f^2 \equiv -\frac{1}{2e^2} \int_{\Sigma_g} d\mu\, \text{Tr} f^2 ~,
\end{equation}
where $e$ is the gauge coupling constant, $g_{ij}$ is the metric of $\Sigma_g$. The scalar $f$ is the Lie-algebra valued zero-from which is given in terms of the field strength as,
\begin{equation}
F_{ij}^a = \sqrt{\text{det } g_{ij}}\, \epsilon_{ij} f^a ~.
\end{equation}
The $\mu$ appearing in the action is the Riemannian measure or the area form and can be used to define the total area of the surface $\Sigma_g$ as,
\begin{equation}
\mathcal{A} = \int_{\Sigma_g} d\mu ~.
\end{equation}
The Yang-Mills partition function is given as
\begin{equation}
Z_{\text{YM}}(\Sigma_g;e^2 \mathcal{A}) = \int DA \, e^{-S_{\text{YM}}} ~,
\end{equation}
where the Feynman path integral is over the space of connections. Note that since the action is invariant under the scaling $e^2 \rightarrow te^2$, $\mu \rightarrow \mu/t$ for any $t \in \mathbb{R}$, the partition function only depends on the invariant combination $e^2\mathcal{A}$. When $e^2\mathcal{A}=0$, the theory becomes a topological field theory (which we shall denote as TYM theory) with the action given as,
\begin{equation}
S_{\text{TYM}} = -i \int_{\Sigma_g} \text{Tr}\, \phi F ~,
\label{YM2topaction}
\end{equation}
where $\phi$ acts as a Lagrange multiplier setting $F(A) = 0$. There is a beautiful prescription to evaluate the partition functions of a Riemann surface in the TYM theory which is along the lines of lattice gauge theory, and is achieved by triangulating the Riemann surface.\footnote{The computation of the partition function is independent of the choice of triangulation. The most convenient way is to cover the surface using polygons. For example, an orientable surface of genus $g$ can be built by considering a single $4g$-sided polygon and gluing together its sides appropriately.} These partition functions have been explicitly computed in the literature (see for example \cite{witten1991} and the references therein) and are given as,
\begin{equation}
Z_{\text{TYM}}(\Sigma_g) = \sum_R \frac{1}{(\text{dim}\,R)^{2g-2}} ~,
\label{YMactiontop}
\end{equation}
where $R$ is an irreducible representation of the gauge group $G$, which we assume to be SU($N$). There is a direct link between the TYM theory and the large $k$ limit of the $3d$ Chern-Simons theory. For example, the above formula for the partition function can be thought of as the large $k$ limit of the Verlinde formula for the number of SU$(N)_k$ conformal blocks on $\Sigma_g$. For the group SU(2), the partition function for the Riemann surfaces with genus $g>1$ will be given as,
\begin{equation}
Z_{\text{TYM}}(\Sigma_g) = \sum_{a=0}^{\infty} \frac{1}{(a+1)^{2g-2}} = \zeta(2g-2) ~.
\end{equation}
We see that these partition functions are given as zeta functions at even integers and they have a similar form as the universal part of the large $k$ limit of the Rényi entropies of torus links appearing in the earlier subsections. Referring to the \eqref{Renyi2pieces}, we can interpret the universal part $\mathcal{R}_m^{\text{uni}}$ of the large $k$ limit of the Rényi entropy of $T_{p,pn}$ link as following:
\begin{equation}
\boxed{\mathcal{R}_m^{\text{uni}}(T_{p,pn}) = \begin{cases} \dfrac{1}{1-m}\ln\left[\dfrac{Z_{\text{TYM}}(\Sigma_{mp+1-2m})}{Z_{\text{TYM}}(\Sigma_{p-1})^m}\right] &\text{,\, for $n=1$} \\[0.5cm] \dfrac{1}{1-m}\ln\left[\dfrac{Z_{\text{TYM}}(\Sigma_{mp+1-m})}{Z_{\text{TYM}}(\Sigma_{p})^m}\right] &\text{,\, for $n>1$} \end{cases}} ~.
\label{RmunivandYMZ}
\end{equation}
Note that for $n=1$, we have $p \geq 3$ since Hopf link is excluded from this. For $n>1$, we have $p \geq 2$. It can be further shown that $\mathcal{R}_m^{\text{uni}}$ is equal to the Rényi entropy associated with certain states prepared in TYM theory as discussed below. 
%............................................
%............................................
\subsection{$\mathcal{R}_m^{\text{uni}}$ as the Rényi entropy of certain states in topological Yang-Mills theory}
The quantum state associated with the Riemann surfaces in TYM theory are known in the literature. These states live in the physical Hilbert space $\mathcal{H}_{\text{TYM}}$ of the theory which consists of $L^2$-class functions on $G$. Thus a natural basis of the Hilbert space is the representation basis and is provided by the characters of the irreducible unitary representations. We will label the basis states as
\begin{equation}
\mathcal{H}_{\text{TYM}} = \left\{ \ket{R}\,: \text{$R$ is an irrep of the gauge group $G$} \right\} ~.
\end{equation}  
The orthonormality of the basis is ensured by the well known fact in group theory that the irreducible characters of a group provide an orthonormal basis for the class functions on $G$. Now consider a Riemann surface with genus $g$ and $y$ number of holes or boundaries. We shall denote such a surface by $\Sigma_{g,y}$. In the following calculations, we will see that we only need the Euler characteristic $\chi \equiv (2-2g-y)$ of the surface. When the desired quantities only depend on $\chi$, we will use the notation $\Sigma^{\chi}_{g,y}$ to denote a Riemann surface with  the Euler characteristic $\chi$. Given such a surface, the associated quantum state $\ket{\Sigma^{\chi}_{g,y}}$ lives in the following tensor product of Hilbert spaces:
\begin{equation}
\ket{\Sigma^{\chi}_{g,y}} \in \bigotimes_{i=1}^y \mathcal{H}_{\text{TYM}} ~.
\end{equation}  
These states can be explicitly computed along the lines of \cite{witten1991} and we simple quote the result here:
\begin{equation}
\ket{\Sigma^{\chi}_{g,y}} = \sum_{R} \frac{1}{(\text{dim}\,R)^{2g-2+y}} \bigotimes_{i=1}^y \ket{R} = \sum_{R} \frac{1}{(\text{dim}\,R)^{-\chi}} \ket{R,R,\ldots,R} ~,
\label{stateTYM}
\end{equation}
where the sum is over the irreducible representations of the gauge group $G$. This state has a GHZ-like structure where the Rényi entropies will be independent of the choice of the bi-partition of the Hilbert space (similar to the case of torus link states in $3d$ Chern-Simons theory). It is straightforward to compute the unnormalized reduced density matrix, which we shall denote as $\omega$. For the group SU(2) and the bi-partition $(x|y-x)$, we can write it as
\begin{equation}
\omega = \sum_{a=0}^{\infty} \frac{1}{(a+1)^{-2\chi}}  \left(\bigotimes_{i=1}^x \ket{a}\right) \otimes \left(\bigotimes_{i=1}^x \bra{a}\right) ~.
\end{equation}
The spectrum of this unnormalized reduced density matrix will be given as,
\begin{equation}
\lambda_a(\omega) = \frac{1}{(a+1)^{-2\chi}} ~,
\end{equation}
where each $\lambda_a$ is an eigenvalue of $\omega$ with $a \geq 0$. Thus the $m^{\text{th}}$ Rényi entropy for the state $\ket{\Sigma^{\chi}_{g,y}}$ in the $2d$ TYM theory is given as,
\begin{equation}
\mathcal{R}_m(\Sigma^{\chi}_{g,y}) = \frac{1}{1-m} \ln\left( \frac{\text{Tr}[\omega^m]}{\text{Tr}[\omega]^m} \right)= \frac{1}{1-m} \ln\left[ \frac{\sum_{a=0}^{\infty} \lambda_a^m}{(\sum_{a=0}^{\infty} \lambda_a)^m} \right] ~.
\end{equation} 
Noting that $\chi < 0$ for $y \geq 2$, the above result will be given as:\footnote{We would like to remark that this Rényi entropy can also be computed using the replica trick without the explicit knowledge of the state in \eqref{stateTYM}. This computation using the replica trick for a Riemann surface with boundaries was studied in \cite{Dwivedi:2020jyx} in the context of (1+1)-$d$ Chern-Simons theory where the Hilbert space was given in the representation basis similar to $\mathcal{H}_{\text{TYM}}$. Thus the calculations of \cite{Dwivedi:2020jyx} can be straightforwardly applied and we will arrive at the same result in \eqref{RETYM}.}
\begin{equation}
\boxed{\mathcal{R}_m(\Sigma^{\chi}_{g,y}) = \frac{1}{1-m} \ln\left[ \frac{\zeta(-2\chi m)}{\zeta(-2\chi)^m} \right]} ~.
\label{RETYM}
\end{equation}
Comparing it with the \eqref{Renyi2pieces}, we can write,
\begin{equation}
\boxed{\mathcal{R}_m^{\text{uni}}(T_{p,pn}) = \begin{cases} \mathcal{R}_m(\Sigma^{2-p}_{g,y}) &\text{,\, for $n=1$} \\ \mathcal{R}_m(\Sigma^{1-p}_{g,y}) &\text{,\, for $n>1$} \end{cases}} ~.
\label{RmuniandRYM}
\end{equation}
Thus we see that the universal part in the large $k$ limit of the Rényi entropy of $T_{p,pn}$ link in the $3d$ Chern-Simons theory is equal to the Rényi entropy of the above mentioned states in the $2d$ topological Yang-Mills theory.
%............................................
%............................................
\subsection{$\mathcal{E}^{\text{uni}}$ and volume of moduli space of flat connections on Riemann surfaces}
The $2d$ TYM theory is also closely connected with the geometry of the moduli space $\mathcal{M}_g$ of flat connections on $\Sigma_g$ defined as \cite{Verlinde:1988sn}:
\begin{equation}
\mathcal{M}_g = \text{Hom}(\pi_1(\Sigma_g), G)/G ~,
\end{equation}
where the group acts by conjugation. This moduli space is a manifold with real dimension:
\begin{equation}
\text{dim}\,\mathcal{M}_g = (2g-2)\,\text{dim}\,G ~.
\end{equation} 
The moduli space $\mathcal{M}_g$ has a natural symplectic structure $\omega$ and thus a natural volume form $\theta$:
\begin{equation}
\theta = \frac{\omega^n}{n!} \quad;\quad n = \frac{\text{dim }\mathcal{M}_g}{2} ~.
\end{equation} 
The symplectic volume of $\mathcal{M}_g$ is given by integrating the volume form:
\begin{equation}
\text{vol}(\mathcal{M}_g) = \int_{\mathcal{M}_g} \theta ~.
\end{equation} 
Exploiting the connection with $3d$ Chern-Simons theory, the explicit expression for the symplectic volume of $\mathcal{M}_g$ is given by the following large $k$ limit of the current algebra \cite{witten1991}:
\begin{equation}
\text{vol}(\mathcal{M}_g) = \lim_{k \to \infty}  \left(\sum_{\alpha}\frac{k^{-n}}{(\mathcal{S}_{0 \alpha})^{2g-2}}\right) ~,
\end{equation} 
where the sum $\alpha$ is the summation over all the integrable representations of SU$(N)_k$. For the present work, we are interested in the volume computed for SU(2) group. In this case, we are looking for the following limit:
\begin{equation}
\text{vol}(\mathcal{M}_g) = \lim_{k \to \infty} \frac{k^{3-3g}\,(k+2)^{g-1}}{2^{g-1}} \sum_{\alpha=0}^k \left(\sin{\frac{\pi(\alpha+1)}{k+2}}\right)^{2-2g} ~.
\end{equation}
The crucial step in extracting the large $k$ behavior is to note that the sine function appearing above is the same for $\alpha$ as well as $k-\alpha$:
\begin{equation}
\sin{\frac{\pi(\alpha+1)}{k+2}} = \sin{\frac{\pi(k-\alpha+1)}{k+2}} ~.
\end{equation} 
Thus the following asymptotic expansion 
\begin{equation}
\sin{\frac{\pi(\alpha+1)}{k+2}} \sim \frac{\pi(\alpha+1)}{k+2}
\end{equation} 
is valid for both fixed $\alpha$ as well as fixed $k-\alpha$. The two regions of  $\alpha \ll k$ or $k-\alpha \ll k$ give equal contribution and
we arrive at the following limit
\begin{equation}
\text{vol}(\mathcal{M}_g) = \lim_{k \to \infty} 2\frac{\pi^{2-2g}\,k^{3-3g}\,(k+2)^{g-1}}{2^{g-1}\,(k+2)^{2-2g}} \sum_{\alpha=0}^{\infty} (\alpha+1)^{2-2g} ~.
\end{equation}
Thus the volume of $\mathcal{M}_g$ for a Riemann surface $\Sigma_g$ of genus $g \geq 2$ will have the following form for the SU(2) group:
\begin{equation}
\text{vol}(\mathcal{M}_g) = \frac{\zeta(2g-2)}{2^{g-2}\, \pi^{2g-2}}   ~.
\end{equation}
Hence we see that the universal part of the large $k$ limit of the entanglement entropy \eqref{EEunivtop} of $T_{p,pn}$ link can be written in terms of the symplectic volume of the moduli space of flat connections as,
\begin{equation}
\boxed{\mathcal{E}^{\text{uni}} = \begin{cases} \ln \text{vol}(\mathcal{M}_{p-1}) - (p-2)\dfrac{\text{vol}'(\mathcal{M}_{p-1})}{\text{vol}(\mathcal{M}_{p-1})} -\ln 2 &\text{,\, for $n=1$} \\[0.5cm] \ln \text{vol}(\mathcal{M}_{p}) - (p-1)\dfrac{\text{vol}'(\mathcal{M}_{p})}{\text{vol}(\mathcal{M}_{p})}-\ln 2 &\text{,\, for $n>1$} \end{cases}} ~,
\label{EEuniYM}
\end{equation}
where we have defined,
\begin{equation}
\text{vol}'(\mathcal{M}_{g}) = \left. \frac{d\, \text{vol}(\mathcal{M}_{x})}{dx} \right|_{x=g}  ~.
\end{equation}
The universal part in the large $k$ limit of the minimum entropy \eqref{Rminunivtop} of $T_{p,pn}$ link can similarly be written as,
\begin{equation}
\boxed{\mathcal{R}_{\text{min}}^{\text{uni}} = \begin{cases} \ln \text{vol}(\mathcal{M}_{p-1})-\ln 2 &\text{, for $n=1$} \\ \ln \text{vol}(\mathcal{M}_{p})-\ln 2 &\text{, for $n>1$} \end{cases}} ~.
\label{RminuniYM}
\end{equation}
%............................................
%............................................
%...................SECTION ends..............
%...................SECTION ends..............
\section{Conclusion and discussion}
\label{sec6}
In this work, we studied the multi-boundary entanglement structure of the state associated with the torus link complement $S^3 \backslash T_{p,q}$ in the set-up of SU(2)$_k$ Chern-Simons theory. These states can be explicitly constructed using the topological tools, and the probability amplitudes are precisely the Chern-Simons partition functions of $S^3 \backslash T_{p,q}$. Our main focus was to study the asymptotic behavior of the Rényi entropies associated with such states in the semiclassical limit of large $k$.

With enough analytic and numerical analysis, we proposed (Proposition 1) that in the large $k$ limit, the Rényi entropies, including the entanglement entropy, converge to a finite value for any generic torus link. To compute the limiting value of the entropies, we calculated the traces of $m^{\text{th}}$ power of the unnormalized reduced density matrix ($\text{Tr}[\sigma^m]$) for various values of $k$. For the torus links of type $T_{p,pn}$, assuming that $k = n\mathbb{Z} + \ell$ where $\ell$ is a fixed integer between 0 and $n-1$, we find that the traces $\text{Tr}[\sigma^m]_{\ell}$ can be expressed as a finite degree polynomial in $k$ for each value of $\ell$. The important point is that the degree of the polynomial is proportional to $m$ as mentioned in \eqref{degreePolypnlink}. As a result, the quantity $\text{Tr}[\sigma^m] / \text{Tr}[\sigma]^m$ and hence the $m^{\text{th}}$ Rényi entropy converges to a finite value in the limiting case of $k \to \infty$ and is determined by the leading order coefficient ($C_{\text{lead}}^m$) of $k$ in $\text{Tr}[\sigma^m]$. 

For the links of type $T_{p,p}$, we show that $C_{\text{lead}}^m$ has a multiplicative factor which is equal to the Riemann zeta function at positive even integer (see \eqref{CleadforTpp}). For links $T_{p,pn}$ with $n>1$, we provide empirical evidence that the $C_{\text{lead}}^m$ contains the Riemann zeta function evaluated at positive even integer \eqref{CleadforTppn}. Since this factor is universal, it contributes as a universal part ($\mathcal{R}_m^{\text{uni}}$) in the large $k$ limit of the Rényi entropy as mentioned in 
\eqref{Renyi2pieces}. The remaining factor in $C_{\text{lead}}^m$ is denoted as $a_m$ which has an intricate dependence on the linking number $n$. This term contributes to the large $k$ limit of the Rényi entropy, which we termed as non-universal part or the linking part ($\mathcal{R}_m^{\text{link}}$). At the moment, we do not have a close-form expression for $a_m$ and we evaluated it on case-by-case basis for several links. The large $k$ limit of the entanglement entropy and the minimum entropy can be easily obtained by taking the $m \to 1$ and $m \to \infty$ limits of the limiting values of $m^{\text{th}}$ Rényi entropy. Because of the presence of $\mathcal{R}_m^{\text{uni}}$ term, the entanglement entropy and the minimum entropy will also have universal parts: $\mathcal{E}^{\text{uni}} = \lim_{m \to 1} \mathcal{R}_m^{\text{uni}}$ and $\mathcal{R}_{\text{min}}^{\text{uni}} = \lim_{m \to \infty} \mathcal{R}_m^{\text{uni}}$ as given in \eqref{EEunivtop} and \eqref{Rminunivtop} respectively. We have also studied the large $k$ and large $n$ limits of the Rényi entropies of $T_{p,pn}$ link when $k = n\mathbb{Z} + \ell$ and propose (Proposition 3) that the entropies converge to a fixed value for a given finite value of $\ell$. %In other words, the non-universal part $\mathcal{R}_m^{\text{link}}$ converges to a finite value as $n \to \infty$ and we present the analytical results for some simpler cases. 

There is an intimate relationship between the (2+1)-$d$ Chern-Simons theory and $2d$ rational conformal field theory. This relation has been extensively exploited in the canonical quantization of $3d$ Chern-Simons theory on $\Sigma_g \times \mathbb{R}$, which associates a Hilbert space $\mathcal{H}_{\Sigma_g}$ to $\Sigma_g$ whose basis is in one-to-one correspondence with the irreducible representations of the symmetry group of the $2d$ rational CFT. Moreover, it is well known that the $2d$ Yang-Mills theory is closely related to the large $k$ limit of the $3d$ Chern-Simons theory and hence to the large $k$ limit of the two-dimensional current algebra. %For example, the $2d$ topological Yang-Mills partition functions computed for Riemann surfaces is precisely the large $k$ limit of the Verlinde formula for the number of conformal blocks appearing in the current algebra of a compact group $G$ at level $k$ (eq.(\ref{YMactiontop})). 
We observe that the large $k$ limit of the multi-boundary Rényi entropy in $3d$ Chern-Simons theory may have a relation with the amplitudes or partition functions computed in $2d$ topological Yang-Mills theory. In fact, we see that the universal part $\mathcal{R}_m^{\text{uni}}$ appearing in the large $k$ limit of the $m^{\text{th}}$ Rényi entropy for $T_{p,pn}$ torus link can be written in terms of the TYM partition functions as given in \eqref{RmunivandYMZ}. Further, we computed the Rényi entropies of certain states associated with Riemann surfaces with boundaries in the $2d$ TYM theory and find that they are equal to $\mathcal{R}_m^{\text{uni}}$ as mentioned in \eqref{RmuniandRYM}. Moreover, the universal term in the large $k$ limits of the entanglement entropy and the minimum entropy of torus link $T_{p,pn}$ can be written in terms of the volume of the moduli space of flat connections on Riemann surfaces. The exact relation is given in \eqref{EEuniYM} and \eqref{RminuniYM}, respectively. It would be interesting to explore the physical interpretation of these results and to study this connection for the large $k$ limits of Rényi entropies of generic link states in $3d$ Chern-Simons theory. Moreover it would be nice if an interpretation of the non-universal or linking part $\mathcal{R}_m^{\text{link}}$ can be given in connection with the $2d$ topological Yang-Mills theory.

In this work, we have restricted to SU(2)$_k$ Chern-Simons theory, but we believe that the results presented here can be generalized to any simple or semi-simple compact gauge group. It would be worthwhile to verify this claim by doing explicit computations. In particular, extending this result to SU$(N)_k$ and analyzing the large $k$ and large $N$ behavior of the Rényi entropies would be fascinating. We hope to report this study in future work. 
%...................SECTION ends..............
%...................SECTION ends..............

\vspace{0.5cm}
\textbf{Acknowledgements} 
The authors would like to express sincere gratitude to P. Ramadevi for helpful discussions. They are also grateful to Piotr Sulkowski, Mauricio Romo, Andrei Mironov, and Bhabani Prasad Mandal for useful correspondence. VKS would like to thank Milosz Panfil, Helder Larraguivel and Sibasish Banerjee for discussion and correspondence. The work of SD is supported by the NSFC grant No. 11975158.
%...................APPENDIX ends..............
%...................APPENDIX ends..............

\appendix
\section{Adams coefficients for SU(2) group}
\label{appA}
The Adams operation on the character associated with a SU(2) representation is given as:
\begin{equation}
(\mathcal{A}^y \chi_{\beta})(t) = \chi_{\beta}(t^y) = \sum_{\gamma} X_{\beta \gamma}(y)\, \chi_{\gamma}(t) ~,
\end{equation} 
where the representations $\beta$ and $\gamma$ are:
\begin{equation}
\beta = \underbrace{\tiny\yng(4)}_{\beta} \quad;\quad \gamma = \underbrace{\tiny\yng(4)}_{\gamma} ~.
\end{equation}
The character is given as,
\begin{equation}
\chi_{\beta}(t) = \frac{t^{\beta+2}-t^{-\beta}}{t^2-1} ~.
\end{equation}
Thus the Adams coefficients are given by the following expansion:
\begin{equation}
\frac{t^{\beta y+2y}-t^{-\beta y}}{t^{2y}-1} = \sum_{\gamma} X_{\beta \gamma}(y)\, \frac{t^{\gamma+2}-t^{-\gamma}}{t^2-1}  ~.
\end{equation}
For $y=1$, it is trivial to see that:
\begin{equation}
X_{\beta \gamma}(1) = \delta_{\beta \gamma} ~.
\end{equation}
For $y>1$, the coefficients can be given as,
\begin{align}
X_{\beta \gamma}(y) = 
\begin{cases} 
{\begin{cases}
1, & \text{if } \gamma \equiv y[\beta]_2 \,\, (\text{mod } 2y) \\[0.3cm] 
-1, & \text{if } \gamma \equiv (2^{[\beta+1]_2})y-2 \,\, (\text{mod } 2y) \\
\end{cases}} & \text{, for  } 0 \leq \gamma \leq y \beta \\[0.9cm]
0 & \text{, otherwise}
\end{cases} ~,
\end{align}
where the notation $[x]_N$ denotes $x$ modulo $N$. The above formula for $y>1$ can be conveniently coded in a computational software by writing it in a compact form:
\begin{equation}
X_{\beta \gamma}(y) = \left( \delta([\gamma]_{2y},\, y[\beta]_2) - \delta([\gamma]_{2y},\, (2^{[\beta+1]_2})y-2) \right)\,\theta(y \beta - \gamma) ~,
\label{Adams-formula}
\end{equation}
where $\delta$ and $\theta$ are Kronecker delta and unit step functions respectively defined as,
\begin{equation}
\delta(a,b) = \begin{cases} 1, & \text{if } a = b \\ 
0, & \text{if } a \neq b
\end{cases} \quad;\quad \theta(a) = \begin{cases} 1, & \text{if } a \geq 0 \\ 
0, & \text{if } a < 0
\end{cases} ~.
\end{equation}
%...................SECTION ends..............
%...................SECTION ends..............
\newpage
\section{Trace of powers of unnormalized reduced density matrix}
\label{appB} 
As discussed earlier, the sequence of trace $\text{Tr}[\sigma^m]$ computed for various values of $k$ can be expressed as a finite degree polynomial in $k$. In this appendix, we list the functional form of $\text{Tr}[\sigma^m]$ as polynomials in $k$ for some of the $T_{p,pn}$ links for small values of $p$ and $n$.
%\begin{table}[h]
%\begin{center}
	\captionsetup{width=15cm}
\begin{longtable}{|c|c|}
		\hline
		\rowcolor{Gray}
		  $m$ & $\text{Tr}[\sigma^m]$\\
\hline
$1$ & 
\scalebox{0.70}{
\parbox[t]{20cm}
	  {$(k^3+6 k^2+11 k+6)/6$}
	  }\\ \hline
$2$ & 
\scalebox{0.70}{
\parbox[t]{20cm}
	  {$(k^6+12 k^5+70 k^4+240 k^3+469 k^2+468 k+180)/180$}
	  }\\ \hline
$3$ & 
\scalebox{0.70}{
\parbox[t]{20cm}
	  {$(2 k^9+36 k^8+309 k^7+1638 k^6+5964 k^5+15624 k^4+29041 k^3+35622 k^2+25164 k+7560)/7560$}
	  }\\ \hline
$4$ & 
\scalebox{0.70}{
\parbox[t]{20cm}
	  {$(3 k^{12}+72 k^{11}+832 k^{10}+6080 k^9+31254 k^8+119136 k^7+346896 k^6+784320 k^5+1374143 k^4+1804792 k^3+1648872 k^2+914400 k+226800)/226800$}
	  }\\ \hline 
$5$ & 
\scalebox{0.70}{
\parbox[t]{20cm}
	  {$(2 k^{15}+60 k^{14}+873 k^{13}+8138 k^{12}+54262 k^{11}+273988 k^{10}+1082884 k^9+3416424 k^8+8701440 k^7+17991424 k^6+30165683 k^5+40426238 k^4+41675256 k^3+30690288 k^2+14119920 k+2993760)/2993760$}
	  }\\ \hline
$6$ & 
\scalebox{0.70}{
\parbox[t]{20cm}
	  {$(1382 k^{18}+49752 k^{17}+873084 k^{16}+9895296 k^{15}+81045999 k^{14}+509035044 k^{13}+2540569928 k^{12}+10306388352 k^{11}+34482000696 k^{10}+95998901856 k^9+223438524192 k^8+435343953408 k^7+708573989483 k^6+955901529348 k^5+1048123499796 k^4+897142366944 k^3+557243407440 k^2+220136616000 k+40864824000)/40864824000$}
	  }\\ \hline
$7$ & 
\scalebox{0.70}{
\parbox[t]{20cm}
	  {$(60 k^{21}+2520 k^{20}+51782 k^{19}+690916 k^{18}+6707268 k^{17}+50340264 k^{16}+303124011 k^{15}+1500376842 k^{14}+6206010476 k^{13}+21694123048 k^{12}+64567822800 k^{11}+164342407008 k^{10}+358378327104 k^9+669261532032 k^8+1067438382839 k^7+1445833515394 k^6+1644443777292 k^5+1535905988136 k^4+1130334911568 k^3+609839007840 k^2+211775860800 k+35026992000)/35026992000$}
	  }\\
\end{longtable}
%\end{center}
%\end{table}
%...................Table ends..............
%...................Table ends..............
\newpage
\captionsetup{width=15cm}
	\begin{longtable}{|c|c|c|}
		\hline
		\rowcolor{Gray}
		  $m$ & $\text{Tr}[\sigma^m]_{\ell=0}$ & $\text{Tr}[\sigma^m]_{\ell=1}$\\
\hline
$1$ & 
\scalebox{0.70}{
\parbox[t]{9.4cm}
	  {$(k^2+4 k+4)/4$}
	  } &
		\scalebox{0.70}{
\parbox[t]{9.4cm}
	  {$(k^2+4 k+3)/4$}
	  } \\ \hline
$2$ & 
\scalebox{0.70}{
\parbox[t]{9.4cm}
	  {$(k^4+8 k^3+26 k^2+40 k+24)/24$}
	  } &
		\scalebox{0.70}{
\parbox[t]{9.4cm}
	  {$(k^4+8 k^3+26 k^2+40 k+21)/24$}
	  } \\ \hline
$3$ & 
\scalebox{0.70}{
\parbox[t]{9.4cm}
	  {$(2 k^6+24 k^5+125 k^4+360 k^3+608 k^2+576 k+240)/240$}
	  } & 
\scalebox{0.70}{
\parbox[t]{9.4cm}
	  {$(2 k^6+24 k^5+125 k^4+360 k^3+608 k^2+576 k+225)/240$}
	  }\\ \hline
$4$ & 
\scalebox{0.70}{
\parbox[t]{9.4cm}
	  {$(17 k^8+272 k^7+1960 k^6+8288 k^5+22498 k^4+40208 k^3+46400 k^2+31872 k+10080)/10080$}
	  } & 
\scalebox{0.70}{
\parbox[t]{9.4cm}
	  {$(17 k^8+272 k^7+1960 k^6+8288 k^5+22498 k^4+40208 k^3+46400 k^2+31872 k+9765)/10080$}
	  } \\ \hline 
$5$ & 
\scalebox{0.70}{
\parbox[t]{9.4cm}
	  {$(62 k^{10}+1240 k^9+11415 k^8+63600 k^7+237426 k^6+620760 k^5+1152460 k^4+1503200 k^3+1323072 k^2+714240 k+181440)/181440$}
	  } & 
\scalebox{0.70}{
\parbox[t]{9.4cm}
	  {$(62 k^{10}+1240 k^9+11415 k^8+63600 k^7+237426 k^6+620760 k^5+1152460 k^4+1503200 k^3+1323072 k^2+714240 k+178605)/181440$}
	  } \\ 
		$6$ & 
\scalebox{0.70}{
\parbox[t]{9.4cm}
	  {$(1382 k^{12}+33168 k^{11}+371668 k^{10}+2568720 k^9+12190431 k^8+41850864 k^7+106619524 k^6+203256240 k^5+288142712 k^4+296818368 k^3+211540608 k^2+94141440 k+19958400)/19958400$}
	  } & 
\scalebox{0.70}{
\parbox[t]{9.4cm}
	  {$(1382 k^{12}+33168 k^{11}+371668 k^{10}+2568720 k^9+12190431 k^8+41850864 k^7+106619524 k^6+203256240 k^5+288142712 k^4+296818368 k^3+211540608 k^2+94141440 k+19802475)/19958400$}
	  } \\
\end{longtable}
%...................Table ends..............
%...................Table ends..............
\newpage
\captionsetup{width=15cm}
	\begin{longtable}{|c|c|}
		\hline
		\rowcolor{Gray}
		  $m$ & $\text{Tr}[\sigma^m]_{\ell=0}$ \\
\hline
$1$ & 
\scalebox{0.70}{
\parbox[t]{20cm}
	  {$(k^5+10 k^4+48 k^3+128 k^2+176 k+96)/96$}
	  } \\ \hline
$2$ & 
\scalebox{0.70}{
\parbox[t]{20cm}
	  {$(17 k^{10}+340 k^9+3200 k^8+18560 k^7+73528 k^6+208544 k^5+432800 k^4+659200 k^3+713600 k^2+493056 k+161280)/161280$}
	  } \\ \hline
$3$ & 
\scalebox{0.70}{
\parbox[t]{20cm}
	  {$(691 k^{15}+20730 k^{14}+298745 k^{13}+2736890 k^{12}+17806218 k^{11}+87117052 k^{10}+331220120 k^9+997468560 k^8+2403054016 k^7+4644213376 k^6+7173711360 k^5+8764467200 k^4+8296204800 k^3+5821627392 k^2+2727475200 k+638668800)/638668800$}
	  } \\ \hline
$4$ & 
\scalebox{0.70}{
\parbox[t]{20cm}
	  {$(929569 k^{20}+37182760 k^{19}+721763240 k^{18}+9028138080 k^{17}+81548196388 k^{16}+565164764288 k^{15}+3117744284000 k^{14}+14019993713280 k^{13}+52211725087392 k^{12}+162705997166336 k^{11}+426950045000960 k^{10}+946102771993600 k^9+1770058757592576 k^8+2786458678452224 k^7+3665382787563520 k^6+3983759752560640 k^5+3516549828403200 k^4+2453674966843392 k^3+1288989776609280 k^2+459421640294400 k+83691159552000)/83691159552000$}
	  } \\ \hline 
$5$ & 
\scalebox{0.70}{
\parbox[t]{20cm}
	  {$(443861162 k^{25}+22193058100 k^{24}+541759923750 k^{23}+8586865730900 k^{22}+99169927393627 k^{21}+887987642127054 k^{20}+6405603349553080 k^{19}+38189732646180080 k^{18}+191570722110918192 k^{17}+818967433277429984 k^{16}+3011343925430410240 k^{15}+9586019865527439360 k^{14}+26533147583071931392 k^{13}+64017372567540660224 k^{12}+134747145683716915200 k^{11}+247224261170439127040 k^{10}+394362442113450115072 k^9+544459934753380368384 k^8+646118017978660290560 k^7+652584577366774251520 k^6+553194871523541319680 k^5+385810757017109397504 k^4+214795675157641297920 k^3+90680278724090265600 k^2+26147604398211072000 k+3892643213082624000)/3892643213082624000$}
	  } \\
\end{longtable}
%...................Table ends..............
%...................Table ends..............
\newpage
\captionsetup{width=15cm}
	\begin{longtable}{|c|c|}
		\hline
		\rowcolor{Gray}
		  $m$ & $\text{Tr}[\sigma^m]_{\ell=0}$ \\
\hline
$1$ & 
\scalebox{0.70}{
\parbox[t]{20cm}
	  {$(k^8+16 k^7+122 k^6+568 k^5+1784 k^4+3904 k^3+5728 k^2+4992 k+1920)/1920$}
	  } \\ \hline
$2$ & 
\scalebox{0.70}{
\parbox[t]{20cm}
	  {$(691 k^{16}+22112 k^{15}+341910 k^{14}+3382120 k^{13}+23928668 k^{12}+128367648 k^{11}+540406944 k^{10}+1823310720 k^9+4992520896 k^8+11163941888 k^7+20408607232 k^6+30384322560 k^5+36459115520 k^4+34418532352 k^3+24202174464 k^2+11243520000 k+2554675200)/2554675200$}
	  } \\ \hline
$3$ & 
\scalebox{0.70}{
\parbox[t]{20cm}
	  {$(6404582 k^{24}+307419936 k^{23}+7212882585 k^{22}+109960850252 k^{21}+1221918490044 k^{20}+10528133611296 k^{19}+73061415963952 k^{18}+418834480964544 k^{17}+2018642938977600 k^{16}+8282785973344256 k^{15}+29192013458371584 k^{14}+88921596860952576 k^{13}+235035702220312576 k^{12}+540189862721519616 k^{11}+1079904079191146496 k^{10}+1875137574084804608 k^9+2818814158083981312 k^8+3648759762398478336 k^7+4034770046946050048 k^6+3768000311277649920 k^5+2920475038284251136 k^4+1823804233151938560 k^3+867544428796968960 k^2+279776464050585600 k+45527990796288000)/45527990796288000$}
	  } \\
\end{longtable}
%...................Table ends..............
%...................Table ends..............
\newpage
\captionsetup{width=15cm}
	\begin{longtable}{|c|c|}
		\hline
		\rowcolor{Gray}
		  $m$ & $\text{Tr}[\sigma^m]_{\ell=0}$ \\
\hline
$1$ & 
\scalebox{0.70}{
\parbox[t]{20cm}
	  {$(2 k^2+9 k+9)/9$}
	  } \\ \hline
$2$ & 
\scalebox{0.70}{
\parbox[t]{20cm}
	  {$(k^4+10 k^3+45 k^2+99 k+81)/81$}
	  } \\ \hline
$3$ & 
\scalebox{0.70}{
\parbox[t]{20cm}
	  {$(13 k^6+201 k^5+1395 k^4+5535 k^3+13932 k^2+21384 k+14580)/14580$}
	  } \\ \hline
$4$ & 
\scalebox{0.70}{
\parbox[t]{20cm}
	  {$(123 k^8+2556 k^7+24374 k^6+139608 k^5+531027 k^4+1390284 k^3+2546316 k^2+3096792 k+1837080)/1837080$}
	  } \\ \hline
$5$ & 
\scalebox{0.70}{
\parbox[t]{20cm}
	  {$(671 k^{10}+17465 k^9+212040 k^8+1585050 k^7+8118243 k^6+29961225 k^5+81559710 k^4+164146500 k^3+242395416 k^2+248793120 k+132269760)/132269760$}
	  } \\  \hline 
$6$ & 
\scalebox{0.70}{
\parbox[t]{20cm}
	  {$(50443 k^{12}+1576338 k^{11}+23244639 k^{10}+214264270 k^9+1378946349 k^8+6552349254 k^7+23697184797 k^6+66205639170 k^5+143380778508 k^4+238589563608 k^3+299807676864 k^2+269490188160 k+130947062400)/130947062400$}
	  } \\  \hline
$7$ & 
\scalebox{0.70}{
\parbox[t]{20cm}
	  {$(1793613 k^{14}+65400237 k^{13}+1134746977 k^{12}+12434765889 k^{11}+96324154927 k^{10}+559341413631 k^9+2518923623931 k^8+8974394295867 k^7+25562724058872 k^6+58378909621032 k^5+106420528114992 k^4+152732418746544 k^3+168783802265088 k^2+136113519225600 k+61283225203200)/61283225203200$}
	  } \\
\end{longtable}
%...................Table ends..............
%...................Table ends..............
\newpage
\captionsetup{width=15cm}
	\begin{longtable}{|c|c|}
		\hline
		\rowcolor{Gray}
		  $m$ & $\text{Tr}[\sigma^m]_{\ell=0}$ \\
\hline
$1$ & 
\scalebox{0.70}{
\parbox[t]{20cm}
	  {$(4 k^5+44 k^4+225 k^3+630 k^2+891 k+486)/486$}
	  } \\ \hline
$2$ & 
\scalebox{0.70}{
\parbox[t]{20cm}
	  {$(82 k^{10}+1836 k^9+19380 k^8+126656 k^7+568743 k^6+1840608 k^5+4391550 k^4+7735824 k^3+9660465 k^2+7584516 k+2755620)/2755620$}
	  } \\ \hline
$3$ & 
\scalebox{0.70}{
\parbox[t]{20cm}
	  {$(50443 k^{15}+1696143 k^{14}+27413094 k^{13}+282249542 k^{12}+2070615675 k^{11}+11473422543 k^{10}+49678386912 k^9+171529970436 k^8+477666441846 k^7+1077469887780 k^6+1965284371944 k^5+2874657220272 k^4+3302123258286 k^3+2828501188884 k^2+1601656885800 k+441946335600)/441946335600$}
	  } \\ \hline
$4$ & 
\scalebox{0.70}{
\parbox[t]{20cm}
	  {$(28621321 k^{20}+1283282060 k^{19}+27933235660 k^{18}+392298592120 k^{17}+3986308449262 k^{16}+31159197860672 k^{15}+194485364208340 k^{14}+993305797449720 k^{13}+4220275410567333 k^{12}+15083230334979924 k^{11}+45671969368167480 k^{10}+117627988727667600 k^9+257944863156367464 k^8+480730087678245696 k^7+757664557239516480 k^6+1001094981300725760 k^5+1093048628668183620 k^4+960234837982767648 k^3+642076128802067040 k^2+289704983328028800 k+65151728794152000)/65151728794152000$}
	  } \\
\end{longtable}
%...................Table ends..............
%...................Table ends..............
\newpage
\captionsetup{width=15cm}
	\begin{longtable}{|c|c|}
		\hline
		\rowcolor{Gray}
		  $m$ & $\text{Tr}[\sigma^m]_{\ell=0}$ \\
\hline
$1$ & 
\scalebox{0.70}{
\parbox[t]{20cm}
	  {$(52 k^8+892 k^7+7165 k^6+34848 k^5+113598 k^4+256608 k^3+385965 k^2+341172 k+131220)/131220$}
	  } \\ \hline
$2$ & 
\scalebox{0.70}{
\parbox[t]{20cm}
	  {$(100886 k^{16}+3472156 k^{15}+57732270 k^{14}+614616820 k^{13}+4686614651 k^{12}+27146199372 k^{11}+123651210474 k^{10}+452443426380 k^9+1346820586650 k^8+3282225070608 k^7+6554836660752 k^6+10682518723200 k^5+14041874067813 k^4+14486585805864 k^3+11052880157304 k^2+5508704829600 k+1325839006800)/1325839006800$}
	  } \\ \hline
$3$ & 
\scalebox{0.70}{
\parbox[t]{20cm}
	  {$(15972720439 k^{24}+824632538331 k^{23}+20807184603678 k^{22}+341234997979528 k^{21}+4081886662012026 k^{20}+37896693534761250 k^{19}+283740837307609864 k^{18}+1757649813435140508 k^{17}+9170586103998750111 k^{16}+40820190978393087003 k^{15}+156445882219395529830 k^{14}+519617707289047175508 k^{13}+1502125271607470890536 k^{12}+3788679691043384240736 k^{11}+8343242069151251438496 k^{10}+16025273691628354011648 k^9+26770578484650894573696 k^8+38701953298342496542464 k^7+48053470617611601236172 k^6+50654639487367916561808 k^5+44489881287426083082192 k^4+31489125692025265209216 k^3+16873295891648127024960 k^2+6051341775902304672000 k+1076567166594567648000)/1076567166594567648000$}
	  } \\
\end{longtable}
%...................Table ends..............
%...................Table ends..............
\newpage
\captionsetup{width=15cm}
	\begin{longtable}{|c|c|}
		\hline
		\rowcolor{Gray}
		  $m$ & $\text{Tr}[\sigma^m]_{\ell=0}$ \\
\hline
$1$ & 
\scalebox{0.70}{
\parbox[t]{20cm}
	  {$(k^2+4 k+4)/4$}
	  } \\ \hline
$2$ & 
\scalebox{0.70}{
\parbox[t]{20cm}
	  {$(k^4+8 k^3+32 k^2+64 k+48)/48$}
	  } \\ \hline
$3$ & 
\scalebox{0.70}{
\parbox[t]{20cm}
	  {$(k^6+12 k^5+70 k^4+240 k^3+544 k^2+768 k+480)/480$}
	  } \\ \hline
$4$ & 
\scalebox{0.70}{
\parbox[t]{20cm}
	  {$(17 k^8+272 k^7+2128 k^6+10304 k^5+34048 k^4+78848 k^3+131072 k^2+147456 k+80640)/80640$}
	  } \\ \hline
$5$ & 
\scalebox{0.70}{
\parbox[t]{20cm}
	  {$(31 k^{10}+620 k^9+6090 k^8+37920 k^7+165648 k^6+530880 k^5+1276160 k^4+2298880 k^3+3096576 k^2+2949120 k+1451520)/1451520$}
	  } \\  \hline 
$6$ & 
\scalebox{0.70}{
\parbox[t]{20cm}
	  {$(691 k^{12}+16584 k^{11}+196064 k^{10}+1488960 k^9+8067048 k^8+32833152 k^7+103254272 k^6+254115840 k^5+490817536 k^4+736395264 k^3+846987264 k^2+707788800 k+319334400)/319334400$}
	  } \\  \hline
$7$ & 
\scalebox{0.70}{
\parbox[t]{20cm}
	  {$(5461 k^{14}+152908 k^{13}+2113566 k^{12}+18920720 k^{11}+122154032 k^{10}+600984384 k^9+2325800048 k^8+7209762560 k^7+18070660608 k^6+36687659008 k^5+60048142336 k^4+78095646720 k^3+79199207424 k^2+59454259200 k+24908083200)/24908083200$}
	  } \\
\end{longtable}
%...................Table ends..............
%...................Table ends..............
\newpage
\captionsetup{width=15cm}
	\begin{longtable}{|c|c|}
		\hline
		\rowcolor{Gray}
		  $m$ & $\text{Tr}[\sigma^m]_{\ell=0}$ \\
\hline
$1$ & 
\scalebox{0.70}{
\parbox[t]{20cm}
	  {$(k^5+10 k^4+48 k^3+128 k^2+176 k+96)/96$}
	  } \\ \hline
$2$ & 
\scalebox{0.70}{
\parbox[t]{20cm}
	  {$(17 k^{10}+340 k^9+3284 k^8+19904 k^7+83776 k^6+256256 k^5+582656 k^4+987136 k^3+1194752 k^2+912384 k+322560)/322560$}
	  } \\ \hline
$3$ & 
\scalebox{0.70}{
\parbox[t]{20cm}
	  {$(691 k^{15}+20730 k^{14}+303860 k^{13}+2869880 k^{12}+19486248 k^{11}+100671472 k^{10}+408969440 k^9+1332175680 k^8+3517229056 k^7+7556724736 k^6+13188096000 k^5+18552995840 k^4+20621076480 k^3+17185370112 k^2+9494323200 k+2554675200)/2554675200$}
	  } \\ \hline
$4$ & 
\scalebox{0.70}{
\parbox[t]{20cm}
	  {$(929569 k^{20}+37182760 k^{19}+730937720 k^{18}+9358419360 k^{17}+87362939728 k^{16}+631454540288 k^{15}+3665187299840 k^{14}+17495053486080 k^{13}+69782338409472 k^{12}+235097062793216 k^{11}+673579725578240 k^{10}+1647275938447360 k^9+3441490862800896 k^8+6130226365988864 k^7+9263672637521920 k^6+11775309040844800 k^5+12417719352360960 k^4+10585552883023872 k^3+6899909622497280 k^2+3043731072614400 k+669529276416000)/669529276416000$}
	  } \\ \hline
$5$ & 
\scalebox{0.70}{
\parbox[t]{20cm}
	  {$(221930581 k^{25}+11096529050 k^{24}+273617920680 k^{23}+4419378970480 k^{22}+52430090077256 k^{21}+485906328343632 k^{20}+3654186845278400 k^{19}+22873556251020160 k^{18}+121318152073766016 k^{17}+552271446774906112 k^{16}+2178131904547051520 k^{15}+7493078411709911040 k^{14}+22590565275172499456 k^{13}+59866129347061276672 k^{12}+139652619111664189440 k^{11}+286758828600306565120 k^{10}+517503436878025588736 k^9+818170590709564833792 k^8+1127367936372700610560 k^7+1343674637314791833600 k^6+1370394879135205294080 k^5+1176574306768615636992 k^4+826263494053488230400 k^3+448130557296613785600 k^2+166223087687172096000 k+31141145704660992000)/31141145704660992000$}
	  } \\
\end{longtable}
%...................Table ends..............
%...................Table ends..............
\newpage
\captionsetup{width=15cm}
	\begin{longtable}{|c|c|}
		\hline
		\rowcolor{Gray}
		  $m$ & $\text{Tr}[\sigma^m]_{\ell=0}$ \\
\hline
$1$ & 
\scalebox{0.70}{
\parbox[t]{20cm}
	  {$(k^8+16 k^7+122 k^6+568 k^5+1784 k^4+3904 k^3+5728 k^2+4992 k+1920)/1920$}
	  } \\ \hline
$2$ & 
\scalebox{0.70}{
\parbox[t]{20cm}
	  {$(691 k^{16}+22112 k^{15}+345320 k^{14}+3477600 k^{13}+25226008 k^{12}+139643968 k^{11}+610312384 k^{10}+2150114560 k^9+6181580416 k^8+14591182848 k^7+28301545472 k^6+44929187840 k^5+57727068160 k^4+58427523072 k^3+43865063424 k^2+21543321600 k+5109350400)/5109350400$}
	  } \\ \hline
$3$ & 
\scalebox{0.70}{
\parbox[t]{20cm}
	  {$(3202291 k^{24}+153709968 k^{23}+3630145302 k^{22}+56023401544 k^{21}+633485614440 k^{20}+5581054788288 k^{19}+39787577367200 k^{18}+235384909886592 k^{17}+1176103128596352 k^{16}+5025774318260224 k^{15}+18533800918791168 k^{14}+59358067505049600 k^{13}+165791833035075584 k^{12}+404785500205056000 k^{11}+864437334327607296 k^{10}+1612943773102440448 k^9+2621945888635355136 k^8+3694827960098881536 k^7+4480012998151634944 k^6+4621620650262921216 k^5+3982518982525059072 k^4+2773034434490793984 k^3+1465096561848483840 k^2+518640787626393600 k+91055981592576000)/91055981592576000$}
	  } \\
\end{longtable}
%...................Table ends..............
%...................Table ends..............

%***********************Acknowledgements ends above***********************
%***********************Bibliography starts below*******************
\bibliographystyle{JHEP}
\bibliography{EELinks} 
%***********************Bibliography ends below*******************

\end{document}